%% file: main_TWC_revised.tex
\documentclass[journal, 10pt, twocolumns]{IEEEtran}

\usepackage{cite}
\usepackage{amsmath,amssymb,amsfonts,amsthm, empheq}
\usepackage{algorithm}
\usepackage{algorithmic}
\usepackage{setspace}
\usepackage{graphicx}
\usepackage{textcomp}
\usepackage{xcolor}
\usepackage{mathtools}
\usepackage{graphicx}
\usepackage{booktabs}
\usepackage{makecell}
\usepackage[font={small}]{caption}
\usepackage{subcaption}
\usepackage{cleveref}
\usepackage{optidef}
\usepackage{enumitem}
\usepackage{cite}
\usepackage{breqn}
\usepackage{mathrsfs}
\usepackage{accents}
\usepackage{acronym}
\usepackage{multirow}
\usepackage{soul}
\usepackage{bm}
\usepackage{url}
\usepackage{wrapfig}
\usepackage{todonotes}
\usepackage{verbatim}
\usepackage{pdfpages}
\usepackage{siunitx}
\usepackage{tikz}
\usepackage{adjustbox}

\usepackage{pgfplots}
\pgfplotsset{compat=1.18}
\usepgfplotslibrary{statistics}
\usepackage{tikz}
\usepackage{pgfplotstable}
\usepackage{xcolor}

\usetikzlibrary{shapes, shapes.misc, arrows, positioning, calc, backgrounds, angles, quotes, patterns, pgfplots.groupplots, arrows.meta, decorations.markings}

\newcommand{\algoname}{MOSAIC}

\DeclareMathOperator*{\argmin}{arg\,min}
\DeclareMathOperator*{\argmax}{arg\,max}

\graphicspath{{./figures/}}
\input{acronyms.tex}

\newcommand{\eq}[1]{Eq.~\eqref{#1}}
\newcommand{\eqs}[2]{Eqs.~(\ref{#1}-\ref{#2})}
\newcommand{\fig}[1]{Fig.~\ref{#1}}
\newcommand{\tab}[1]{Tab.~\ref{#1}}
\newcommand{\secref}[1]{Section~\ref{#1}}

\newcommand{\rev}[1]{{\color{blue}#1}}
\renewcommand{\rev}[1]{{\color{black}#1}}

\newcommand{\mytexttilde}{{\raise.17ex\hbox{$\scriptstyle\mathtt{\sim}$}}}

\IEEEaftertitletext{\vspace{-2.5\baselineskip}}

\newcounter{remark}[section]

\begin{document}
\bstctlcite{IEEEexample:BSTcontrol}

\title{Hierarchical Coherent Imaging of Composite Anisotropic Moving Targets in ISAC}

\author{Jacopo Pegoraro,~\IEEEmembership{Member,~IEEE}, Dario Tagliaferri,~\IEEEmembership{Member,~IEEE}, Joerg Widmer,~\IEEEmembership{Fellow,~IEEE}
\thanks{Jacopo Pegoraro was with the Department of Information Engineering, University of Padova, Italy. 
He is now with the Microwave Sensing, Signals and Systems (MS3) group,
Delft University of Technology, 2628 CD Delft, The Netherlands (email: J.Pegoraro@tudelft.nl).
Dario Tagliaferri is with the Department of Electronics, Information, and Bioengineering, Politecnico di Milano, Italy. Joerg Widmer is with IMDEA Networks, Madrid, Spain.

This work was partially supported by the Smart Networks and Services Joint Undertaking (SNS JU) under the European Union's Horizon Europe research and innovation programme, project MultiX (Grant Agreement No 101192521).
}
}

\maketitle

\begin{abstract}
In \ac{isac} networks, distributed devices can cooperate to produce radio \textit{images} of the surrounding environment by exploiting phase-coherent signal processing.
However, existing imaging methods are not well-suited for \textit{composite moving targets} with multiple independently moving extended parts. 
This is due to simplistic isotropic scattering models and the lack of methods to compensate for distinct Doppler shifts from each component, which leads to image defocusing.
We propose \algoname{}, the first hierarchical imaging method for composite moving targets using distributed \acp{ue} and a single \ac{isac} \ac{bs}. 
\algoname{} generates high-resolution images of each target part and estimates its velocity vector. 
\textit{Coherent} imaging is performed within selected clusters of \acp{ue} observing a locally isotropic scattering from each part, while cluster-specific images are combined \textit{non-coherently} across wide angles to improve the reconstruction. 
To mitigate Doppler-induced defocusing, Doppler components are pre-compensated before coherent imaging, turning a limitation into an additional means of resolving multiple target parts. 
This also enables low-complexity velocity estimation by associating Doppler frequencies across \acp{ue}. 
\rev{Simulations show a 35\% to 45\% improvement in image quality compared to existing methods, in terms of Wasserstein distance, and dm/s-level velocity estimation accuracy.}
\end{abstract}

\begin{IEEEkeywords}
Integrated Sensing and Communication, coherent imaging, distributed sensing, anisotropy, synchronization.
\end{IEEEkeywords}

\section{Introduction}

\acf{isac} is a new paradigm in which wireless communication systems are augmented with sensing functionalities to perceive the surrounding environment~\cite{10188491}. 
Recently, the research interest has shifted from single \ac{isac} terminals to \ac{disac}, where networks of multiple devices cooperate to improve sensing accuracy, resolution, and overall reliability~\cite{han2026mimo}. 
\ac{disac} networks can concurrently leverage monostatic and bistatic sensing modalities, depending on whether the sensing transmitter and receiver are co-located or not, as well as generic multistatic configurations~\cite{Masouros2024_networked_ISAC_interference}.

However, most of the existing literature does not fully exploit the potential of \ac{disac}, focusing on detecting targets and estimating their location and velocity. 
Inspired by airborne \ac{sar} systems~\cite{moreira2013tutorial}, some works have proposed to integrate \textit{radio imaging} into \ac{disac} networks~\cite{IIAC_lightweight,
negosanti2026ofdm}, which consists of retrieving a 2D or 3D map of the complex reflectivity of the environment.
From such maps (or \textit{images}), one cannot only detect and localize targets, but also identify their shape and size, enabling fine-grained sensing applications. 
While target localization and velocity estimation focus on maximizing the sensing \textit{accuracy}, the radio imaging performance is mainly determined by the image \textit{resolution}.
The latter depends on the sensing signal bandwidth, but also on the geometry of the \ac{disac} network~\cite{manzoni2024wavefield}.
Improving the image resolution requires processing sensing signals at multiple \ac{disac} devices in a \textit{phase-coherent} way, typically using over-the-air synchronization methods~\cite{pegoraro2024jump, tagliaferri2024cooperative}. 

\begin{figure}
    \centering
\includegraphics[width=\linewidth]{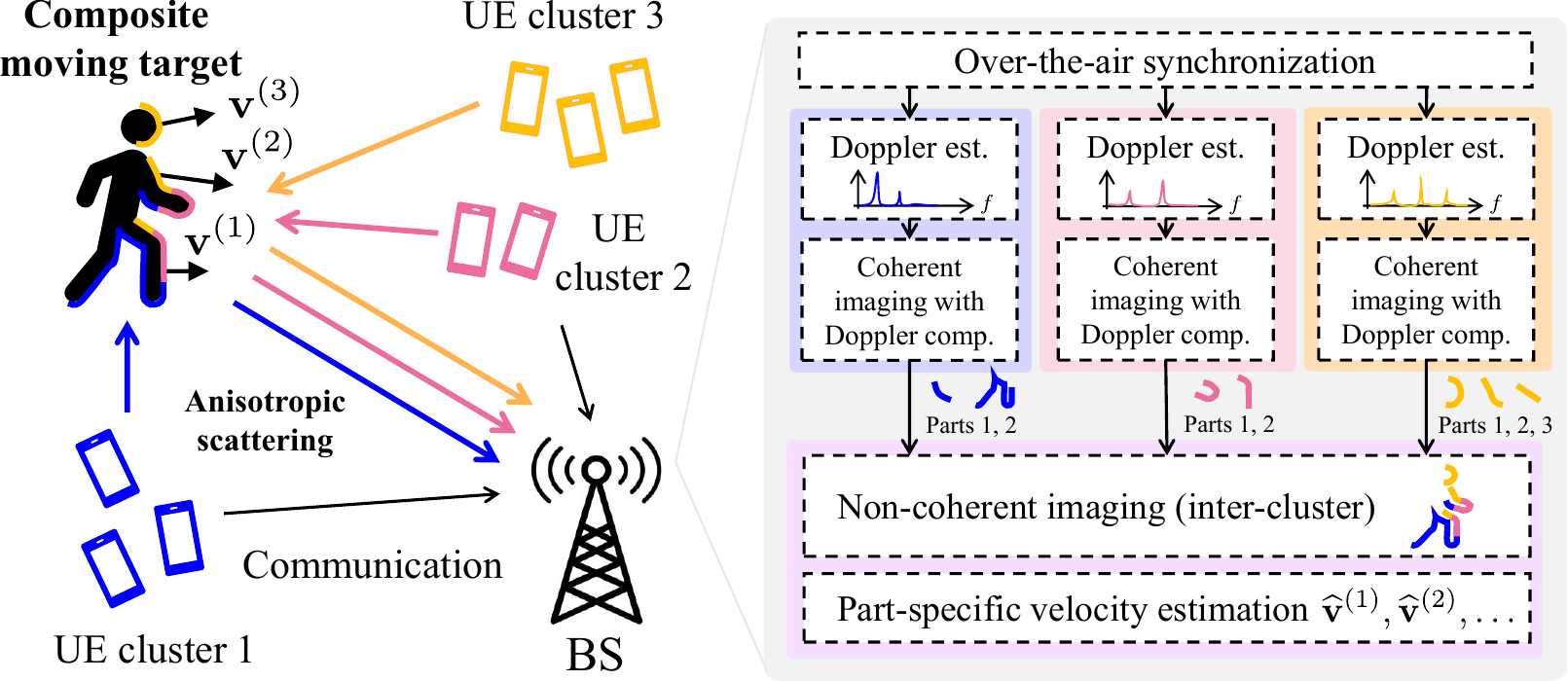}
    \caption{Overview of the proposed imaging system (\algoname{}).}
    \label{fig:system_model}
\end{figure}

Existing works on imaging in \ac{isac}, such 
as~\cite{IIAC_lightweight,negosanti2026ofdm,zhi2025nearfield_distributed,Gui_ISAC_ISAR}, 
rest on two overly restrictive assumptions about the sensing targets.
First, they model scattering as uncorrelated and isotropic at each image 
pixel~\cite{zhi2025nearfield_distributed}, or assume the presence of multiple 
independent point scatterers that radiate isotropically across a given range of 
incident signal angles~\cite{negosanti2026ofdm}.
These models are well-suited to \ac{sar} systems that observe targets from large distances and over relatively small angular apertures.
They are, however, not applicable to \ac{disac} networks, where observation angles 
can vary substantially due to the distributed placement of nodes, and where pixel 
sizes can reach centimeter scale, giving rise to spatially correlated scattering.
Second, most of these works assume static targets~\cite{IIAC_lightweight,zhi2025nearfield_distributed}, thereby neglecting the Doppler-induced 
phase shifts caused by motion.
This precludes their use in dynamic sensing scenarios, where targets of 
interest undergo complex motion, often comprising independently moving components.

Anisotropy and target motion have been \textit{separately} addressed in the \ac{sar} and \ac{isar} 
literature~\cite{Moses2004WideAngleSAR,MartorellaISAR_2014,Li2018Motion,Salvetti2019Multiview}.
Works that account for target anisotropy~\cite{Moses2004WideAngleSAR} restrict 
themselves to static scenes, while works on moving-target imaging either 
approximate targets as isotropic to enable coherent multistatic 
imaging~\cite{MartorellaISAR_2014,Li2018Motion,Salvetti2019Multiview} or resort 
to non-coherent imaging~\cite{Testa2023Decentralized}.
However, in many terrestrial \ac{disac} applications, such as human body 
imaging~\cite{Nuria_Proc_IEEE_2024}, the target consists of multiple extended 
parts, each moving with a distinct velocity in an articulated or 
non-articulated fashion.
The resulting scattering is anisotropic at the level of each target part and 
exhibits a \textit{multi-modal} Doppler spectrum that varies across nodes of the 
\ac{disac} network.
Such realistic targets have not been addressed in the \ac{isac} or \ac{sar} 
literature, and are the focus of this work.

In this paper, we design and validate \algoname{} - Moving Object Sensing under Anisotropy by Imaging with Coherent and non-coherent fusion, whose block diagram is shown in~\fig{fig:system_model}.
\algoname{} is the first algorithm to perform high-resolution imaging of composite anisotropic moving targets from distributed asynchronous \ac{isac} networks, leveraging \ac{ofdm} pilot preambles (e.g., DMRS in 3GPP~\cite{electronics13112028}) received at a \ac{bs} from multiple \acp{ue} without dedicated sensing signals. 
To handle the anisotropic scattering of composite targets, we derive an analytical anisotropy model as a function of the observation angle.
This is used to group \acp{ue} sharing similar viewpoints into clusters that perform \textit{coherent} imaging. 
The resulting images are then combined \textit{non-coherently} across clusters, making \algoname{} the first hierarchical imaging algorithm for \ac{disac}. 
At the core of \algoname{} lies a modified backprojection algorithm~\cite{manzoni2024wavefield} that pre-compensates the Doppler shift observed by each \ac{ue} cluster and integrates coherent images across slow-time, yielding separate \textit{Doppler-specific} images of each moving part. 
This mitigates clutter and inter-part interference while producing a richer output than existing methods, which collapse all contributions into a single Doppler-compensated image~\cite{noviello2015focused}. 
Finally, by exploiting the cluster geometry, \algoname{} estimates the velocity vector of each target part via an association algorithm that links Doppler shifts across clusters. 
This is an additional feature that differentiates it from existing methods~\cite{zhi2025nearfield_distributed,tagliaferri2024cooperative,negosanti2026ofdm}, achieving orders of magnitude lower complexity than exhaustive velocity search approaches~\cite{moving_target_sar,MartorellaISAR_2007}. 
Extensive numerical simulations show that \algoname{} improves image quality by 35\% to 45\% compared to state-of-the-art methods~\cite{noviello2015focused} in terms of Wasserstein distance and $2$ to $3$ times in terms of contrast, while estimating each part's velocity vector with dm/s-level accuracy.

The paper is organized as follows.
\secref{sec:related_work} introduces the related work.
\secref{sec:system_model} outlines the signals and system model, including the composite target scattering model and the contribution of the Doppler shift of the moving target parts. 
\secref{sec:preprocessing} details the signal preprocessing steps.

\secref{sec:ue-clustering} presents the analysis of the target anisotropy and derives the \ac{ue} clustering rule.
\secref{sec:imaging_moving} presents \algoname{}, while \secref{sec:results} describes the numerical results from our simulations.
\secref{sec:conclusion} concludes the paper. 

\section{Related work}\label{sec:related_work}

\textit{Imaging in \ac{isac}.} 
Very few attempts have been made to integrate imaging within communication systems.
Works \cite{ManzoniCOSMIC,IIAC_lightweight,bazzi2025isac_imaging_csi_raytracing,Alkhateeb23_imaging_comm,tornielli2025enabling_nlos_imaging} focus on single-device imaging, addressing waveform design~\cite{ManzoniCOSMIC}, lightweight linear image formation techniques~\cite{IIAC_lightweight}, imaging using ray tracing~\cite{bazzi2025isac_imaging_csi_raytracing}, and imaging using beam sweeping by a \ac{bs} in downlink~\cite{Alkhateeb23_imaging_comm,tornielli2025enabling_nlos_imaging}. 
The extension to \ac{disac} systems, exploiting the aperture given by multiple \ac{isac} devices, has been covered in~\cite{IIAC_3D_imaging,tong2022environment_sensing_occlusion,zhi2025nearfield_distributed,li2024networked}.
The work in \cite{zhi2025nearfield_distributed} details a \ac{disac} system to enable multistatic phase-coherent imaging of the environment based on compressed sensing.
Targets are modeled as a collection of infinitesimal scattering points with independent and isotropic reflectivities.
Works \cite{IIAC_3D_imaging,tong2022environment_sensing_occlusion, li2024networked} propose to exploit the signals exchanged between distributed \acp{ue} and one or more \acp{bs}, cast as a non-linear computational imaging problem. 

The previous works share one or more of the following three limitations: \textit{(i)} they model targets as the collection of isotropic point scatterers, whose reflectivity does not change with the observation angle, \textit{(ii)} they assume the scattering response from each pixel or point scatterer is independent from the others~\cite{zhi2025nearfield_distributed}, \textit{(iii)} they only consider \textit{static} targets. 
In typical \ac{disac} systems, however, \textit{(i)} and \textit{(ii)} are unrealistic, since the high spreading of devices in space with respect to the target makes the scattering response anisotropic and correlated.
In this paper, we model each target part as a collection of \textit{patches}, smaller than the signal wavelength, that produce \textit{angle-dependent} scattering responses 
affecting multiple image pixels, yielding a physics-consistent scattering model compliant with the Huygens principle~\cite{knott2004radar}. 
\algoname{} handles this by combining coherent and non-coherent imaging, deriving an angular region within which \ac{isac} devices see a coherent scattering response.

Lastly, assuming static targets, \textit{(iii)}, significantly limits the applicability of existing methods to dynamic and realistic conditions.
High-resolution imaging of \textit{moving targets} holds great promise for \ac{isac}, as dynamic objects are the main targets of interest in many 6G applications, e.g., human gesture recognition, crowd sensing, among others~\cite{Nuria_Proc_IEEE_2024}. 
In the \ac{isac}-based imaging context, the existing literature on moving target imaging is scarce and focused on restrictive target models (e.g., rigid targets) and application scenarios~\cite{Gui_ISAC_ISAR,negosanti2026ofdm}, which we discuss in the last part of this section. 

\textit{Imaging of Anisotropic Targets.}
Owing to the typical geometry of air- or space-borne \ac{sar} systems, the scattering from objects of interest is typically modeled as a set of \textit{isotropic} point scattering centers.
However, when the observation aperture is too wide, the isotropic scattering assumption does not hold~\cite{sambon2025electromagnetic, Moses2004WideAngleSAR,Ash2014WideAngleSARModels}, and coherent imaging produces corrupted reconstructions of the reflectivity map.
Therefore, some works propose to non-coherently fuse sub-aperture images~\cite{CetinKarl2001FeatureEnhancedSAR}, while others explicitly characterize the angular-dependent scattering of anisotropic targets in a sparsity-based computational imaging framework~\cite{Wang2021EnhancedAnisotropicSAR}.
In both cases, the model considers attributed scattering centers parameterized via geometrical diffraction theory~\cite{gerry1999parametric,PotterMoses1997AttributedScatteringCenters}, whose positions are assumed \textit{independent of the observation angle}.
While this holds for typical \ac{sar} geometries, it fails in \ac{disac} terrestrial networks, where devices observe the target from very different angles, leading to \textit{angle-dependent} (migrating) locations of the main scattering centers.
Although the migration of scattering centers has been considered in~\cite{Varshney2006JointImageFormationAnisotropySAR}, the method presented in this work considers static targets, has high computational complexity, and requires careful initialization of the locations of the scattering centers.

\textit{Imaging of Moving Targets.}
Imaging of moving targets has been extensively studied in \ac{sar} and \ac{isar}~\cite{cheney2008imaging,berizzi2001high,noviello2015focused,MartorellaISAR_2014,Li2018Motion,Salvetti2019Multiview,Testa2023Decentralized}, and was recently considered in \ac{isac}~\cite{Gui_ISAC_ISAR,negosanti2026ofdm}.
In \ac{isar}, the synthetic aperture is formed by exploiting target motion as a source of diversity, requiring estimation of motion parameters to compensate for Doppler-induced phase shifts.
The optimal approach is to perform backprojection in four dimensions, including 2D target parts locations and 2D velocity vectors~\cite{cheney2008imaging}, but its complexity is only suited to offline processing.
\cite{berizzi2001high,noviello2015focused} reduce complexity by parameterizing the Doppler phase using the Doppler centroid and rate.
Extensions to multistatic \ac{isar} exist~\cite{MartorellaISAR_2014,Li2018Motion,Salvetti2019Multiview,Testa2023Decentralized}, applying similar Doppler compensation approaches.
The main limitation of these works in the \ac{disac} context is that their Doppler spectrum model is designed for extended \textit{rigid} targets such as ships and aircraft, for which the centroid and rate parametrization is sufficiently accurate.
However, for composite targets with multiple independently moving parts (e.g., human beings, industrial machinery, and robots), the Doppler spectrum is \textit{multimodal}, exhibiting multiple peaks corresponding to the different parts, and the centroid/rate parametrization cannot fully compensate for the Doppler-induced phase shift.
Conversely, \algoname{} applies a target part-specific Doppler compensation, obtaining \textit{separate} images of each part.

\section{System model}\label{sec:system_model}

We consider an \ac{isac} network formed by $M$ \acp{ue} and one \ac{bs}.
In a 2D global reference frame, each \ac{ue} is located at $\mathbf{x}_m \in \mathbb{R}^{2 \times 1}$, $m=1,...,M$, while the \ac{bs} is located in $\mathbf{x}_{\rm bs}\in \mathbb{R}^{2 \times 1}$. 
We consider the \acp{ue} and \ac{bs} locations to be known.
In practice the position of \acp{ue} is estimated by the \ac{bs}, with an accuracy that depends on several factors, mainly on the on-board positioning sensors of the \acp{ue}. 
This aspect is further discussed in the next sections.

Each \ac{ue}-\ac{bs} pair is not only a communication link, but also a bistatic \ac{isac} measurement channel. 
The goal of the \ac{isac} network is to exploit the \ac{ul} communication phase to generate a high-resolution image of a \ac{roi} in which a composite moving target is present. 
The moving target is an extended body, composed of multiple parts, each in motion with a different velocity, as described in~\secref{sec:ch_modeling}. 
The \ac{bs} is equipped with a \ac{ula} of $L$ antennas, while each \ac{ue} has a single antenna.

\subsection{Transmitted signal and \ac{isac} frame structure}\label{sec:Tx_signal}

The carrier frequency of operation is denoted by $f_0$, and the total bandwidth is $B$, shared by all \acp{ue}. 
The \ac{isac} network employs an \ac{ul} pilot \ac{ofdm} signal with $S$ subcarriers and $K$ \ac{ofdm} symbols. 
We stress that pilot signals are already transmitted as part of the communications network operation, and are \textit{reused} by \algoname{}.
Hence, our approach does not introduce additional overhead for sensing, and it does not affect the communication functionality.

The subcarrier spacing is \mbox{$\Delta f = B/S$}.
The time between \ac{ofdm} symbols used for sensing is $T$, analogous to the \ac{pri} in radar, yielding a total duration of the \ac{ul} burst of \mbox{$T_{\rm tot} = K T$}. 
In the following, we assume that each \ac{ue} operates over the full bandwidth $B$.
This can be achieved, e.g., by interleaving the \ac{ofdm} subcarriers used by each \ac{ue} in the frequency domain and then interpolating over the full bandwidth during the channel estimation phase.

Using \ac{isac} terminology, we distinguish variable $t\in [0,T]$, representing \textit{fast} time, from $kT$, representing the time of reception of the $k$-th \ac{ofdm} symbol.
This can be considered as the \textit{slow-time} variable, i.e., the granularity at which the targets' motion manifests on the received (Rx) signal at the \ac{bs}~\cite{richards2010principles}. 
We assume the transmitted (Tx) signal has unit power to simplify the notation.
The Tx baseband signal at the $k$-th \ac{ofdm} symbol is given by $g_m(t,kT)=\sum_{s =1}^S G_{k,s} e^{j 2 \pi s \Delta f t }$.
The upconverted Tx signal is then $g^{\rm RF}_m(t,kT) =g_m(t,kT)e^{j 2 \pi f_0 t}$.

\subsection{Channel model with composite anisotropic targets}\label{sec:ch_modeling}

We model the response of the environment within the \ac{roi} as the anisotropic scattering from the surface of a composite extended target. 
The target includes $N$ different parts.
Each part $n$ is an extended rigid body with purely translational motion with velocity vector~\mbox{$\mathbf{v}_n\in\mathbb{R}^{2 \times 1}$}. 
This assumption holds for composite targets such as machines or a human body over short time spans. 
Extending our method to fast rotations of different target parts and to a 3D spatial representation is an interesting direction for future research.

We adopt a physical optics-based model~\cite{knott2004radar} where the scattering from the surface is modeled as the superposition of reflections from sufficiently small square \textit{patches} with area $A$, each exhibiting an anisotropic \ac{rcs}. 
This approach well represents surface scattering from targets of arbitrary shape, where penetration is negligible.
% ~\cite{garcia2022cramer_rao_extended_vehicular}. 
The $n$-th extended target is composed of $P_n$ patches, with $n=1,..,N$, each located in $\mathbf{x}_{n,p}$ and oriented according to the unit vector $\mathbf{t}_{n,p}$, parallel to the surface of the patch.
The reflectivity and the scattering amplitude of each patch are modeled as~\cite{richards2005fundamentals}
\begin{align}
   \varrho_{n,p,m} =& V_{n,p,m} \sqrt{\Gamma_{n,p,m}} e^{j \varphi_{n,p}}, \label{eq:rho}\\
   \zeta_{n,p,m} = &\frac{\lambda_0}{(4 \pi)^{\frac{3}{2}} \| \mathbf{x}_{\rm bs} - \mathbf{x}_{n,p}\| \| \mathbf{x}_{n,p} - \mathbf{x}_{m}\|}\label{eq:zeta}
\end{align}
where $V_{n,p,m}$ is the visibility function of the patch in $\mathbf{x}_{n,p}$, that equals 1 if the patch is visible from both the $m$-th \ac{ue} and the \ac{bs}, and 0 otherwise, $\Gamma_{n,p,m}$ is the \ac{rcs} of the $p$-th patch of the $n$-th target observed by the $m$-th \ac{ue}, and $\varphi_{n,p}$ is the intrinsic scattering phase of the $(p,n)$-th patch. 

\rev{Define the unit vectors pointing from the patch to the \ac{ue} and from the patch to the \ac{bs}, respectively,
\begin{align}\label{eq:unit-vecs}
    \mathbf{u}_{n,p,m} = \frac{\mathbf{x}_{m} - \mathbf{x}_{n,p} }{\| \mathbf{x}_{n,p} - \mathbf{x}_{m}\|}, \quad \mathbf{u}_{n,p,{\rm bs}} = \frac{\mathbf{x}_{\rm bs} - \mathbf{x}_{n,p} }{\| \mathbf{x}_{n,p}- \mathbf{x}_{\rm bs}\|}.
\end{align}
The \ac{rcs} of the single patch is given by~\cite[Eq. (5.14)]{knott2004radar}
\begin{align}\label{eq:RCS}
    \Gamma_{n,p,m} & = \frac{4 \pi A^2}{\lambda_0^2}\mathrm{sinc}^2\left( \frac{\sqrt{A}}{\lambda_0} \mathbf{t}^{\mathsf{T}}_{n,p}\left(\mathbf{u}_{n,p,m} +
\mathbf{u}_{n,p,{\rm bs}}\right)\right), 
\end{align}
where $\mathrm{sinc}(x) = \sin(\pi x)/\pi x$.}
The \ac{rcs} is direction-dependent, with maximum intensity when the vector along the bisector of the bistatic angle formed by the \ac{ue}, the patch, and the \ac{bs} aligns with the normal vector of the patch.

The above model is valid for \textit{(i)}~targets whose size is larger than the wavelength, which is the case for the considered target sizes and cm-wave carrier frequencies ($3$ to $26$~GHz), and \textit{(ii)}~\textit{single-bounce scattering}, i.e., when the signal emitted by the \ac{ue} undergoes \textit{at most} one scattering contribution from each patch, \textit{(iii)} low or absent surface roughness. 
The patch-based model generalizes recent approaches in the \ac{isac} literature that model extended targets as collections of a few point scatterers, with angle-dependent scattering coefficients~\cite{negosanti2026ofdm}, and is more realistic than considering uncorrelated scattering from each pixel~\cite{zhi2025nearfield_distributed}.

In addition to the scattering from the \ac{roi}, we also consider the presence of 
a \ac{los} propagation path from the $m$-th \ac{ue} to the \ac{bs}, which we identify with index $0$.
The \ac{los} path loss is $\zeta_{0,m} =\lambda_0/(4 \pi D_{0,m})$
where \mbox{$D_{0,m} =  \| \mathbf{x}_{\rm bs} -\mathbf{x}_{m} \|$} is the \ac{los} distance. 
The continuous-time \ac{cir} model for the $m$-th \ac{ue} at the \ac{bs} is the \mbox{$L\times 1$} vector
\begin{multline}\label{eq:channel-simple}
     \mathbf{h}_m(t,kT)
    = \mathbf{a}(\psi_{0,m}) \zeta_{0,m} \delta(t-\tau_{0,m}) + \\
     + \sum_{n=1}^N \sum_{p=1}^{P_n} \mathbf{a}(\psi_{n,p,m}) \zeta_{n,p,m}\, \varrho_{n,p,m} \, \delta(t-\tau_{n,p,m}(kT)),
\end{multline}
where $\psi_{0,m}$ is the \ac{los} \ac{aoa} at the \ac{bs}, from \ac{ue} $m$, $\psi_{n,p,m}$ is the \ac{aoa} from patch $n,p$ from \ac{ue} $m$, $\mathbf{a}(\psi_{0,m}), \mathbf{a}(\psi_{n,p,m}) \in \mathbb{C}^{L \times 1}$ are the \ac{bs} \ac{ula} steering vectors for angles $\psi_{0,m}, \psi_{n,p,m}$, and $\tau_{0,m}, \tau_{n,p,m}(kT)$ are the \ac{tof} values of the \ac{los} and the path from patch $(n,p)$ from \ac{ue} $m$, respectively.
The time-varying \ac{tof} for patch $(n,p)$ is 
\begin{equation}\label{eq:TOF}
    \tau_{n,p,m}(kT) = \frac{\| \mathbf{x}_{n,p} +\mathbf{v}_n k T -  \mathbf{x}_{m} \| + \|\mathbf{x}_{\rm bs} - \mathbf{x}_{n,p} - \mathbf{v}_n k T \| }{c},
\end{equation}
which is a function of the translational velocity of the target.
The \ac{tof} for the \ac{los} is
$\tau_{0,m} = D_{0,m}/c$.

\rev{In this work, we do not explicitly model clutter contributions from scatterers other than the target of interest. Indeed, the imaging procedure is restricted to the \ac{roi}, so that clutter sources located outside it are spatially filtered by the backprojection process, except for possible sidelobe leakage into the \ac{roi} (see \secref{sec:bp}). 
All reflectivity components from within the \ac{roi} are considered to be part of the target of interest regardless of whether they are static or moving.}

\begin{figure*}[!t]
    \begin{align}
        \mathbf{y}_{m}(t,kT) 
       &=  \mathbf{a}(\psi_{0,m}) \zeta_{0,m} \, g_m(t-\tau_{0,m} + \alpha_m(kT), kT) e^{-j2\pi f_0\tau_{0,m}}e^{j\Omega_{m}(t,kT)}\label{eq:Rx_signal}\\
        & +  \sum_{n=1}^N \sum_{p=1}^{P_n} \mathbf{a}(\psi_{n,p,m}) \zeta_{n,p,m} \varrho_{n,p,m} g_m(t-\tau_{n,p,m}(kT) + \alpha_m(kT) ,kT) e^{-j2 \pi f_0 \tau_{n,p,m}(kT)}e^{j\Omega_{m}(t,kT)}+\mathbf{w}_{m}(t,kT).\nonumber
    \end{align}
\hrulefill
\end{figure*}

\subsection{Received signal}\label{sec:Rx_signal}

The baseband Rx signal at the \ac{bs} is the convolution of the passband Tx signal $g^{\rm RF}_m(t,kT)$ with the \ac{cir} in~\eq{eq:channel-simple}, followed by downconversion.
Due to the asynchrony between the \ac{ue} and \ac{bs} \acp{lo} that generate the carrier signal, the Rx signal from \ac{ue} $m$ is affected by relative, slow-time varying \ac{to}, $\alpha_m(kT)$, \ac{cfo} (normalized to the carrier frequency), $\vartheta_m(kT)$, and \ac{pn}, $\xi_{m}(kT)$.
Combined, they introduce a \ac{po} on the Rx signal, equal to
\begin{equation}\label{eq:po}
    \Omega_m(t, kT) = 2\pi f_0\left[\alpha_m(kT)\hspace{-0.05cm} + \hspace{-0.05cm}\vartheta_m(kT)(t\hspace{-0.05cm}-\hspace{-0.05cm}kT)\right] +\xi_{m}(kT).
\end{equation}
The \ac{po} must be compensated for before combining signals from different \acp{ue} to produce coherent images of the target.

The $L \times 1$ Rx signal vector from the $m$-th \ac{ue}, after demodulation, is reported in \eq{eq:Rx_signal}, where $\mathbf{w}_{m}(t,kT)\sim\mathcal{CN}(\mathbf{0}, \sigma_w^2\mathbf{I})$ is a complex Gaussian noise vector and $\mathbf{I}$ the identity matrix.

\rev{The first term is due to the LoS channel, and the second is the scattering by the composite extended target in the \ac{roi}.}
The baseband signal $g_m(t,kT)$ is affected by the propagation delay of each path: $\tau_{0, m}$ for the \ac{los}, and $\tau_{n,p,m}(kT)$ for the patches composing the target parts.
Moreover, each path is further delayed by the \ac{to} $\alpha_m(kT)$, which is common to all paths.
Time distortions due to \ac{cfo} and \ac{po} are negligible in the baseband signal~\cite{tagliaferri2024cooperative}.
 
Assuming no range migration occurs, as is typical in the considered case study, we can rewrite the phase $-2\pi f_0\tau_{n,p,m}(kT)$ in~\eq{eq:Rx_signal} as the sum of a static component, $-2\pi f_0\tau_{n,p,m}$, due to the location of the $(n,p)$-th patch at the beginning of the slow-time processing interval, and a Doppler component $2\pi f^{\rm D}_{n,p,m} kT$ where the Doppler shift of $(n,p)$-th patch seen by \ac{ue} $m$ is obtained
% using~\eq{eq:TOF} and~\eq{eq:unit-vecs}
as~\cite{richards2010principles}
\begin{equation}\label{eq:doppler-freq}
    f^{\rm D}_{n,p,m} = \frac{f_0}{c} \left(\mathbf{u}_{n,p,m} +
\mathbf{u}_{n,p,{\rm bs}}\right)^{\mathsf{T}} \mathbf{v}_n.
\end{equation}
The Doppler shift is a function of velocity $\mathbf{v}_n$, and of the specific \ac{ue} and patch, hence different \ac{ue}-\ac{bs} pairs observe different Doppler shifts associated with the same target part.

\section{Preprocessing of the received signal}\label{sec:preprocessing}

The Rx signal in \eq{eq:Rx_signal} is pre-processed to extract the \ac{cir} at each \ac{ofdm} symbol. 
This is achieved by the following steps: \textit{(i)}~the Rx signal is first transformed to the frequency domain; 
\textit{(ii)}~for each \ac{ofdm} symbol, the \ac{cfr} is estimated at the pilot subcarrier indices using a conventional least squares approach~\cite{Sturm2011} \textit{(iii)}~\ac{cfr} is interpolated to obtain the complete \ac{cfr} over the $S$ subcarriers and $K$ \ac{ofdm} symbols~\cite{BarnetoFullDuplex}, \textit{(iv)}~the \ac{cir} is obtained as the inverse discrete Fourier transform of the interpolated \ac{cfr}. 
A common assumption in communication and \ac{isac} systems is that the non-normalized \ac{cfo} is smaller than the subcarrier spacing, i.e., $|f_0 \vartheta_{m}(kT)| \ll \Delta f$, so it does not compromise the estimation of the \ac{cir}. 

The estimated \ac{cir} for \ac{ue} $m$ is denoted by $\widehat{\mathbf{h}}_{m}(t, kT)$
and its expression is omitted for conciseness.
Its structure is identical to that of the Rx signal in \eq{eq:Rx_signal}, but the baseband signals $g_m$ are replaced by 
delayed sinc functions whose mainlobe width is inversely proportional to the system bandwidth.
We directly give the expression of the \ac{cir} after synchronization in~\eq{eq:CIR_corrected}, which is obtained after the processing steps described in the following section.

\subsection{Over-the-air multistatic synchronization}\label{sec:sync}

\rev{The estimated \ac{cir} is affected by clock synchronization errors that prevent the correct formation of images~\cite{tagliaferri2024cooperative}. 
Using the \ac{los} path between each \ac{ue} and the \ac{bs}, we adapt our previous work in~\cite{pegoraro2024jump} to synchronize each \ac{ue} with the \ac{bs}.
The idea behind such method is to leverage the \ac{los} path to estimate the \ac{to} and \ac{po}, since these affect all paths in the same way. 
Once estimated, the \ac{to} and \ac{po} are removed from the reflected paths from the \ac{roi}.

Before proceeding to the actual synchronization procedure, we apply digital beamforming at the \ac{bs} (Rx) to enhance the \ac{los} path, multiplying the  \ac{cir} estimated by \ac{ue} $m$, $\widehat{\mathbf{h}}_{m}(t,kT)$, by the steering vector $\mathbf{a}(\psi_{0,m})$, matched to the known \ac{aoa} of the \ac{los} path with \ac{ue} $m$, denoted by $\psi_{0,m}$.
Such \ac{aoa} is known since it only depends on the available locations of the \ac{bs} and considered \ac{ue}.
Hence, we obtain the scalar \ac{cir} $h_{m}(t,kT) = \mathbf{a}(\psi_{0,m})^{\mathsf{H}}\widehat{\mathbf{h}}_{m}(t,kT)$ where the \ac{snr} of the \ac{los} path has increased due to the beamforming operation.

From the scalar \ac{cir}, we estimate a delay correction term that we use to remove the effect of the \ac{to}.
The correction term is approximately equal to the difference of the~\ac{los} propagation delay and~\ac{to}, and is obtained as 
\begin{equation}\label{eq:to-est}
    \widetilde{\tau}_m(kT)
= \argmax_{\tau}
\lvert h_m(\tau, kT)\rvert^2 \approx \tau_{0, m} - \alpha_m(kT).
\end{equation}
We have assumed that the~\ac{los} signal is stronger than the scattered echoes and it is resolvable in time, so the effect of~\ac{nlos} paths is negligible in the estimation. 
The \acp{ue} that do not have a \ac{los} path to the \ac{bs} are ignored in the imaging process.
The proposed method can be readily modified to apply to cases where the \ac{los} path is not the strongest path detected by the \ac{bs}, although this is unlikely after the Rx beamforming operation.
To do so, one can replace \eq{eq:to-est} with a detection rule that selects the \textit{first} path that is sufficiently strong to reach the \ac{bs}, since the \ac{los} path has the shortest propagation distance.

To remove the \ac{po}, we estimate the phase of the \ac{cir} from the \ac{los} path as 
\begin{equation}
    \widetilde{\phi}_{m}(kT) = \angle h_{m}( \widetilde{\tau}_m(kT), kT)
    \approx -2\pi f_0 \tau_{0,m} + \Omega_m(t,kT),
\end{equation}
which contains the \ac{po} that affects the $m$-th \ac{ue}-\ac{bs} link, and can be used with the delay correction term $\widetilde{\tau}_m(kT)$ to obtain the corrected \ac{cir}. 

\ac{to} and \ac{po} are compensated for by counter-rotating the estimated \ac{cir} by the \ac{los} phase, $\widetilde{\phi}_{m}(kT)$, and aligning the time origin with the \ac{los} echo using $\widetilde{\tau}_m(kT)$.
This is achieved by computing
\begin{equation}\label{eq:topo-correction}
    \Delta \widetilde{\mathbf{h}}_{m}(t, kT) = \widehat{\mathbf{h}}_{m}(t+\widetilde{\tau}_{m} (kT), kT) e^{-j \widetilde{\phi}_{m}(kT)}.
\end{equation}

\eq{eq:topo-correction} approximately removes the delay shift introduced by the \ac{to} and cancels out the \ac{po}, yielding
\begin{multline}\label{eq:CIR_corrected}
\Delta \widetilde{\mathbf{h}}_{m}(t, \hspace{-0.05cm}kT)
    \approx \mathbf{a}(\psi_{0,m}) \zeta_{0,m} \, \mathrm{sinc}\left( B t\right)+ 
   %e^{-j 2\pi f_0\Delta\tau_{0,m} }
   \\
    + \sum_{n=1}^N \sum_{p=1}^{P_n} \mathbf{a}(\psi_{n,p,m}) \zeta_{n,p,m} \varrho_{n,p,m}  \mathrm{sinc}\left( B (t \hspace{-0.08cm}- \hspace{-0.08cm}\Delta\tau_{n,p,m}(kT)\right) \cdot \\
    \cdot e^{-j 2\pi f_0 \Delta\tau_{n,p,m}}e^{j2 \pi f^{\rm D}_{n,p,m} kT} +\widetilde{\mathbf{z}}_{m}(t,kT),
\end{multline}
where $\Delta\tau_{n,p,m}=\tau_{n,p,m}-\tau_{0,m}$ is the relative \ac{tof}, with respect to the \ac{los} one, 
and $\widetilde{\mathbf{z}}_{m}(t,kT)$ is the noise vector.
}

\subsection{Formation of low-resolution images}\label{sec:bp}
 
From the corrected \ac{cir} in~\eq{eq:CIR_corrected}, the \ac{bs} forms $M$ images, one for each \ac{ue}, using the backprojection method, which is a low-complexity linear image formation technique implementing a matched filter in space and time~\cite{manzoni2024wavefield}. 
Backprojection generates the complex map of the environment reflectivity over the \ac{roi}, which is discretized over a finite grid of pixels, each identified by the location of its center~$\mathbf{x}\in \mathbb{R}^{2}$. 

Let us consider a single pixel $\mathbf{x}$ in the~\ac{roi}. 
Backprojection compensates for \textit{(i)}~the propagation carrier phase and \textit{(ii)}~the antenna-dependent phase shift represented by the steering vectors $\mathbf{a}_m(\psi_{\mathbf{x}})$, where $\psi_{\mathbf{x}}$ is the \ac{aoa} at the \ac{bs} of a reflection from pixel $\mathbf{x}$.
The propagation carrier phase is defined as $- 2 \pi f_0 \Delta \tau_{m}(\mathbf{x})$, where $\Delta \tau_{m}(\mathbf{x})= \tau_{m}(\mathbf{x})-\tau_{0,m}$ is the relative \ac{tof} for pixel $\mathbf{x}$, and $\tau_{m}(\mathbf{x})$ is given by \eq{eq:TOF} by substituting $\mathbf{x}$ for $\mathbf{x}_{n,p}$ and \mbox{$\mathbf{v} = \mathbf{0}$}, since the target parts' velocities are unknown at this stage.
The image for \ac{ue} $m$ is formed as
\begin{multline}\label{eq:BP}
    I_m(\mathbf{x},kT) =    
 \mathbf{a}_{m}^{\mathsf{H}}(\psi_{\mathbf{x}}) \Delta \widetilde{\mathbf{h}}_{m}(\Delta \tau_{m}(\mathbf{x}), kT)e^{j 2 \pi f_0 \Delta \tau_{m}(\mathbf{x})} \\
 \approx \sum_{n=1}^N \sum_{p=1}^{P_n}  \zeta_{n,p,m} \varrho_{n,p,m} \chi_m\left(\mathbf{x}-\mathbf{x}_{n,p,m}\right)e^{j 2 \pi f_{n,p,m}^{\rm D} kT },
\end{multline}
where $\chi_m(\mathbf{x})$ is the \ac{psf} representing the imaging system (\ac{ue} and \ac{bs}) response to a point scatterer whose reflectivity is a delta function.
Note that $I_m(\mathbf{x},kT)$ is affected by a noise term which is a function of $\widetilde{\mathbf{z}}_{m}(t,kT)$, but we omit it in~\eq{eq:BP} and in the following equations involving images for conciseness.
\eq{eq:BP} shows that the resulting image is the convolution of the ideal image, $I^{\rm ideal}_m(\mathbf{x}) = \sum_{n=1}^N \sum_{p=1}^{P_n} \varrho_{n,p,m} \delta(\mathbf{x}-\mathbf{x}_{n,p})$ with the \ac{psf} affected by Doppler shift.
The \ac{psf} defines the \textit{effective resolution} of single-\ac{ue} images since its main lobe width is determined by the system's bandwidth, the \ac{bs} number of antennas, and \ac{ue} location.
For a more detailed derivation of the \ac{psf}, we refer the reader to~\cite{manzoni2024wavefield}. 

\rev{Phase-coherent fusion across different \ac{ue}-\ac{bs} pairs also requires sufficiently accurate knowledge of the relative \ac{ue}-\ac{bs} geometry. 
In fact, \ac{ue} localization errors affect \eq{eq:TOF}, perturbing the bistatic propagation delay, used in backprojection, resulting in spatially dependent phase errors that cannot, in general, be represented by a common timing or phase offset. Therefore, geometric and/or phase calibration may be required before coherently combining the \ac{ue} images. 
Existing approaches exploit strong static scatterers or clutter components as environmental references to estimate and compensate for relative geometric inconsistencies among sensing nodes~\cite{tagliaferri2024cooperative}. 
The development of a dedicated calibration procedure is outside the scope of this work, but a detailed analysis of the impact of \ac{ue} localization error on \algoname{} without calibration procedures is provided in~\secref{sec:loc-error-impact}.}

\section{Angular anisotropy and \ac{ue} clustering}\label{sec:ue-clustering}

In this section, we analyze the scattering response from a target part, based on the channel model presented in \secref{sec:ch_modeling}.
Then, we obtain an approximate expression of the scattering phase from the part's surface and use it to derive a clustering rule for \acp{ue} to enable coherent imaging, based on their observation angle to the target.

We assume that each target part has a \textit{locally} convex shape, i.e., it has a convex surface within the angular interval of observation of those \acp{ue} in the \ac{isac} network that perform coherent imaging.
In this case, the image in \eq{eq:BP} 
is dominated by the backscattering from the \textit{specular point} over the contour of the $n$-th part of the target.
This is a consequence of the combination of the responses of the different patches on the target's surface.
The specular point has the property that vector $(\mathbf{u}_{n,p,m} + \mathbf{u}_{n,p,\rm bs}) $, given by the \ac{ue} and \ac{bs} locations, and the surface normal vector of the patch are aligned.
Therefore, each \ac{ue} observes a different specular point, due to its different location.
This means that the specular point \textit{migrates} around the surface of the object when changing $\mathbf{u}_{n,p,m}$. 
This prevents coherent imaging using all the available \acp{ue}, since \acp{ue} that are far from each other will observe different points.

\subsection{Insights on the migration of the specular point}\label{sec:anisotropy-insights}

To gain insight into the migration of the specular point, let us consider the simplified case of a cylinder target of radius $R$, centered in $\mathbf{o}$.
We consider the contour of the target to be composed of infinitesimally small patches located at \mbox{$\mathbf{x}(\theta) = \mathbf{o} + R\mathbf{n}(\theta)$}, where $\theta$ is the angle spanning the contour and $\mathbf{n}(\theta) = [\cos \theta, \sin \theta]^{\mathsf{T}}$ is the outward normal vector to the contour in $\theta$. 
The object is illuminated by a bistatic pair, with \ac{ue} $1$ in $\mathbf{x}_{1}$ and the \ac{bs} in $\mathbf{x}_{\rm bs}$. 
We assume for simplicity that the scattering from the object comes only from the contour, with a reflection coefficient $\varrho(\theta)$ that follows the general model in \eq{eq:rho} using location $\mathbf{x}(\theta)$, with incidence and reflection angles $\psi_{m}(\theta)$ and  $\psi_{\rm bs}(\theta)$.
Similarly, the propagation loss is described by $\zeta(\theta)$, given by \eq{eq:zeta}.
Define the bistatic path length as $D(\mathbf{x}(\theta)) \triangleq D(\theta) = \| \mathbf{x}(\theta) - \mathbf{x}_{1} \| + \| \mathbf{x}_{\rm bs}  - \mathbf{x}(\theta) \|$.
Define $\mathbf{x}(\theta_1^*)$ as the specular reflection point on the target boundary that corresponds to angle $\theta_1^*$, the specular reflection angle for \ac{ue}~$1$.
In Appendix~\ref{sec:app-a}, we show that the phase of the image obtained by the \ac{bs} via \eq{eq:BP} computed at a point $\mathbf{x}(\theta)$, in the neighborhood of the specular point, is
\begin{equation}\label{eq:phase_difference_extended_target}
\begin{split}
    \angle I_1(\theta) \approx \frac{2\pi f_0}{c} R \cos\left(\frac{\beta_1^*}{2}\right) ( \theta - \theta_1^*)^2 + \Theta(\theta_1^*)
\end{split}
\end{equation}
where $\Theta(\theta_1^*) = \angle\varrho(\theta_1^*) - \frac{\pi}{4} \mathrm{sign}\left(D''(\theta_1^*)\right)$, $\angle\varrho(\theta_1^*)$ is the intrinsic scattering phase of the material composing the target, and $\beta_1^*$ is the bistatic angle of \ac{ue}~$1$ corresponding to the specular angle~$\theta_1^*$.
In $\mathbf{x}(\theta_1^*)$, the phase of the image equals $\Theta(\theta_1^*)$, which contains the object's scattering phase.

Consider now a different \ac{ue}~$2$ with a bistatic angle $\beta^*_2 =\beta^*_1+ \Delta \beta$.
The specular point for \ac{ue} $2$ is $\theta_2^* = \theta^*_1 + \Delta \theta$ with $\Delta \theta = \Delta \beta/2$, due to a rotation of the normal vector.
As a consequence, the image obtained by \ac{ue}~$2$, computed in $\mathbf{x}(\theta_1^*)$ has a phase of $ \angle I_2(\theta_1^*) \approx 2\pi f_0 R \cos\left(\beta_2^*/2\right) \Delta \theta^2/c + \Theta(\theta_2^*)$.
We assume that the scattering phase due to the target is constant across small angle variations, i.e., $\Theta(\theta_1^*)\approx \Theta(\theta_2^*)$. 
This assumption is reasonable since the intrinsic reflectivity of the material composing the object is constant apart from non-idealities.
A coherent combination of the images obtained by \ac{ue}~$1$ and \ac{ue}~$2$ via backprojection is therefore affected by a residual phase discrepancy $\Delta\phi = \angle I_2(\theta_1^*) - \angle I_1(\theta_1^*)= \pi f_0 R \cos(\beta_2^*/2) \Delta \beta^2/(2c)$.
When \mbox{$R \rightarrow 0$}, the phase difference decreases, reducing to the case of a point-like isotropic target, while it diverges for \mbox{$R \rightarrow \infty$}, approaching a planar target.
As a result, our model is suitable to approximate the anisotropy of target parts of any shape (considering their \textit{local} curvature radius) due to both angle-dependent reflection and migration of the scattering centers.

\subsection{Anisotropy-based \ac{ue} clustering}\label{sec:anis-clust}

The phase discrepancy $\Delta\phi$ between images obtained by different \acp{ue} is quadratically related to the difference between their bistatic angles $\Delta \beta$.
\algoname{} uses this relation to cluster \acp{ue} with similar bistatic angles into groups that can perform coherent imaging since their phase discrepancy is small.
This approach can be implemented at the \ac{bs} side with no additional information, since it only requires the knowledge of the locations of the \acp{ue} and the \ac{roi}.
However, since the target is not a cylinder in general, and its location within the \ac{roi} is unknown, we make the following simplifications: \textit{(i)}~we assume that the maximum local curvature radius of the target, $R_{\max}$, is known for a given target type (e.g., human body), \textit{(ii)}~the \ac{roi} center is used to compute the bistatic angles instead of the target contour location. 

\begin{figure}[t!]
	\begin{center}   
		\centering
	\subcaptionbox{}[0.45\columnwidth]{\includegraphics[width=0.45\columnwidth]{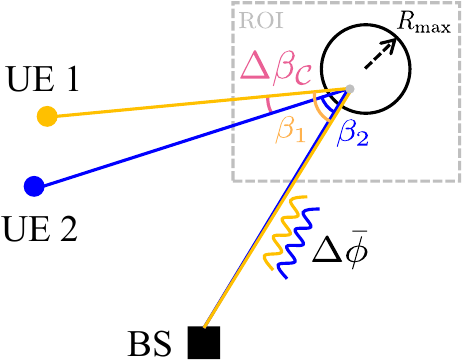}}
        \subcaptionbox{}[0.53\columnwidth]{\input{figures_resubmission/delta_psi_plot.tex}}
		\caption{(a) - Visual representation of the approximate scattering geometry of \eq{eq:phase-diff}. (b) - Angular threshold to ensure scattering phase coherence obtained from \eq{eq:phase-diff} using \mbox{$f_0=7$~GHz}, \mbox{$R_{\max}=0.5$~m}. Different colors represent different values of the bistatic angle.}
		\label{fig:angle-clusters}
	\end{center}
\end{figure}

We define the maximum tolerable phase discrepancy, $\Delta\bar{\phi}$, among the images obtained using a group of \acp{ue} above which they are considered non-coherent. 
By inverting~\eq{eq:phase_difference_extended_target} we obtain the maximum difference among the bistatic angles of the \acp{ue} in cluster $\mathcal{C}$ such that the corresponding phase difference does not exceed $\Delta\bar{\phi}$, i.e., 
\begin{equation}\label{eq:phase-diff}
   \Delta \beta_{\mathcal{C}} =  \sqrt{\frac{2c\Delta\bar{\phi}}{\pi f_0 R_{\rm max}  \cos(\bar{\beta}_{\mathcal{C}}/2)}   }.
\end{equation}
$\bar{\beta}_{\mathcal{C}} = \sum_{m \in \mathcal{C}}\beta_m / |\mathcal{C}|$ is the average bistatic angle of the \acp{ue} in~$\mathcal{C}$, with~$\beta_m$ the bistatic angle of \ac{ue}~$m$ with respect to the center of the \ac{roi}, which is used to approximate the unknown $\beta^*$ in~\eq{eq:phase_difference_extended_target}, and $|\mathcal{C}|$ the number of \acp{ue} in $\mathcal{C}$.
This approximation is accurate when the size of the \ac{roi} is much smaller than its distance from the \ac{bs}.
The different quantities are shown in \fig{fig:angle-clusters}a.
In \fig{fig:angle-clusters}b, we plot \eq{eq:phase-diff} for different values of the bistatic angle $\bar{\beta}_{\mathcal{C}}$, using $R_{\max}=0.5$~m and $f_0=7$~GHz. 
For a maximum phase discrepancy $\Delta \bar{\phi} = 45^{\circ}$, the difference in the observation angles of different \acp{ue} can be at most around $15^{\circ}$, demonstrating the importance of taking anisotropy into account, even in the case of a cylindrical target.

In \algoname{}, we group the $M$ \acp{ue} into~$Q$ clusters, denoted by
$\{\mathcal{C}_q\}_{q=1}^Q$.
The clustering is performed by ensuring that within $\mathcal{C}_q$ the following condition is met
\begin{equation}\label{eq:clustering}
    \max_{m \in \mathcal{C}_q} \beta_m - \min_{m \in \mathcal{C}_q} \beta_m \leq \Delta \beta_{\mathcal{C}},
\qquad \forall q = 1,\dots,Q,
\end{equation}
i.e., the difference among the bistatic angles of the \acp{ue} in $\mathcal{C}_q$ does not exceed the threshold $\Delta \beta_{\mathcal{C}}$.
In cluster $q$, there are $M_q$ \acp{ue} indexed by $i=1,...,M_q$.
We adopt a greedy approach, starting from the \ac{ue} with the largest bistatic angle and adding new \acp{ue} to its cluster as long as \eq{eq:clustering} is satisfied. 
Then, a new cluster is initiated, and the process is repeated.

\rev{In \fig{fig:rmax-error}a, we analyze the practical case where the true $R_{\max}$ of the target is unknown and only a prior estimate of $\widehat{R}_{\max}$ is available.
We take $R_{\max}=0.5$~m (dashed vertical line), and vary $\widehat{R}_{\max} \in [0.1, 100]$~m.
For $\widehat{R}_{\max} > R_{\max}$, the maximum target radius is overestimated, leading to a lower, more conservative value of the angular threshold $\widehat{\Delta \beta}_{\mathcal{C}}$ since targets that approach a flat surface are less coherent.
Underestimating the correct radius, i.e., $\widehat{R}_{\max} < R_{\max}$, leads to using a too large angular threshold $\widehat{\Delta \beta}_{\mathcal{C}}$, allowing coherent combination among \acp{ue} that may observe incoherent views of the target.
In \secref{sec:clust-thr-impact}, we show the impact of using mismatched values of the angular threshold, $\widehat{\Delta \beta}_{\mathcal{C}} \neq \Delta \beta_{\mathcal{C}}$, on the imaging performance.

In \fig{fig:rmax-error}b, we analyze instead the case where the target center $\mathbf{o}$ does not coincide with the \ac{roi} center, where the displacement, $\epsilon$, is on the $x$-axis.  
This causes an error in the computation of the bistatic angle $\bar{\beta}_{\mathcal{C}}$, which affects \eq{eq:phase-diff}.
Different curves correspond to different values of distance between the \ac{ue} and the target contour $\|\mathbf{x}_1 - \mathbf{x}\| \in \{3,5,10\}$. 
\eq{eq:phase-diff} is robust to a displacement of the target center, since even for large $\epsilon$ of up to $2$~m the resulting error in  $\bar{\beta}_{\mathcal{C}}$ is of few degrees, and it improves when the \ac{ue} is far from the target.
}

\rev{Note that the clustering criterion in \eq{eq:clustering} is designed to preserve scattering coherence from extended targets and does not explicitly optimize the coherent \ac{psf} of the resulting distributed cluster aperture (\secref{sec:intra-coherent-img}). Nevertheless, by restricting coherent processing to \acp{ue} with similar bistatic observation angles, it indirectly limits the effective coherent aperture and can mitigate the formation of strong spatial sidelobes compared with unrestricted coherent combination of widely separated \acp{ue}.}

\rev{
\begin{figure}[t!]
	\begin{center}   
		\centering
	\subcaptionbox{}[0.45\columnwidth] {\input{figures_resubmission/plot_v2_absolute}}
        \subcaptionbox{}[0.53\columnwidth]{\input{figures_resubmission/center_displacement}}
		\caption{\rev{Analysis of the impact of a mismatched prior estimate of $R_{\max}$ (a), and displaced \ac{roi} center with respect to the target, using \mbox{$f_0=7$~GHz} and \mbox{$\Delta \bar{\phi}=45^\circ$}.
        (a)~Values of $\widehat{\Delta \beta}_{\mathcal{C}}$ obtained from \eq{eq:phase-diff} using a prior estimate $\widehat{R}_{\max} \in [0.1, 100]$~m.
        The vertical dashed line corresponds to the value of the real radius $R_{\max}=0.5$~m, and the horizontal lines show the corresponding value of the correct $\Delta \beta_{\mathcal{C}}$. Different colors represent different values of the bistatic angle.
        (b)~Values of $\widehat{\Delta \beta}_{\mathcal{C}}$ obtained from \eq{eq:phase-diff} when the \ac{roi} center is displaced by $\epsilon$ from the target center $\mathbf{o}$. Different colors represent different values of the \ac{ue}-target distance. }}
		\label{fig:rmax-error}
	\end{center}
\end{figure}
}

\subsection{Impact of Doppler on intra-cluster coherent imaging}\label{sec:intra-coherent-img}

After the \ac{bs} is synchronized with each \ac{ue}, existing approaches in the literature would generate $M_q$ images for each cluster, $I_{q}(\mathbf{x}, kT)$, assuming static targets within the \ac{roi}~\cite{tagliaferri2024cooperative}.
This is done by obtaining images for each \ac{ue}-\ac{bs} pair via \eq{eq:BP}, and combining them coherently, i.e., $I_q(\mathbf{x},kT)= \sum_{i=1}^{M_q} I_{q,i}(\mathbf{x},kT)$.
The coherent image is characterized by a \ac{psf} \mbox{$\chi_q(\mathbf{x}) = \sum_{i=1}^{M_q} \chi_{q,i}(\mathbf{x})$}, which exhibits a much finer resolution due to the wider spatial aperture across multiple \acp{ue}, but also a much higher sidelobe level due to the sparse \ac{ue} deployment in space.
The resolution of the $q$-th cluster is $\rho_{x,q}$ and $\rho_{y,q}$ along $x$ and $y$, respectively, defined as the width of the mainlobe of the \ac{psf}.

However, this approach is only applicable to \textit{static} targets.
In the presence of a \ac{ue}- and target-specific Doppler shift, as per \eq{eq:doppler-freq}, single images $I_{q,i}(\mathbf{x},kT)$ do not add up coherently in the true targets' locations due to the different Doppler shifts. 
Hence, $I_q(\mathbf{x},kT)$ shows a significant signature for the static targets, having \mbox{$f_{n,p,q,i}^{\rm D}=0$}, while for moving targets it is defocused due to Doppler-induced phase errors.

\section{Composite moving target imaging method}\label{sec:imaging_moving}

\begin{figure}[t!]
\centering
\includegraphics[width=\linewidth]{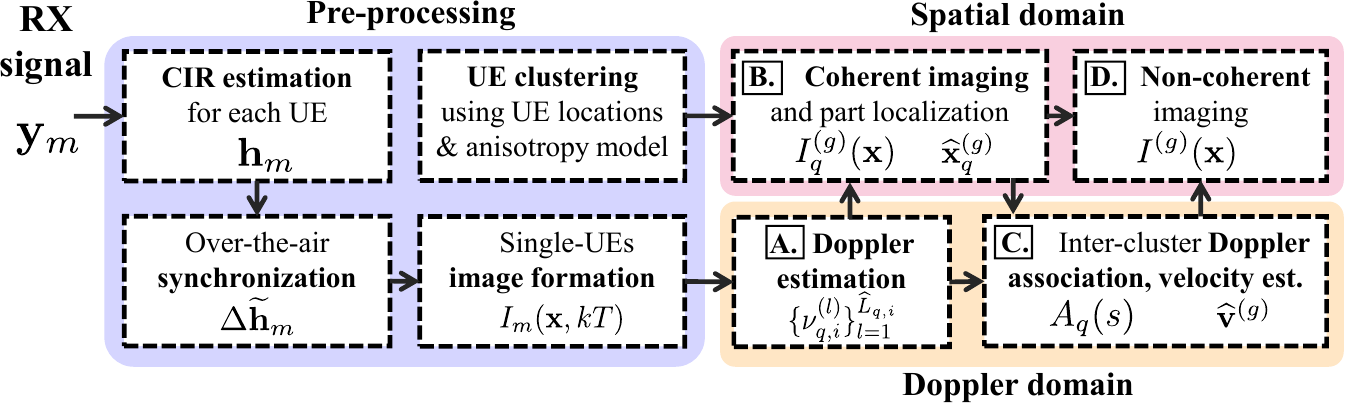}
\caption{Block diagram of \algoname{}.}
\label{fig:block-diagram}
\end{figure}

The proposed algorithm follows $4$~steps, as shown in \fig{fig:block-diagram}: A) Doppler spectrum estimation from low-resolution images and intra-cluster Doppler peaks association, B) intra-cluster Doppler pre-compensation and coherent high-resolution imaging, with localization of each target part, C) inter-cluster Doppler peaks association and target part-wise velocity vector estimation, and D) non-coherent global image formation using all clusters. 
The steps are detailed in the following sections. 

\subsection{Doppler estimation and intra-cluster association}
\label{sec:doppler-low-res}

The low-resolution images $I_{q,i}(\mathbf{x}, kT)$ obtained at the \ac{bs} for each \ac{ue} with~\eq{eq:BP} are affected by a
residual linear phase contributions $ 2\pi f^{\rm D}_{n,p,q,i} kT$ due to the Doppler shift of each patch of each target part. 
However, images $I_{q,i}(\mathbf{x}, kT)$ do not have, in general, enough resolution to distinguish the single target parts nor the individual patches, due to the limited bandwidth of each~\ac{ue} and antenna size of the~\ac{bs}. 
Hence, the estimation of the individual Doppler shifts $f^{\rm D}_{n,p,q,i}$ for each patch $(n, p)$ and \ac{ue} $(q,i)$ is infeasible.
However, we can still estimate the main Doppler contribution from each part as the Doppler shift of the main scattering points (specular reflections) in the \ac{roi}. 

\subsubsection{Doppler estimation}\label{sec:doppler-est}

We compute the non-coherent average of the Doppler spectra of the pixels in the \ac{roi}
\begin{equation}\label{eq:spectrum-low-res}
    \bar{S}_{q,i}(\nu ) = \frac{1}{N_{\mathbf{x}}}\sum_{\mathbf{x} \in \rm ROI} \left|\sum_{k=0}^{K-1} I_{q,i}(\mathbf{x}, kT) W(k) e^{- j 2 \pi \nu k T}\right|^2,
\end{equation}
where $N_{\mathbf{x}}$ is the number of pixels and $W(k)$ is a window function (e.g., Hamming window) to suppress sidelobes.
The inner sum in~\eq{eq:spectrum-low-res} is the Doppler spectrum for pixel $\mathbf{x}$, with~$\nu$ being the discrete Doppler frequency variable. 
Pixels with no target signature will not contribute to $\bar{S}_{q,i}(\nu)$, while pixels with moving or static targets will produce Doppler spectra that add up incoherently. 
From $\bar{S}_{q,i}(\nu )$, we estimate a set of Doppler frequency shifts for each \ac{ue}-\ac{bs} pair using standard methods e.g., \ac{cfar}~\cite{richards2010principles}. 
If the $N$ moving target parts are resolvable in the Doppler domain $\bar{S}_{q,i}(\nu)$ exhibits $N$ peaks corresponding to the Doppler shifts of specular reflection points on target parts. 
In practice, the number of detected peaks is affected by the \ac{snr}, the performance of the peak detection algorithm, and the Doppler resolution $\Delta f^{\rm D} = 1 / (KT)$.
% \rev{If some Doppler peaks are undetected by \ac{ue} $(q,i)$ due to a lack of sufficient resolution at this stage, they can still be resolved by other \acp{ue} or clusters due to the geometric diversity provided by different \ac{ue} locations.
% If two target parts are unresolvable in the Doppler domain by all \ac{ue} clusters, they will be treated as the same target part by \algoname{}, but they would still be \textit{spatially} resolvable in the final image provided that enough spatial resolution is available.}
We denote the set of Doppler peaks as $\{\nu_{q,i}^{(\ell)}\}_{\ell=1}^{\widehat{N}_{q,i}}$, where $\widehat{N}_{q,i}$ is the number of estimated Doppler peaks by the $(q,i)$-th \ac{ue}-\ac{bs} pair.

\subsubsection{Intra-cluster Doppler association}\label{sec:intra-doppler-assoc}

Within the same cluster, \acp{ue} observe slightly different Doppler spectra, due to the different bistatic geometry and specular point location on each target. 
This may lead to the detection of a different number of peaks for each \ac{ue}-\ac{bs} pair.
However, the locations of the detected Doppler peaks corresponding to a certain moving target part will be \textit{similar} across \acp{ue} in the same cluster, since the observation angles are similar by construction.
This is a consequence of the clustering procedure in \secref{sec:ue-clustering}. 

For each cluster $\mathcal{C}_q$, with $q=1,...,Q$, we identify the \ac{ue} in $\mathcal{C}_q$ that has detected the most Doppler peaks, assigning it index \mbox{$i=1$} for convenience.
This \ac{ue} determines how many Doppler peaks are identified in cluster $\mathcal{C}_q$.
We assign indexes $g=1, \dots, N_q$ to the Doppler peaks of \ac{ue}~$1$, and we interpret $N_q$ as an estimate of the number of target parts according to the $q$-th cluster.

Then, we assign the Doppler shifts of other \acp{ue} in $\mathcal{C}_q$ to those of \ac{ue}~$1$ using a minimum distance criterion.
This is done via an association function $A_{q,i}(\cdot)$ that maps Doppler peak $\ell$ of \ac{ue}~$i$ with peak $g$ of \ac{ue} $1$ defined as
\begin{equation}\label{eq:intra-assoc}
    A_{q, i}(\ell) = \argmin_{g} \left|\nu_{q,i}^{(\ell)}-  \nu^{(g)}_{q,1}\right|.
\end{equation}
To avoid inconsistent associations, we set a maximum distance threshold, \mbox{$\nu_{\max}=10 \Delta f^{\rm D}$}, and leave those Doppler peaks unassociated for which the minimum distance in \eq{eq:intra-assoc} exceeds $10$ times the Doppler resolution.
In case of multiple associations to the same peak of \ac{ue}~1, we selecting the closest one and discard the other, which remains unassociated.

We define the inverse of the association function $\ell_{q,i}(g)= A_{q,i}^{-1}(g)$. 
The above association rule works effectively as long as the \acp{ue} in a cluster all share similar bistatic angles with respect to the target. 
While this is true within each cluster by definition, due to \eq{eq:clustering}, it may not hold across different clusters.
In \secref{sec:inter-doppler-assoc}, we take a more sophisticated approach to associate Doppler peaks in such case.

\rev{In the intra-cluster association phase, the Doppler resolution is determined by the reference \ac{ue}, which detects the most Doppler peaks.
Since \acp{ue} within the same cluster have similar spatial locations, it is unlikely that \acp{ue} $i \neq 1$ can resolve Doppler peaks that are unresolved by \ac{ue}~$1$.
If this happens, the two corresponding target parts will \textit{both} appear in the same image of the reference \ac{ue}, and they will contribute to the cluster image obtained in the next section.
}
\vspace{-0.3cm}

\subsection{Doppler-compensated intra-cluster coherent imaging}\label{sec:doppler-precompensation}

With the estimated Doppler peaks, we form intra-cluster coherent images of \textit{single} target parts, $I_{q}^{(g)}(\mathbf{x})$, by 
% \eq{eq:BP_withcompensation}
\begin{align}\label{eq:BP_withcompensation}
    I_{q}^{(g)}(\mathbf{x})  = & \sum_{k=0}^{K-1}\sum_{i = 1}^{M_q}I_{q, i}(\mathbf{x}, kT) e^{-j2\pi \nu_{q,i}^{(\ell_{q,i}(g))} kT}\\
   \approx  & K\sum_{i = 1}^{M_q} \sum_{p=1}^{P_g} \varrho_{g,p,q,i} \chi_{q,i}\left(\mathbf{x}\hspace{-0.05cm}-\hspace{-0.05cm}\mathbf{x}_{g,p,q,i}\right)+ \label{eq:BP_withcomp_full}\\ 
   + & \sum_{k=0}^{K-1} \sum_{n\neq g} \sum_{p=1}^{P_n} \varrho_{n,p,q,i} \chi_{q,i}\left(\mathbf{x}\hspace{-0.05cm}-\hspace{-0.05cm}\mathbf{x}_{n,p,q,i}\right) e^{j 2 \pi \varepsilon^{(g)}_{n,p,q,i} kT }.\nonumber
\end{align}
In \eq{eq:BP_withcompensation}, we perform Doppler compensation, canceling out the phase contribution due to the Doppler shift of the $g$-th target part seen by $(q,i)$-th \ac{ue}-\ac{bs} pair. 
We use the inverse association function $\ell_{q,i}(g)$ to represent the index of the Doppler peak observed by pair $(q, i)$ corresponding to part~$g=1, \dots, N_q$.
The result of such Doppler-compensated image formation is shown in \eq{eq:BP_withcomp_full}. 
Having canceled out the Doppler of the \mbox{$g$-th} target part, the latter is correctly focused (first term), while other target parts (second term) retain a residual Doppler shift $\varepsilon^{(g)}_{n,p,q,i} = f_{n,p,q,i}^{\rm D} -  \nu_{q,i}^{(\ell_{q,i}(g))}$.

This residual term makes the contribution of the other target parts \textit{vanish} in the intra-cluster image $I_{q}^{(g)}(\mathbf{x})$, thanks to the summation along slow-time: For sufficiently large $|\varepsilon^{(g)}_{n,p,q,i}|$ and $KT$, the contributions of the other parts add up destructively.
Therefore, while the $g$-th target part is focused, other target parts are filtered out, effectively leveraging Doppler as an additional means to resolve different target parts. 

This allows estimating the location of the main scattering point of each target part according to the $q$-th \ac{ue} cluster, as the peak of the target-part specific image $I_{q}^{(g)}(\mathbf{x})$, 
\begin{equation}\label{eq:scatter-local}
    \widehat{\mathbf{x}}^{(g)}_q = \argmax_{\mathbf{x}} \left|I_{q}^{(g)}(\mathbf{x})\right|^2.
\end{equation}
The locations $\widehat{\mathbf{x}}^{(g)}_q$ are used in the next section to relate  Doppler shifts with velocity vectors for each target part. 

\subsection{Inter-cluster Doppler association and velocity estimation}\label{sec:inter-assoc}

At this stage, the \ac{bs} has $\sum_{q=1}^{Q} N_q$ high-resolution images of the target parts and, for each of those, an estimate of the location of the main scattering center, $\widehat{\mathbf{x}}^{(g)}_q$. 

We use the images and the locations to estimate the velocity vector of each target part by exploiting the geometry of each bistatic triangle formed by the \ac{ue}, \ac{bs}, and target part location~$\widehat{\mathbf{x}}^{(g)}_q$.

As a first step, we identify the correspondences between the Doppler peaks observed by the \acp{ue} across the different clusters, i.e., perform an inter-cluster association of the Doppler shifts. 
Compared to the intra-cluster case (\secref{sec:intra-doppler-assoc}), the inter-cluster association is more complex since \ac{ue} clusters can have very different observation angles, leading to significantly different Doppler shifts.

\subsubsection{Inter-cluster association}\label{sec:inter-doppler-assoc}
To solve the problem, we take the \ac{ue} cluster with the highest number of detected Doppler peaks as a reference and assign it index $q=1$. 
Due to this choice, we have that $N_1 \geq N_q, \forall q$.
We perform the association among \textit{cluster pairs} formed by the reference cluster and each of the other clusters, i.e., $\{(1, q)\}_{q=2}^Q$. 
For each pair $(1,q)$, we associate the $N_q$ Doppler peaks (target parts) of cluster $\mathcal{C}_q$ with the $N_1$ Doppler peaks of the reference cluster.
To associate Doppler peak $r$ in the reference cluster with Doppler peak $s$ in cluster $\mathcal{C}_q$, we first collect all the Doppler shifts observed by each of the two clusters in vector form.
We define the $M_q \times 1$ vectors $\bm{\nu}_q^{(g)} =[\nu_{q, 1}^{(\ell_{q,1}(g))}, \dots, \nu_{q, M_q}^{(\ell_{q,M_q}(g))}]^{\mathsf{T}}$
and the vectors defining the inward bisector of the bistatic angle formed by the $(q,i)$-th \ac{ue}, $g$-th target part, and \ac{bs},
\begin{equation}
    \mathbf{u}_{q,i}^{(g)} =  \frac{\mathbf{x}_{q,i}-\widehat{\mathbf{x}}^{(g)}_q}{\|\mathbf{x}_{q,i}-\widehat{\mathbf{x}}^{(g)}_q\|} +  \frac{\mathbf{x}_{\rm bs}-\widehat{\mathbf{x}}^{(g)}_q }{\|\mathbf{x}_{\rm bs}-\widehat{\mathbf{x}}^{(g)}_q\|}.
\end{equation}
Vectors $\mathbf{u}_{q,i}^{(g)}$ are stacked into the $M_q \times 2$ matrix $\mathbf{U}_{q}^{(g)} = [\mathbf{u}_{q,1}^{(g)}, \dots, \mathbf{u}_{q,M_q}^{(g)}]^{\mathsf{T}}$.
The association cost between Doppler shift~$r$ in the reference cluster and Doppler shift~$s$ in cluster~$\mathcal{C}_q$, called $c_{r, s}$, is obtained as
\begin{equation}\label{eq:assoc-cost}
    c_{r, s} 
    = \min_{\mathbf{v}\in\mathbb{R}^2} \left\|
    \begin{bmatrix}
       \bm{\nu}_1^{(r)} \\
       \bm{\nu}_q^{(s)}
    \end{bmatrix}
     - \frac{f_0}{c}
     \begin{bmatrix}
         \mathbf{U}_{1}^{(r)}\\
         \mathbf{U}_{q}^{(s)}
     \end{bmatrix} \mathbf{v}
    \right\|^2.
\end{equation}
The cost $c_{r, s}$ measures the minimum error made by a velocity~$\mathbf{v}$ in matching the Doppler shifts measured by the \acp{ue} in the reference and $q$-th clusters when applied to a target part located at $\mathbf{x}_{1}^{(r)}$ when observed by the reference cluster and at $\mathbf{x}_{q}^{(s)}$ when observed by the $q$-th cluster. 
We select the association with the lowest \textit{cumulative} cost across all possible pairs~$(r,s)$. 
Finding the association that yields the lowest total cost is a combinatorial problem called \textit{minimum cost assignment}~\cite{ramshaw2012minimum}. 
We define binary association variables $b_{r, s}\in \{0, 1\}$, collected in the $N_1 \times N_q$ matrix $\mathbf{B}_q$, where we recall that $N_1 \geq N_q$ by construction. 
If $b_{r, s}=1$, the $r$-th Doppler peak of the reference cluster is associated with the $s$-th Doppler peak of cluster $\mathcal{C}_q$. 
Moreover, we collect all costs into a matrix $\mathbf{C}_q$, of the same shape, with elements~$c_{r, s}$.
The problem amounts to finding the $\mathbf{B}_q$ that minimizes 
\begin{align}\label{eq:hung}
    \argmin_{\mathbf{B}_q}& \quad \mathbf{1}_{N_1}^{\mathsf{T}}\left(\mathbf{B}_q \odot \mathbf{C}_q\right)\mathbf{1}_{N_q}\\
    \text{s.t.} \quad& \quad \mbox{(a)} \,\,\,\mathbf{1}_{N_1}^{\mathsf{T}} \mathbf{B}_q = \mathbf{1}_{N_q}\quad \mbox{(b)} \,\left[\mathbf{B}_q\mathbf{1}_{N_q}\right]_r \leq 1 \,\,,\forall r.\label{eq:hung-constr}
\end{align}
The minimization argument is the total sum of costs $c_{r, s}$ for which \mbox{$b_{r, s}=1$}.
Due to constraint \eqref{eq:hung-constr}(a), each of the $N_q$ peaks in cluster $\mathcal{C}_q$ must be associated with exactly one peak in the reference cluster. 
Constraint \eqref{eq:hung-constr}(b) imposes that peaks in the reference cluster that do not match well any of the peaks in $\mathcal{C}_q$ may remain unassociated.
\rev{Due to constraint \eqref{eq:hung-constr}(a), it may happen that a peak in cluster $\mathcal{C}_q$ is associated with a peak in the reference cluster despite having a very large association cost, for example because the corresponding target part is undetected by the reference cluster.
In this case, we cancel the association, i.e., we discard Doppler peak $s$ from cluster $\mathcal{C}_q$ and leave the corresponding peak $r$ in the reference cluster unassociated with cluster $q$.
The value of the threshold for determining a high association cost can be determined based on the Doppler resolution.
In our experiments we use \mbox{$c_{r, s}  > (10\Delta f^{\rm D})^2(M_1 + M_q) $}, i.e., we set a per-\ac{ue} average maximum cost of $10$ times the Doppler resolution, and multiply it by the number of \acp{ue} contributing to the cost~$c_{r, s}$.}

The solution to \eq{eq:hung} can be found efficiently using, e.g., the Hungarian algorithm~\cite{crouse2016on}, and consists of a set of associations between the Doppler peaks of the reference cluster and those of cluster~$\mathcal{C}_q$.
The inter-cluster association function for cluster $\mathcal{C}_q$, $A_q(s)$, returns the index of the Doppler peak of the reference cluster associated with the $s$-th peak of cluster~$\mathcal{C}_q$, $A_q(s) = \argmax \left[\mathbf{B}_q\right]_{:, s}$,
since exactly one element per column of the association matrix $\mathbf{B}_q$ is nonzero (and equal to~$1$).
The inverse function of $A_q(s)$ is $s_q(g) = A^{-1}_q(g)$.

This procedure is repeated for all pairs of clusters $\{(1, q)\}_{q=2}^Q$, and we construct an association among the Doppler peaks (targets) seen by each cluster.

\rev{In the inter-cluster association phase, the Doppler resolution is determined by the reference cluster $\mathcal{C}_1$.
Due to the very possibly different locations of the different clusters, it may happen that cluster $\mathcal{C}_q, q\neq 1$, can resolve Doppler peaks that are unresolved by cluster $\mathcal{C}_1$.
The association phase will not solve this problem, since it takes cluster $\mathcal{C}_1$ as a reference.
However, this does not mean the two parts corresponding to the unresolved Doppler peaks are indistinguishable for \algoname{}.
The two parts will indeed appear in the \textit{same} image of the reference cluster, since they have an indistinguishable Doppler shift, but they can be \textit{spatially} resolved if they occupy sufficiently different spatial locations.
}

\subsubsection{Velocity estimation}\label{sec:vel-est}
After inter-cluster associations have been obtained as described in the previous section, the velocity vectors of each target part are estimated by exploiting the geometric diversity of the \ac{ue} clusters.
To this end, we collect all Doppler shifts (from all clusters) corresponding to the $g$-th target part seen by the reference cluster in a single vector as
\begin{equation}
    \bm{\nu}^{(g)} = \left[\left(\bm{\nu}_{1}^{(g)}\right)^{\mathsf{T}}, \dots, 
         \left(\bm{\nu}_{q}^{(s_q(g))}\right)^{\mathsf{T}}\right]^{\mathsf{T}}.
\end{equation}
Then, the following matrix of direction vectors is constructed
\begin{equation}
    \mathbf{U}^{(g)} = \left[\left(\mathbf{U}_{1}^{(g)}\right)^{\mathsf{T}}, \dots, 
         \left(\mathbf{U}_{q}^{(s_q(g))}\right)^{\mathsf{T}}\right]^{\mathsf{T}}.
\end{equation}
If target part $g$ is associated with a corresponding Doppler peak in all clusters, $\bm{\nu}^{(g)}$ has size $M$ and $\mathbf{U}^{(g)}$ has shape $M\times 2$.
The estimate of the velocity vector of the $g$-th target part is then obtained as
\begin{equation}\label{eq:vel-estimation}
    \widehat{\mathbf{v}}^{(g)}
    = \argmin_{\mathbf{v}} \left\|
       \bm{\nu}^{(g)}
     \hspace{-0.1cm}-\hspace{-0.1cm} \frac{f_0}{c}
     \mathbf{U}^{(g)} \mathbf{v}
    \right\|^2 \hspace{-0.1cm}= \frac{c}{f_0} \left(\mathbf{U}^{(g)} \right)^{\dag}\bm{\nu}^{(g)}.
\end{equation}

\subsection{Non-coherent global image formation}\label{sec:non-coh-image}

We combine images $I_{q}^{(g)}(\mathbf{x})$ non-coherently using the following algorithm.
First, we apply a detection step to the images of each cluster $I_{q}^{(g)}(\mathbf{x})$, by thresholding its squared amplitude and retaining only the strongest image components.
Then, we sum together the resulting thresholded images across clusters, combining them into target-specific images that include information obtained from all the available clusters of \acp{ue}, hence integrating multiple viewpoints.
This is written as
\begin{equation}\label{eq:non-coh-image}
    I^{(g)}(\mathbf{x}) = \sum_{q=1}^{Q}\mathcal{T}_{\eta}\left\{\left|I_{q}^{(g)}(\mathbf{x})\right|^2\right\},
\end{equation}
where $\mathcal{T}_{\eta}\left\{\cdot\right\}$ is a hard thresholding function that sets to zero pixels with intensity lower than $\eta$ times the maximum intensity in the image.
The choice of $\eta$ should achieve a trade-off between only selecting the strongest component of the cluster-specific images and preserving the shape of the object. \rev{The fixed threshold is adopted here for simplicity and it is consistent with our assumption that all reflectivity components in \ac{roi} are of interest. 
More complex approaches with dynamic thresholds could be applied, for example using cluster-dependent thresholds obtained from the sidelobe level of the corresponding theoretical \acp{psf}, or spatially adaptive thresholds based on local clutter or noise statistics if clutter is modeled within the \ac{roi}.}
\eq{eq:non-coh-image} produces hybrid images of each target part, obtained from a mix of coherent and non-coherent processing.
The proximity of \acp{ue} within each cluster is exploited to enhance the resolution via coherent combination, while the spatial distribution of the \ac{ue} clusters is used to observe the parts from possibly very different angles via non-coherent combination.

Additionally, \algoname{} obtains a global image $\widehat{I}(\mathbf{x})$ containing the signature of all the target parts by summing the images of each target non-coherently as $\widehat{I}(\mathbf{x}) = \sum_{g=1}^{N_1} I^{(g)}(\mathbf{x})$.

\subsection{Remark on computational complexity}
The computational complexity of our approach accounts for \textit{(i)} the generation of low-resolution images, whose number of complex multiplications scales as $\mathcal{O}(M N_{\mathbf{x}}  L)$ per slow-time instant, where $N_{\mathbf{x}} $ is the number of pixels,
\textit{(ii)}~the Doppler spectrum estimation via the DFT in \eq{eq:spectrum-low-res} that scales as $\mathcal{O}(N_{\mathbf{x}} K \log_2 K)$ per \ac{ue}, \textit{(iii)}~the Doppler association problem solved for $Q-1$ clusters, whose complexity scales as $\mathcal{O}(QN_1^3)$ and \textit{(iv)}~the generation of $N_1$ pre-compensated images for each cluster, $\mathcal{O}(K N_1 M N_{\mathbf{x}}  L)$. 

The image formation \textit{(iv)} and the Doppler estimation \textit{(ii)} dominate, scaling linearly with the number of pixels $N_{\mathbf{x}}$, while the Doppler association is typically negligible in practical situations since $QN_1^3 \ll  N_{\mathbf{x}} $.
This is because the number of target parts is typically small (fewer than $5$ for most targets of interest).
Conversely, the number of pixels $N_{\mathbf{x}}$ is large in high-resolution imaging even for small \acp{roi}. 
Our approach has lower complexity than an exhaustive search for the velocity vectors, as performed, e.g., in~\cite{moving_target_sar}.
With a grid of $N_{\mathbf{v}} $ velocity candidates, the complexity of such search scales as $\mathcal{O}(KM N_{\mathbf{x}} N_{\mathbf{v}}  L)$. 
For $N_{\mathbf{v}} = 100 \times 100$, \mbox{$N_{\mathbf{x}} = 1000 \times 1000$}, $M=20$,  $L=30$, $K=32$, and $N_1=2$, the complexity of exhaustive velocity search is three orders of magnitude worse than \algoname{}. 

\section{Numerical results}\label{sec:results}

\rev{In this section, we evaluate \algoname{} using numerical simulations, comparing it against existing imaging algorithms.}

\subsection{Simulation scenario and parameters}\label{sec:sim-scenario}

To evaluate the proposed method, we perform numerical simulations including a variable number of Tx \ac{isac} \acp{ue}, and a single Rx \ac{bs} equipped with \mbox{$L=30$} antennas.
 As default parameters, we consider a carrier frequency \mbox{$f_0 = 7$~GHz} and signal bandwidth \mbox{$B=400$~MHz}, with $S=2049$ subcarriers, representing a \ac{fr3} cellular scenario~\cite{bazzi2026coverage}. 
The available bandwidth gives a range resolution of $c/(2 B) = 37.5$~cm in a monostatic sensing scenario. 
When specified, we vary the carrier frequency to evaluate its impact on \algoname{}.
 
The \ac{bs} is located at the center of the reference frame, in~ $[0,0]^{\mathsf{T}}$~m, while the \ac{roi} is centered in~$[10, 10]^{\mathsf{T}}$~m.
\acp{ue} are static, and deployed uniformly at random in half-annulus around the \ac{roi} according to a tunable density measured in \acp{ue}/m$^2$.
The annulus inner and outer radii are $2.5$~m and $7.5$~m, respectively.
\rev{An example realization of the considered scenario is shown in \fig{fig:scenarios-plots}, with randomly placed \acp{ue} within the half-annulus with a density of $0.5$~UEs/m$^2$.}

The locations of the \acp{ue} are assumed to be known perfectly or with a Gaussian localization error, depending on the experiment.
\acp{ue} are clustered using \eqs{eq:phase-diff}{eq:clustering}, using a phase difference threshold $\Delta \bar{\phi} =45^{\circ} $, $R_{\max}=0.3$~m, and the bistatic angle derived from their known location and the center of the \ac{roi}.
In general, $R_{\max}$ should be selected based on the size of the considered target and its shape, e.g., flat or curved.
The non-coherent imaging threshold used in \eq{eq:non-coh-image} is set to \mbox{$\eta=0.95$}, which effectively suppresses sidelobes in cluster-specific images while still allowing good quality object shape reconstruction, although \algoname{} obtains similar performance for $\eta>0.9$.

The preamble repetition interval is $T=3$~ms, with slow-time processing windows of $KT$~seconds with $K=32$, which gives a Doppler resolution of $10.4$~Hz.
We consider the \ac{to} $\alpha_m(kT)$ to be uniformly distributed in~$[0, 10/B]$. 
The \ac{cfo} evolves across slow-time according to a first-order auto-regressive random process. 
The process is initialized as \mbox{$\vartheta_m(0)\sim  \mathcal{N}(0, \sigma_{0}^2)$}, with $\sigma_0 = 10^{-7}$. 
Then, it evolves following $\vartheta_m(kT) = 0.99 \vartheta_m((k-1)T) + 0.01W_k$, with \mbox{$W_k \sim \mathcal{N}(0, \sigma_0^2)$}~\cite{Vannicola83}. 
The \ac{pn} is selected as $\xi_m(kT) \sim \mathcal{N}(0, (\pi/8)^2)$. 
In the following, we consider the sensing \ac{snr} within the \ac{roi} as the ratio between the power of the scattered signal from the \ac{roi} at the \ac{bs} and the noise power. 
The \ac{rcs} of each patch is generated using~\eq{eq:RCS}, rescaled such that the overall \ac{rcs} of the composite target is $0$~dBm$^2$~\cite{Manfredi_RCS_human}, using the human body as a reference.
The composite targets used in the simulations include $2$ or $3$ different moving parts, depending on the experiment.
\rev{\fig{fig:scenarios-plots} shows an example target composed of two circles, both with a radius of $0.3$~m. This is the basic target considered in several of our experiments in the following sections.
\tab{tab:target-params} summarizes the geometrical parameters of the targets used in the simulations.}
Each part is modeled as a circular or elliptical shape composed of square scattering patches with an \ac{rcs} given by~\eq{eq:RCS}.
Each target has a main velocity common to all its parts, while 
each part also has an independent relative velocity, whose vector sum with the main velocity yields its absolute velocity.
\rev{The velocity of each of the target parts is given separately in the following sections.}

\rev{
\begin{table}[t!]
    \centering
    \footnotesize
    \caption{\rev{Shape parameters of the targets used in the simulations.}}
    \setlength{\tabcolsep}{3pt}
    \begin{tabular}{ccccc}
    \toprule
    \textbf{Target} & \textbf{Part} & \textbf{Shape} & \textbf{Parameters} & \textbf{Center} \\
    \midrule
    \multirow{2}{*}{2-parts} & 1 & Circle & radius $0.3$ m&  $[10, 10]^{\mathsf{T}}$ \\
                              & 2 & Circle & radius $0.3$ m&  $ [10.45, 10]^{\mathsf{T}}$  \\
    \midrule
    \multirow{3}{*}{3-parts} & 1 & Circle  & radius $0.1$ m&  $ [9.75, 10]^{\mathsf{T}}$  \\
                              & 2 & Circle  & radius $0.1$ m&  $[10.75, 10]^{\mathsf{T}}$  \\
                              & 3 & Ellipse & axes $0.45$, $0.25$ m & $ [10.25, 10]^{\mathsf{T}}$  \\
        \bottomrule
    \end{tabular}
    \label{tab:target-params}
\end{table}
}

\rev{
\begin{figure}[t!]
\footnotesize
    \begin{center}   
        \centering
            \input{figures_resubmission/scenario_1}
        \caption{\rev{Example of the considered scenario with \ac{ue} density $0.5$~UEs/m$^2$ and angular threshold $15^\circ$. The \ac{bs} is represented by the black  triangle, the target by the two black circles, and the \acp{ue} by the squares, where colors indicate different clusters.}}
        \label{fig:scenarios-plots}
    \end{center}
    % \vspace{-0.3cm}
\end{figure}
}

\rev{
\subsection{Point spread function analysis}\label{sec:psf-analysis}

As a first step to evaluate the imaging performance of \algoname{}, we analyze the \ac{pslr} and spatial resolution of its \ac{psf}.
Since \acp{ue} are randomly distributed with different densities in the considered scenario, the \ac{psf} can significantly change depending on the spatial coverage of the \acp{ue}.
We consider a single static isotropic point target located at $[10, 10]$~m and an \ac{snr} of $20$~dB.

In \fig{fig:boxplot-pslr}, we show the \ac{pslr} (left $y$-axis) and the average spatial resolution (right $y$-axis) obtained by coherent imaging using single clusters of \acp{ue} (Clusters) generated using the approach described in~\secref{sec:anis-clust}, and by \algoname{}, i.e., combining multiple clusters as described in~\secref{sec:non-coh-image}. 
The spatial resolution is averaged over all the clusters in the considered scenario.
In each plot, the \ac{ue} angular clustering threshold $\Delta\beta_{\mathcal{C}}$ in~\eq{eq:phase-diff} is varied from $5^\circ$ to $180^\circ$, which is equivalent to coherently combining all the available \acp{ue}.

The images obtained by the single clusters are affected by sidelobes of the \ac{psf} depending on the \acp{ue} locations, as discussed in~\secref{sec:intra-coherent-img}.
Even for a perfectly isotropic point target, coherently combining \acp{ue} across wider angles does not improve the \ac{pslr}, although it improves spatial resolution in both the $x$ and $y$ axes due to the increased spatial aperture.
Indeed, a wider angular clustering threshold causes higher sidelobes (lower \ac{pslr}) due to the higher sparsity of the \acp{ue} spatial coverage.
Spatial resolution monotonically improves when increasing $\Delta\beta_{\mathcal{C}}$ but does not capture the underlying anisotropy effect, which may degrade imaging performance even if the theoretical resolution improves.

A higher \ac{ue} density improves the \ac{pslr} by leading to more compact \ac{ue} clusters.
\algoname{} effectively improves the \ac{pslr} for all values of $\Delta\beta_{\mathcal{C}}$, by combining images obtained by multiple clusters non-coherently.
Such improvement is due to the low probability of sidelobes observed by different clusters to perfectly overlap.
As a result, sidelobes are not enhanced by non-coherent combination.
Conversely, the contributions of the point target overlap at the same location, increasing the image peak value.

Despite providing important information on the imaging performance, the \ac{pslr} and spatial resolution of the \ac{psf} do not evaluate two fundamental aspects considered in this work: (i)~they do not quantify imaging performance for extended anisotropic targets, since these intrinsically contain multiple contributions from different target parts for which the \ac{pslr} is not well defined and the spatial resolution fails to capture the anisotropy effect, (ii)~they do not consider the Doppler shift of the different target parts and the resulting image defocusing effect.
In the following, we introduce benchmark algorithms and better suited metrics to take these aspects into account.

\begin{figure}[t!]
    \begin{center}   
        \centering
        \begin{subfigure}[b]{0.48\columnwidth}
            \includegraphics[width=\columnwidth]{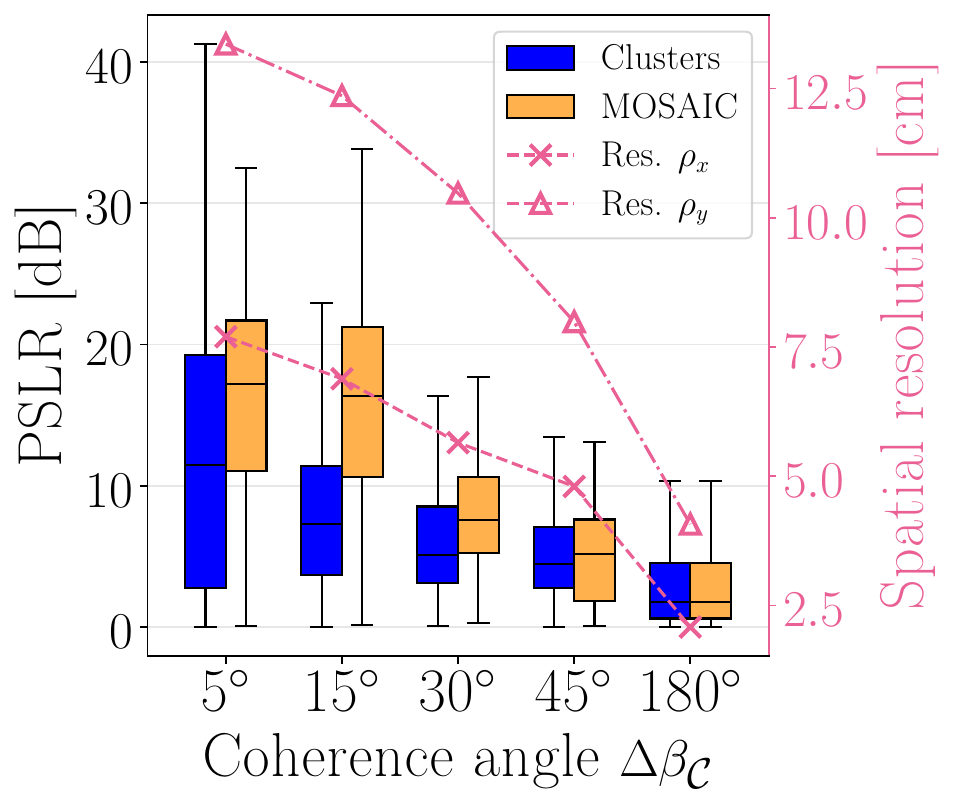}
            \caption{\ac{ue} density $0.5$~UEs/m$^2$.}
            \label{fig:boxplot-pslr-d05}
        \end{subfigure}
        \begin{subfigure}[b]{0.48\columnwidth}
            \includegraphics[width=\columnwidth]{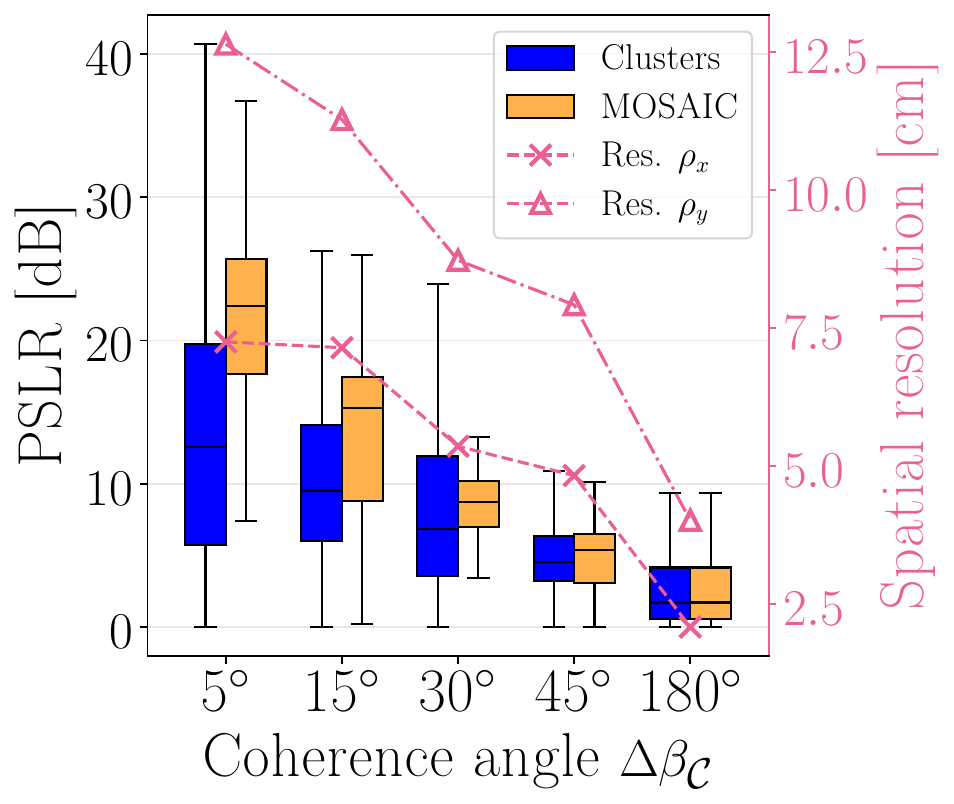}
            \caption{\ac{ue} density $1$~UEs/m$^2$.}
            \label{fig:boxplot-pslr-d1}
        \end{subfigure}
        \caption{\rev{\ac{pslr} and average spatial resolution of the \ac{psf} obtained by coherent imaging using single clusters (Clusters) and \algoname{} with different \ac{ue} density. 
        With $\Delta \beta_{\mathcal{C}}=180^\circ$, only a single cluster is used including all \acp{ue}, so single clusters and \algoname{} coincide.}}
        \label{fig:boxplot-pslr}
    \end{center}
    % \vspace{-0.3cm}
\end{figure}

}

\subsection{Benchmark algorithms for comparison}\label{sec:benchmarks}

We compare our approach to the following three algorithms.

\textbf{1) Doppler centroid and rate pre-compensation (DCR)}. 
This approach is based on \ac{sar} imaging and is presented in~\cite{noviello2015focused}. 
It first estimates the Doppler \textit{centroid}, $f_{q, i}^{\rm DC}$, and \textit{rate}, $f_{q, i}^{\rm DR}$, parameters of the Doppler spectrum observed by each \ac{ue} $(q,i)$.
Then, it applies a modified backprojection algorithm that mitigates the effect of the Doppler spectrum using the compensation term $e^{-j2\pi (f_{q, i}^{\rm DC} + f_{q, i}^{\rm DR}kT)kT }$.
Different from \algoname{},
this approach produces a single image that incorporates the contributions of all target parts.
The image focus highly depends on the quality of the approximation of the Doppler spectrum with centroid and rate parameters, which is well-suited for large objects with a rigid motion, such as ships or aircraft~\cite{noviello2015focused}. 
In our results, we show that this approximation is not suited for composite targets with independent parts.

\textbf{2) Mean Doppler compensation (Mean).} This method applies the backprojection in~\eq{eq:BP_withcompensation} replacing the target part-specific Doppler shift with the mean Doppler shift observed by each cluster of \acp{ue}.
This method assumes that the Doppler shift of the whole target is well approximated by the mean Doppler of its parts, which is only true if the parts move slowly compared to the main velocity. 

\textbf{3) No Doppler pre-compensation (No Doppl.).} This approach does not compensate for the Doppler effect and treats the targets as static. 
Except for the synchronization step, this approach follows the state-of-the-art method proposed in~\cite{tagliaferri2024cooperative}, designed for imaging static targets.
 
We do not compare \algoname{} to algorithms performing exhaustive search of the velocity of each target part due to their prohibitive complexity~\cite{moving_target_sar}.

\subsection{Imaging quality}\label{sec:im-quality}

\begin{figure}[t!]
	\begin{center}   
		\centering
	\includegraphics[width=\columnwidth]{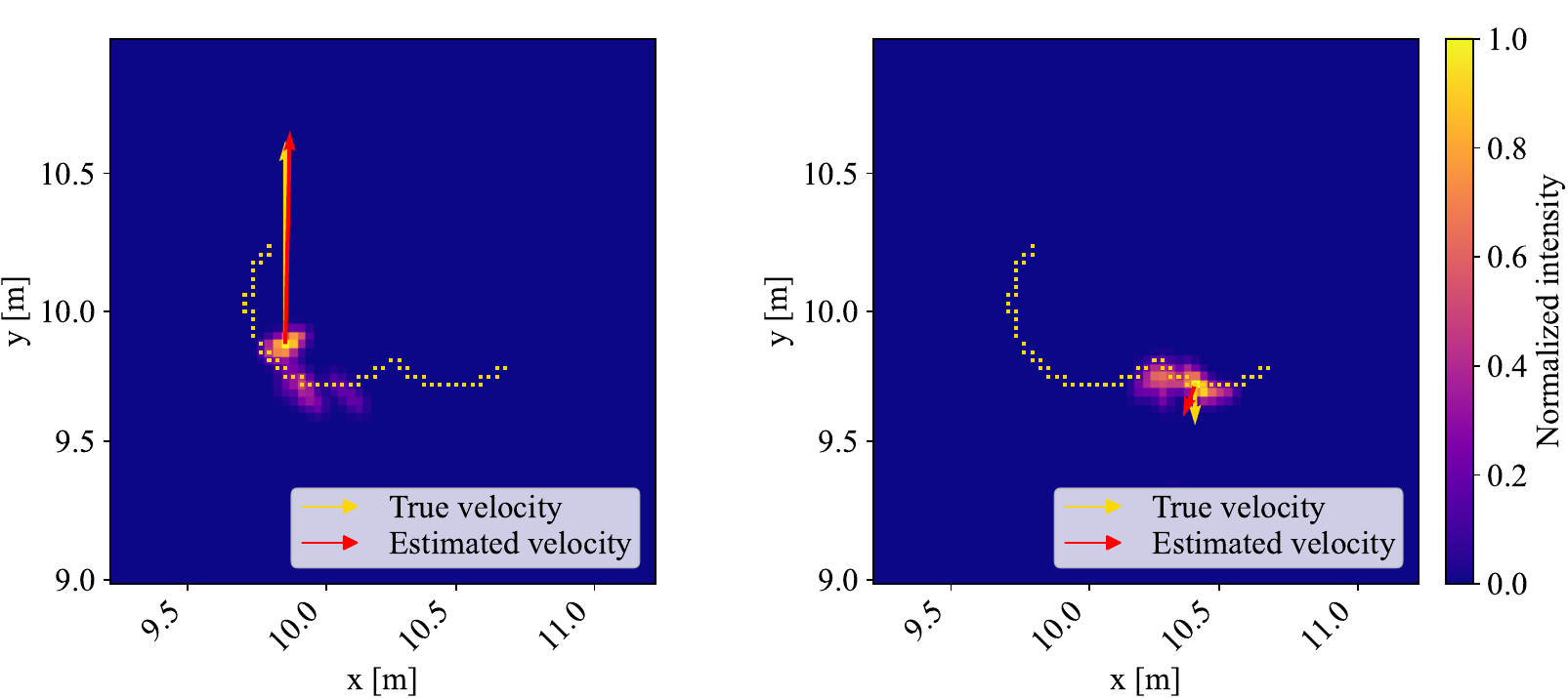}
		\caption{\algoname{} output images for each moving part of the extended target. Only the visible part of the target is represented (yellow dots). The main velocity of the target is $[0, 1]^{\mathsf{T}}$~m/s and the relative velocities of the two parts are $[0, \pm 1.5]^{\mathsf{T}}$~m/s.}
		\label{fig:images-example-split}
	\end{center}
 \vspace{-0.3cm}
\end{figure}

\algoname{} obtains a separate image for each moving part of the targets in the \ac{roi}, i.e., $I^{(g)}(\mathbf{x}), \forall g$.
This is shown in \fig{fig:images-example-split}, where a target composed of two moving circles is considered.
The main velocity of the target is $[0, 1]^{\mathsf{T}}$~m/s and the relative velocities of the two parts are $[0, \pm 1.5]^{\mathsf{T}}$~m/s.
The yellow dots represent the visible contour of the target, i.e., the portion of the target's surface that is not occluded by other target parts.
This serves as ground truth for evaluating image quality.  
In addition to the images, \algoname{} also estimates velocity vectors for each part, $\widehat{\mathbf{v}}^{(g)}, \forall g$, shown as red arrows.
The yellow arrows, instead, indicate the ground-truth velocity vectors.
\algoname{} correctly separates the images of the two moving parts thanks to the Doppler-specific backprojection followed by slow-time integration of \eq{eq:BP_withcompensation}.
Moreover, the non-coherent aggregation of the images obtained by each cluster allows reconstructing the visible region of the contour of each extended moving part, obtaining a more informative image.
Note that some regions of the target contour, although not occluded and hence visible in theory, cannot be reconstructed in the considered \ac{ue} placement scenario due to the absence of a specular reflection point with respect to the \ac{bs}.
This is the case of, e.g., the top left part of the target in~\fig{fig:images-example-split}.

\begin{figure}[t!]
     \centering
    \includegraphics[width=\columnwidth]{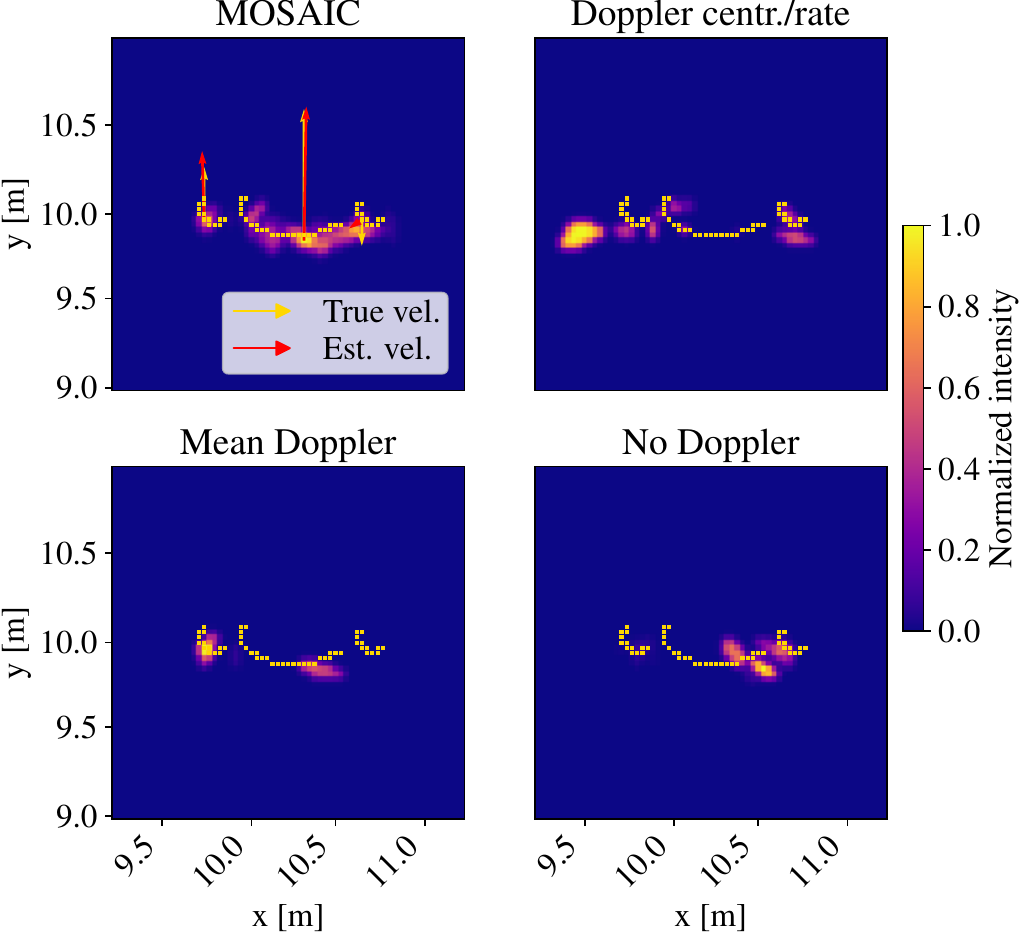}
     \caption{Example images obtained with the $4$ methods with \ac{snr}$\, =20$~dB, \ac{ue} density $1$~\acp{ue}/m$^2$, a composite target including a torso with velocity $[0, 1]$~m/s, and two arms with relative velocities  $[0, \pm1.5]$~m/s.}
     \label{fig:example-image-3trg}
     % \vspace{-3mm}
\end{figure}

\subsubsection{Visual comparison} 
In \fig{fig:example-image-3trg}, we show the global output image $\widehat{I}(\mathbf{x})$ obtained by \algoname{} with a target mimicking a human torso and arms (\fig{fig:example-image-3trg}).
In addition, we show the images obtained by the benchmark algorithms described in \secref{sec:benchmarks}.
We use $20$~dB \ac{snr} and density $1$~\acp{ue}/m$^2$.

\algoname{} consistently provides the best image quality, being able to reconstruct the visible shape of the composite target.
Conversely, the Doppler centroid and rate pre-compensation method produces unreliable results that lead to reconstructing non-existent target components or failing to detect visible parts of the target.
Compensating for just the mean Doppler or neglecting Doppler compensation causes the missed detection of several parts of the target, leading to incomplete images.

Moreover, unlike the three benchmarks, \algoname{} also estimates the velocity vectors of the moving parts, which provide additional information on the target motion.

\subsubsection{Quantitative comparison} 

To evaluate the image quality of the different algorithms, we use the 1-\ac{wd}~\cite{lu2024full} and contrast~\cite{negosanti2026ofdm} metrics.
Compared to other deviation metrics, such as the pixel-wise mean-squared error, the \ac{wd} jointly considers the intensities of each image and the distance between the considered pixels~\cite{lu2024full}.
The contrast is instead the average squared deviation from the mean image intensity, so it measures the level of detail in the image without considering adherence to the ground truth~\cite{negosanti2026ofdm}.

\begin{figure}[t!]
    \begin{center}   
        \centering
        \begin{subfigure}[b]{0.48\columnwidth}
            \includegraphics[width=\columnwidth]{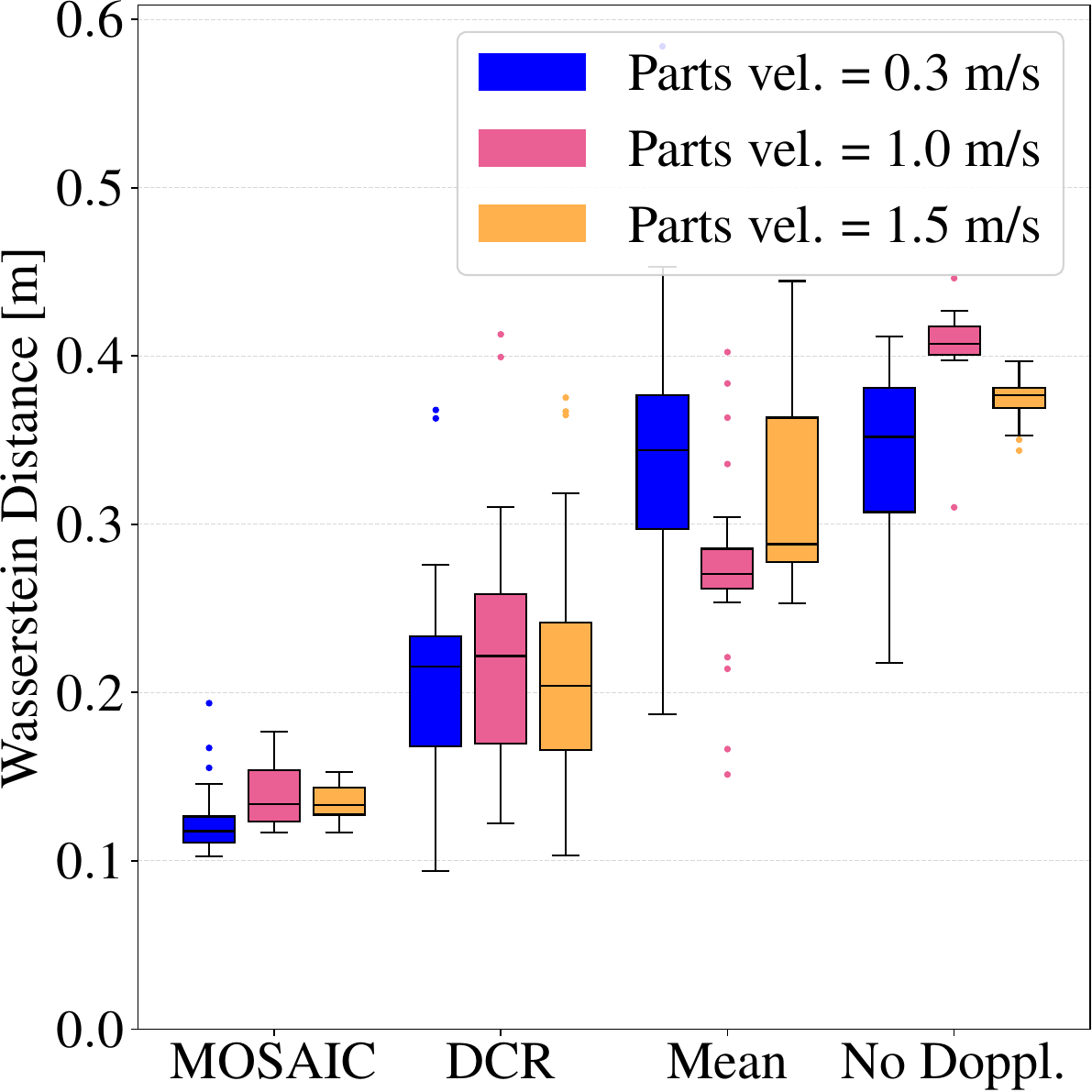}
            \caption{Wasserstein distance.}
            \label{fig:boxplot-wd-pvel-3trg}
        \end{subfigure}
        \begin{subfigure}[b]{0.48\columnwidth}
            \includegraphics[width=\columnwidth]{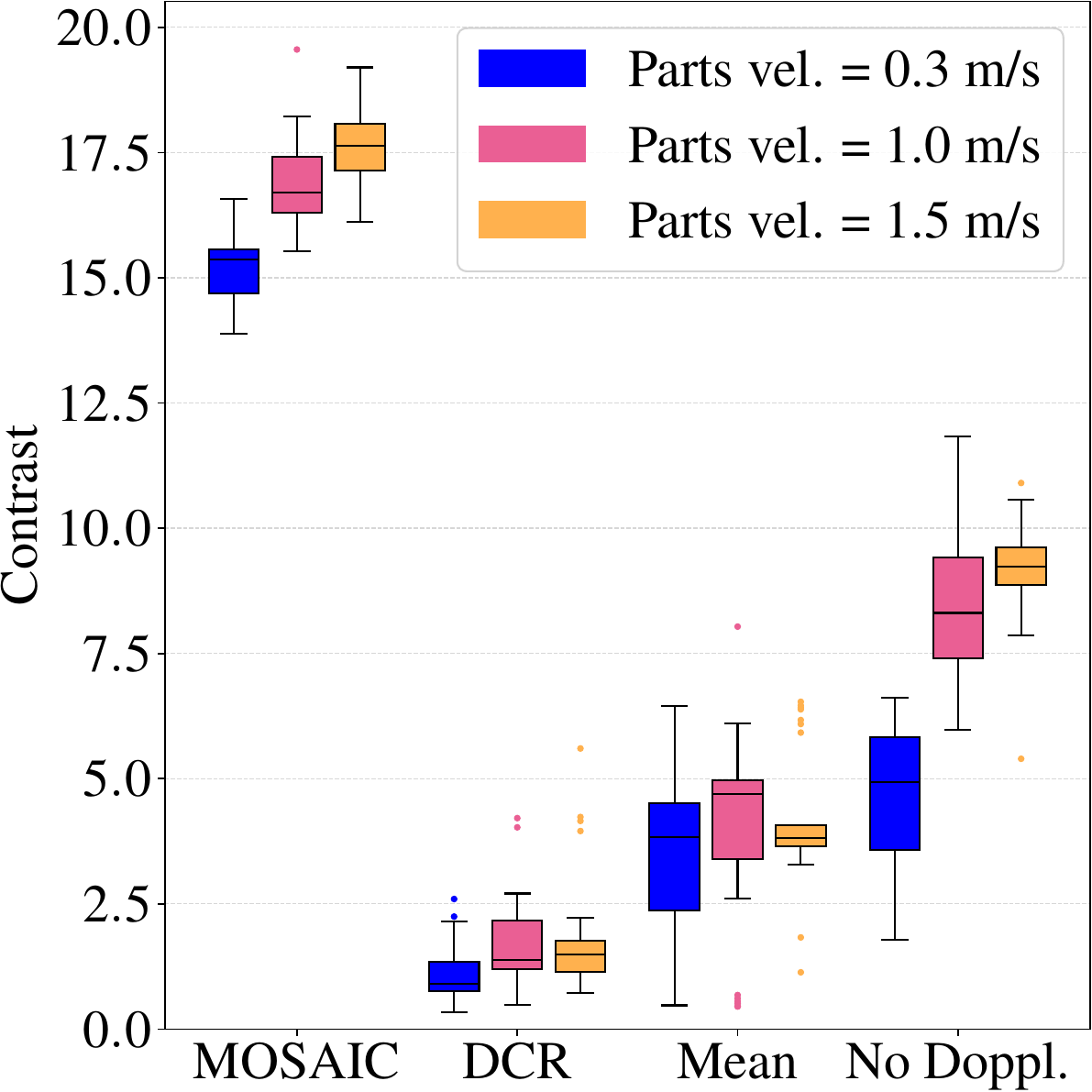}
            \caption{Contrast (higher is better).}
            \label{fig:boxplot-ic-pvel-3trg}
        \end{subfigure}
        \caption{Wasserstein distance and contrast obtained with the $4$ algorithms on the composite target with $3$ parts, with \ac{snr}$\, =20$~dB and \ac{ue} density $1$~\acp{ue}/m$^2$, with different velocities of the target parts.
        The main velocity of the target is $[0, 1]^{\mathsf{T}}$~m/s.}
        \label{fig:boxplot-wd-ic-pvel-3trg}
    \end{center}
    % \vspace{-0.3cm}
\end{figure}

In \fig{fig:boxplot-wd-ic-pvel-3trg}, we show the Wasserstein distance and contrast obtained by \algoname{} and the three benchmarks with a human target with $3$ parts representing a torso and two arms.
Boxplots are obtained over $50$ different realizations.
Different colors correspond to different relative velocities of the two arms with respect to the torso, from $\pm0.3$~m/s to $\pm 1.5$~m/s along the $y$-axis.
The main velocity of the target is $[0, 1]^{\mathsf{T}}$~m/s.
Our approach significantly outperforms the benchmarks in both metrics, achieving almost $2$ times lower Wasserstein distance and over $2$ times higher contrast compared to the closest competitors.
This demonstrates that compensating for Doppler centroid and rate, or just for the mean, is not sufficient in the case of composite targets with independently moving parts.

The reason why the minimum Wasserstein distance obtained by \algoname{} is around $12$~cm is that the top left part of the target does not produce specular reflections and hence cannot be reconstructed by coherent imaging, as noted above.

\subsubsection{Impact of \ac{snr}}

\begin{figure}[t!]
	\begin{center}   
		\centering
	\begin{subfigure}{0.48\columnwidth}\includegraphics[width=\columnwidth]{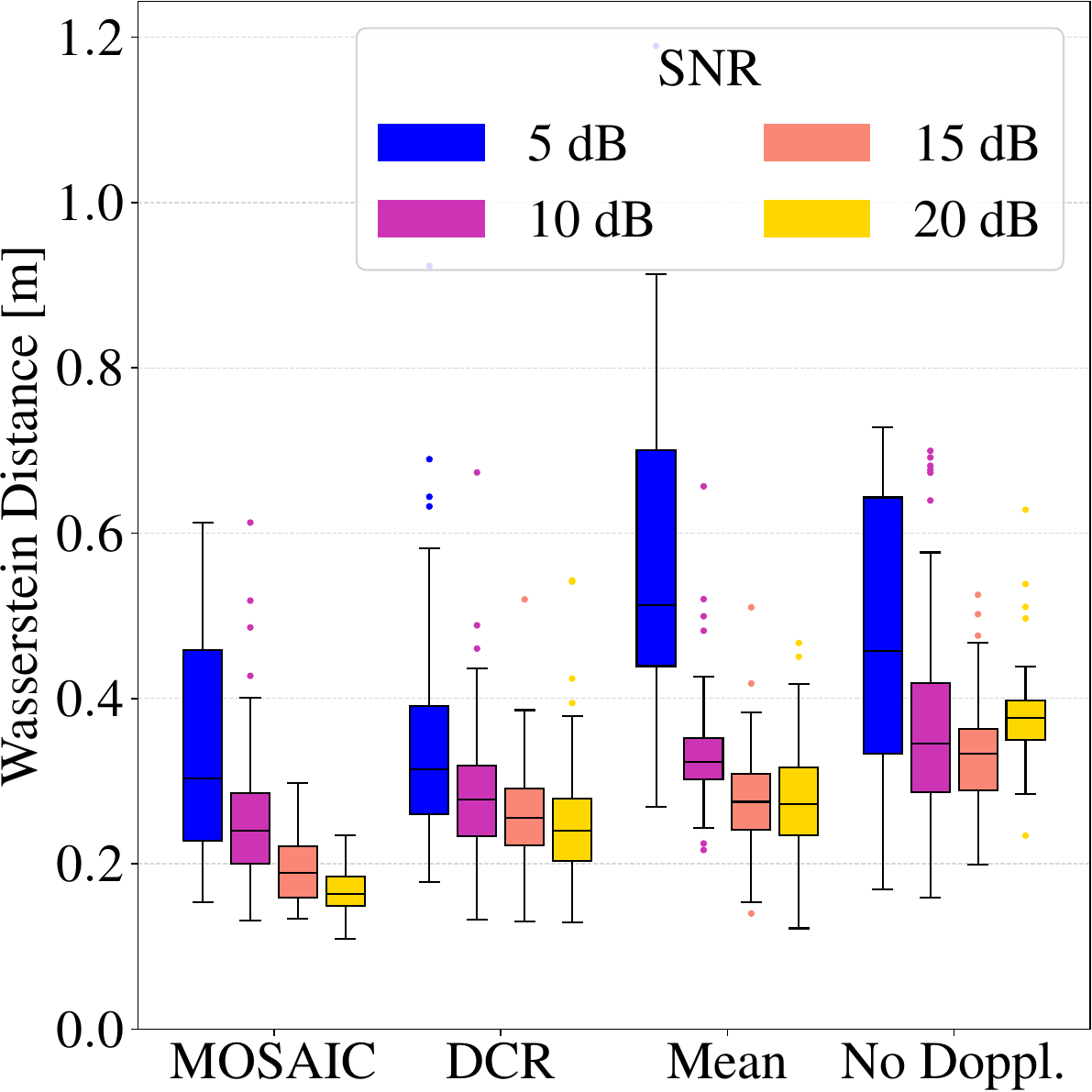}
    \caption{Varying \ac{snr}.}
    \label{fig:boxplot-wd-snr}
    \end{subfigure}
 \begin{subfigure}{0.48\columnwidth}\includegraphics[width=\columnwidth]{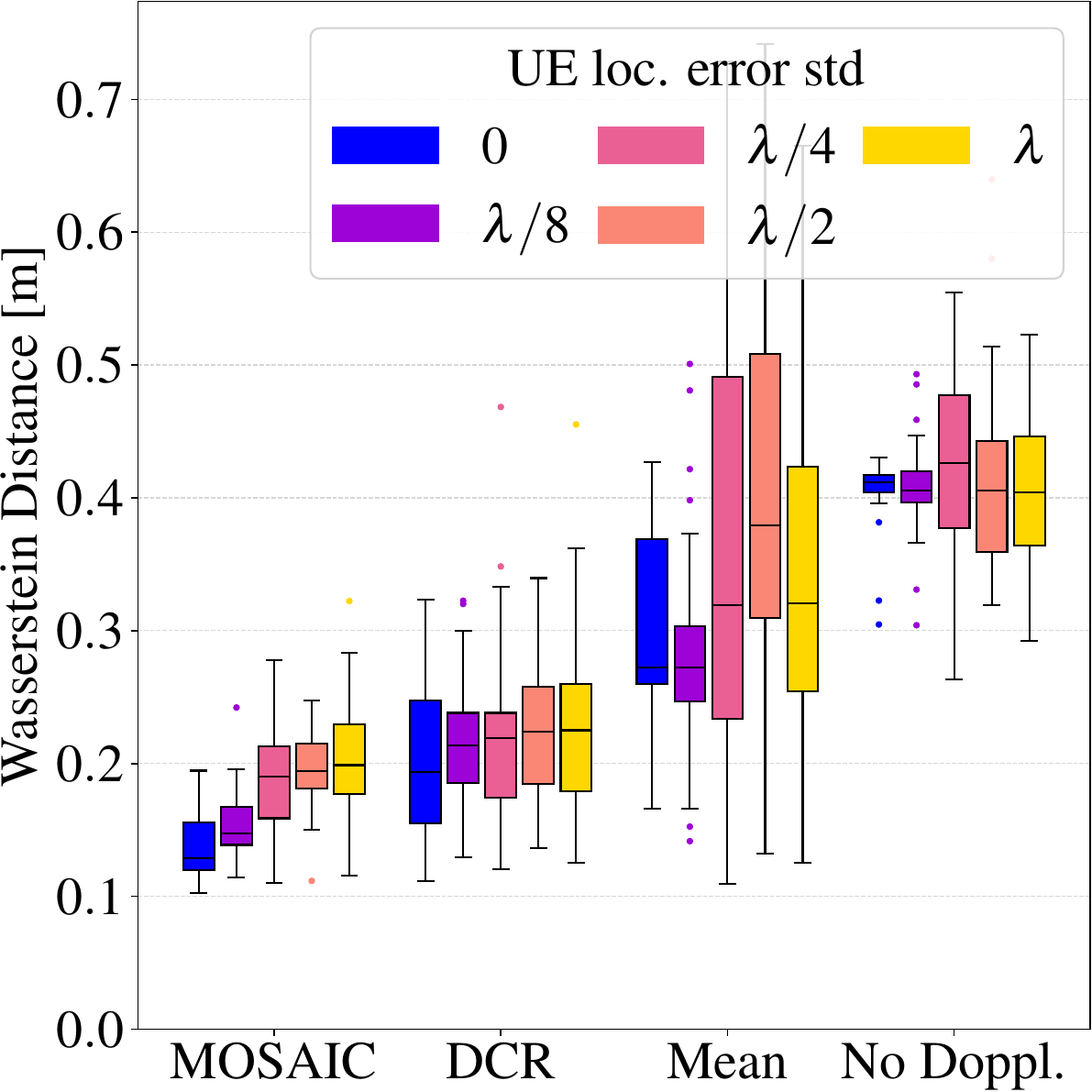}
 \caption{Varying \acp{ue} location error.}
 \label{fig:boxplot-wd-poserror}
 \end{subfigure}
		\caption{Wasserstein distance obtained on the composite target with $2$ parts. We use $0.5$~\acp{ue}/m$^2$ in (a), and $1$~\acp{ue}/m$^2$ in (b). The main velocity of the target is $[0, 1]^{\mathsf{T}}$~m/s and the relative velocities of the two parts are $[0, \pm 1.5]^{\mathsf{T}}$~m/s.}\label{fig:boxplot-wd-snr-poserror}
	\end{center}
 % \vspace{-0.3cm}
\end{figure}

In \fig{fig:boxplot-wd-snr}, we show the Wasserstein distance obtained by the different methods with \ac{snr}$\in \{5, 10, 15, 20\}$~dB.
The main velocity of the target is $[0, 1]^{\mathsf{T}}$~m/s and the relative velocities of the two parts are $[0, \pm 1.5]^{\mathsf{T}}$~m/s.
Our approach significantly outperforms the benchmarks for \ac{snr} larger than or equal to $10$~dB.
With an \ac{snr} lower than or equal to $5$~dB, all methods show degraded performance and generate images with large \ac{wd} that do not match the shape of the target.
This proves that imaging extended targets requires a medium or high \ac{snr} on the scattered signal.
Even in this case, \algoname{} obtains a lower median Wasserstein distance, although its variability is slightly higher than Doppler centroid and rate compensation.

\subsubsection{Impact of different carrier frequencies}

\begin{figure}[t!]
     \centering
    \includegraphics[width=\columnwidth]{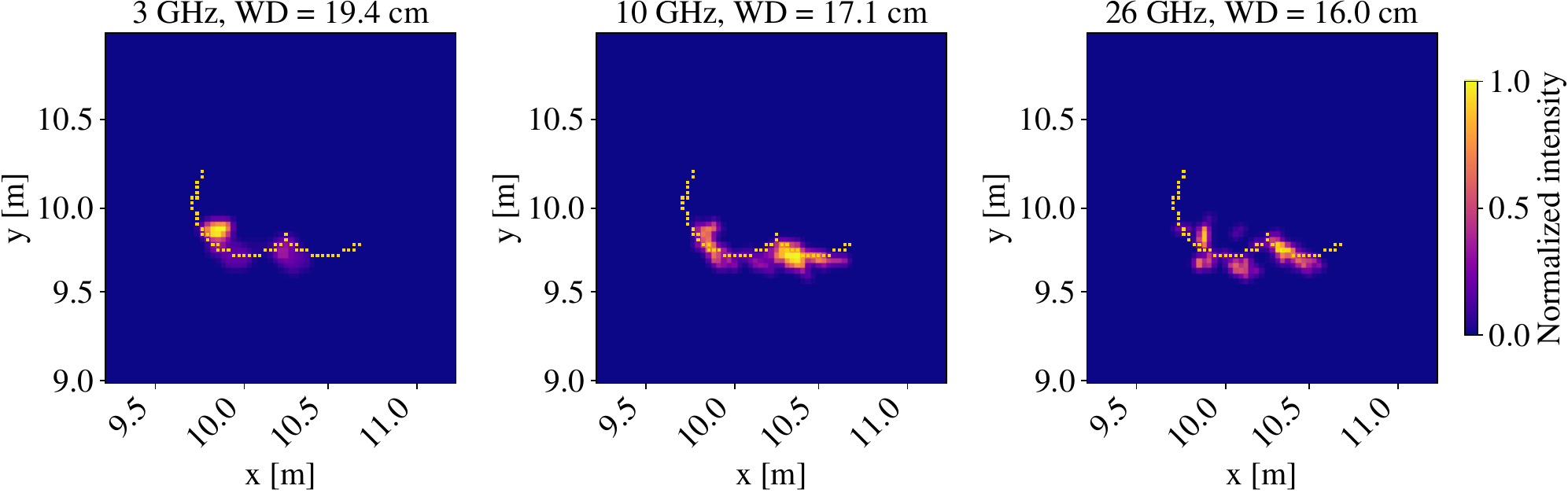}
     \caption{Example images obtained by \algoname{} with \ac{snr}$\, =20$~dB, \ac{ue} density $1$~\acp{ue}/m$^2$, main velocity $[0, 1]$~m/s, and relative velocity of the target parts $[0, \pm1.5]$~m/s for different carrier frequencies ($3,10,26$~GHz) and a bandwidth of $400$~MHz.}
     \label{fig:example-image-carriers}
     % \vspace{-2mm}
\end{figure}

In \fig{fig:example-image-carriers}, we show the imaging results obtained with our approach used with different carrier frequencies, selected among $f_0 \in \{3,10,26\}$~GHz.
We use a composite target with two circle-shaped moving parts.
The obtained images visually match the real shape of the object in all cases.
In general, higher carrier frequencies lead to a more fragmented reconstruction since small phase errors have a greater impact on the carrier phase for smaller wavelengths. 
However, in terms of Wasserstein distance (reported above each image), the result is not necessarily worse, as shown in \tab{tab:carrier-errors}.

\subsubsection{Impact of \ac{ue} localization error}\label{sec:loc-error-impact}

\begin{figure}[t!]
     \centering
    \includegraphics[width=\columnwidth]{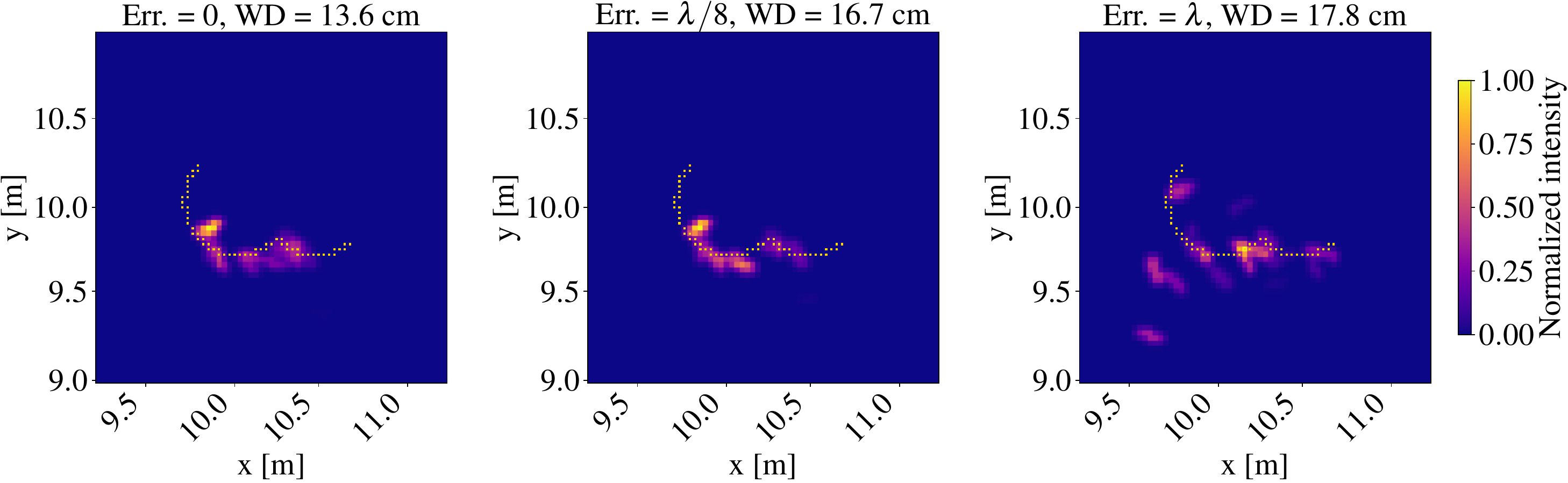}
     \caption{Example images obtained with \algoname{} with \ac{snr}$\, =20$~dB, \ac{ue} density $1$~\acp{ue}/m$^2$, main velocity $[0, 1]$~m/s, and relative velocity of the target parts $[0, \pm1.5]$~m/s. Each image is obtained with a different value of the \ac{ue} localization error STD, and we report the corresponding value of the Wasserstein distance metric.}
     \label{fig:example-image-pvel03}
     % \vspace{-3mm}
\end{figure}

Phase-coherent imaging is sensitive to errors in the localization of the \acp{ue}, since these map to phase errors.
In \fig{fig:example-image-pvel03}, we show the images obtained by our method when adding a localization error to the \ac{ue} positions using the same setup as in the previous paragraph.
The \ac{std} of the localization error is selected as a function of the wavelength, in $\{0, \lambda/8, \lambda\}$.
Thanks to the non-coherent combination of the images obtained by each cluster of \acp{ue}, the global images still approximate the shape of the target even when the localization error has a standard deviation of $\lambda$, which would prevent coherent-only imaging completely.
Nevertheless, an expected degradation is visible.

In \fig{fig:boxplot-wd-poserror}, we report the Wasserstein distance obtained by the $4$ algorithms in the presence of \ac{ue} localization error.
Our approach obtains good or sufficient results even with the maximum localization error (with standard deviation equal to $\lambda$).
Existing automatic registration methods~\cite{tagliaferri2024cooperative} could be applied in this case to align the images obtained by each \ac{ue} before the coherent combination, improving the results.

\begin{table}[t!]
    \centering
    \footnotesize
    \caption{Mean and standard deviation of the \ac{wd} obtained by \algoname{} for different carrier frequencies $f_0$.}
    \setlength{\tabcolsep}{3pt}
    \begin{tabular}{cccccc}
    \toprule
   &$3$ GHz&$7$ GHz&$10$ GHz&$15$ GHz&$26$ GHz\\
   \cmidrule{2-6}
      \ac{wd} [cm] &$22.3\pm0.5$&$15.3\pm0.2$&$14.3\pm0.2$&$14.2\pm0.1$&$15.8\pm0.2$\\

        \bottomrule
    \end{tabular}
    \label{tab:carrier-errors}
\end{table}

\subsubsection{Impact of different \ac{ue} clustering angle thresholds}\label{sec:clust-thr-impact}

\begin{figure}[t!]
    \begin{center}   
        \centering
        \begin{subfigure}[b]{0.48\columnwidth}
            \includegraphics[width=\columnwidth]{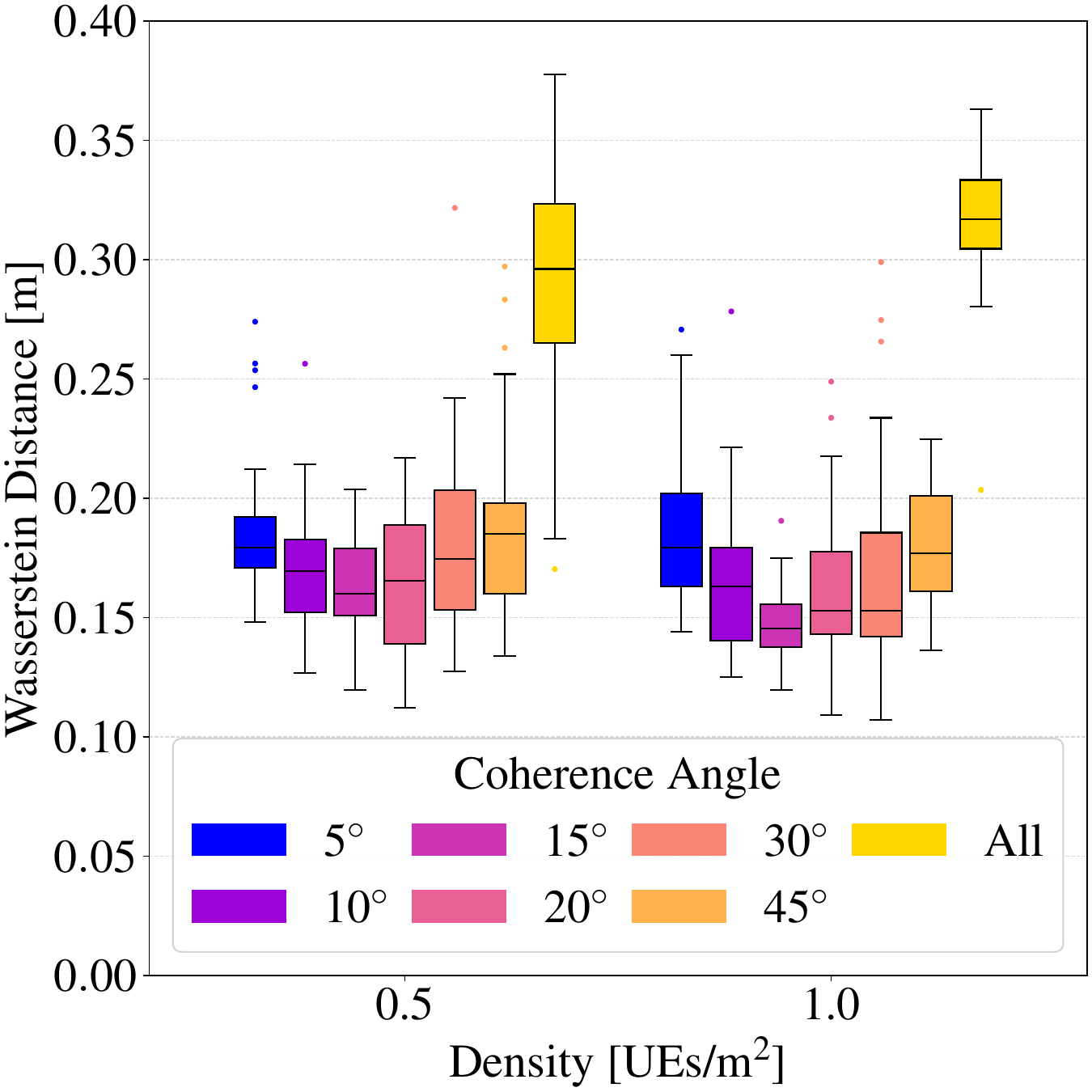}
            \caption{Varying clustering angle.}
            \label{fig:boxplot-wd-angles}
        \end{subfigure}
        \begin{subfigure}[b]{0.495\columnwidth}
            \includegraphics[width=\columnwidth]{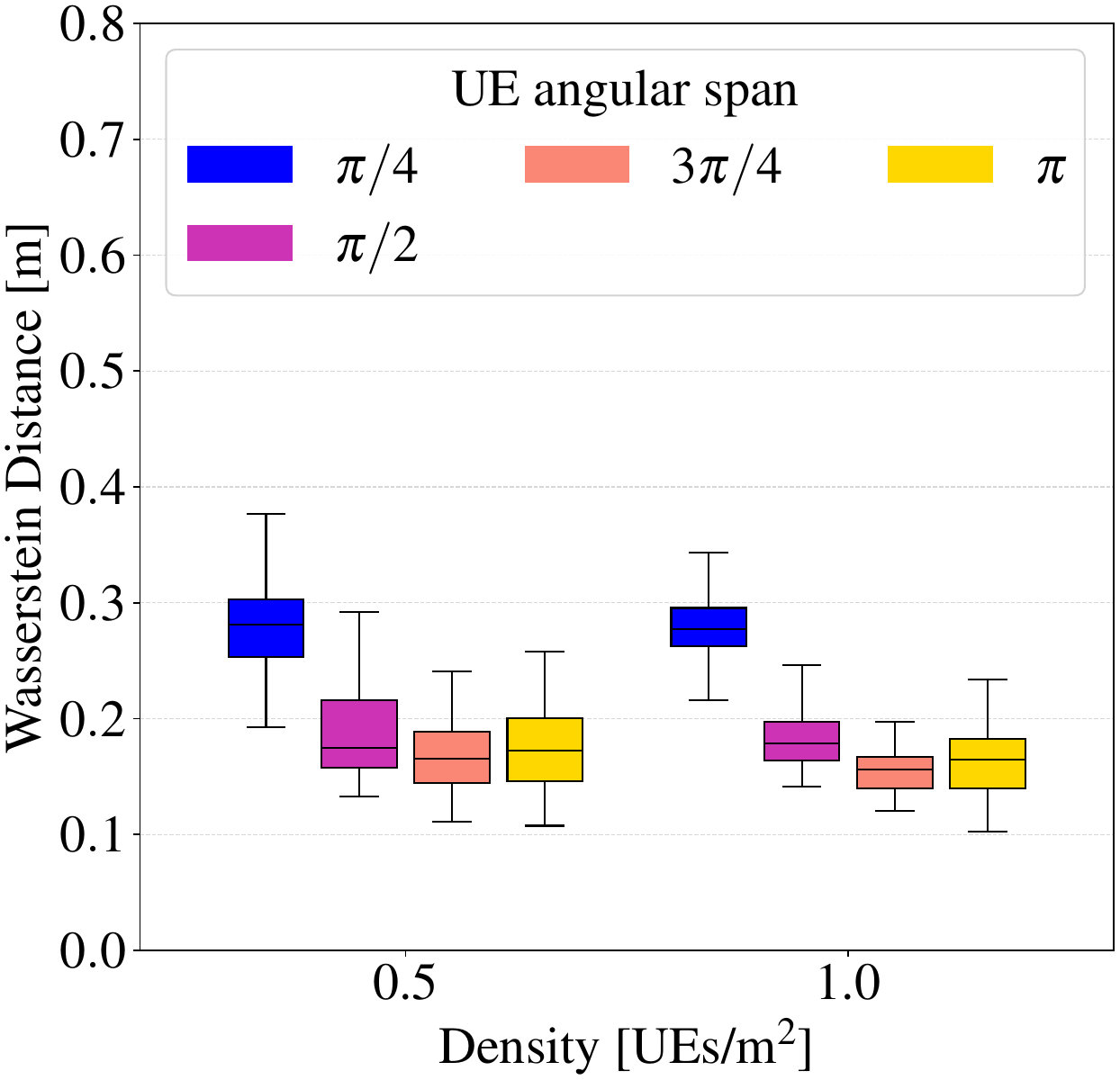}
            \caption{Varying \ac{ue} span.}
            \label{fig:boxplot-wd-uespan}
        \end{subfigure}
        \caption{Wasserstein distance obtained by \algoname{} with \ac{snr}$\, = 20$~dB, $f_0 = 7$~GHz. ``All" indicates the coherent combination of all the available \acp{ue}. The main velocity of the target is $[0, 1]^{\mathsf{T}}$~m/s and the relative velocities of the two parts are $[0, \pm 1.5]^{\mathsf{T}}$~m/s.}
        \label{fig:boxplot-wd-angles-uespan}
    \end{center}
    \vspace{-0.3cm}
\end{figure}

As expressed by \eq{eq:phase-diff}, one of the key parameters of our approach is the angle threshold used to cluster the \acp{ue} based on the coherence of the target parts. 
In \fig{fig:boxplot-wd-angles}, we show the imaging results obtained by our method for different values of the angle threshold used for clustering.
More specifically, we substitute the correct angle threshold obtained via \eq{eq:phase-diff} with $\Delta\beta$, chosen in $\{5^{\circ}, 10^{\circ}, 15^{\circ}, 20^{\circ}, 30^{\circ}, 45^{\circ}\}$,  to show the impact of deviations from the correct value.
Moreover, we also report the Wasserstein distance obtained by coherently combining all the available \acp{ue}, hence using a single cluster, indicated by ``All" in~\fig{fig:boxplot-wd-angles}.
The main velocity is $[0, 1]^{\mathsf{T}}$~m/s and the relative velocities of the two target parts are $[0, \pm 1.5]^{\mathsf{T}}$~m/s.

These results reveal a \textit{trade-off} between image focus and target reconstruction.
For low $\Delta \beta$, images are accurately focused since the target is observed over small angular apertures, where it is coherent, but the resolution gain introduced by coherent processing is limited by the small angular aperture.
Conversely, for wide $\Delta \beta$, the coherent processing result degrades due to the combination of incoherent return signals from the extended target.
The minimum Wasserstein distance is thus obtained for intermediate values of $\Delta \beta$, specifically at \mbox{$\Delta \beta = 15^{\circ}$}, where the algorithm strikes a balance between resolution gain due to coherent processing and the target anisotropy that prevents fully coherent imaging over wide apertures.
\rev{This value is consistent with the theoretical one obtained from \eq{eq:phase-diff}, which gives $\Delta \beta = 15.9^\circ$ for $\bar{\beta}_{\mathcal{C}}=45^\circ$.}
Coherently combining all the \acp{ue} (yellow box) gives degraded images that do not match the real target reflectivity, hence demonstrating the validity of the hierarchical approach.

\subsubsection{Impact of \ac{ue} angular span}

In \fig{fig:boxplot-wd-uespan}, we show the Wasserstein distance obtained while varying the angle span of the \acp{ue} around the \ac{roi}, for different values of the \ac{ue} density.
This demonstrates that the non-coherent image combination in~\eq{eq:non-coh-image} improves the imaging performance, reducing the Wasserstein distance by $40$\% when using a \ac{ue} span of $\pi$ compared to $\pi /4$.
As expected, a higher \ac{ue} density is beneficial and leads to better images.
The Wasserstein distance for \ac{ue} span equal to $\pi$ shows a slightly higher variance compared to $3\pi/4$, due to the possibility that some \acp{ue} at the very edge of the field of view of the \ac{bs} are present and produce degraded images. 

\rev{
\subsubsection{Impact of object parts rotation around a common axis}\label{sec:rotation-impact}

In the previous sections, we analyzed \algoname{}'s performance under the assumption of purely translational motion of each body part.
In \fig{fig:boxplot-rotation} instead, we report the imaging performance of \algoname{} on a target composed of two moving parts with circular shape rigidly rotating around a common axis with angular speed $\omega \in \{1, 2, 5, 10\}$~rad/s, while jointly translating with velocity vector \mbox{$\mathbf{v} = [0, -1]^{\mathsf{T}}$}.
In this scenario, each patch composing the extended target parts has a slightly different velocity vector due to the rotation, producing a Doppler spectrum that exhibits more Doppler spread around the peak components.
Nevertheless, \algoname{} still robustly separates the parts in the Doppler domain and outperforms the benchmark algorithms in image quality.
In the worst case of $\omega = 10$~rad/s, DCR which is the closest competitor achieves over $25\%$ higher Wasserstein distance to the ground truth image.
This demonstrates that, even in case of rigid rotation of the target parts, our approach of detecting and compensating for the Doppler peaks brings a substantial advancement over compensating for the Doppler centroid and rate.

\begin{figure}[t!]
	\begin{center}   
		\centering
{\includegraphics[width=\columnwidth]{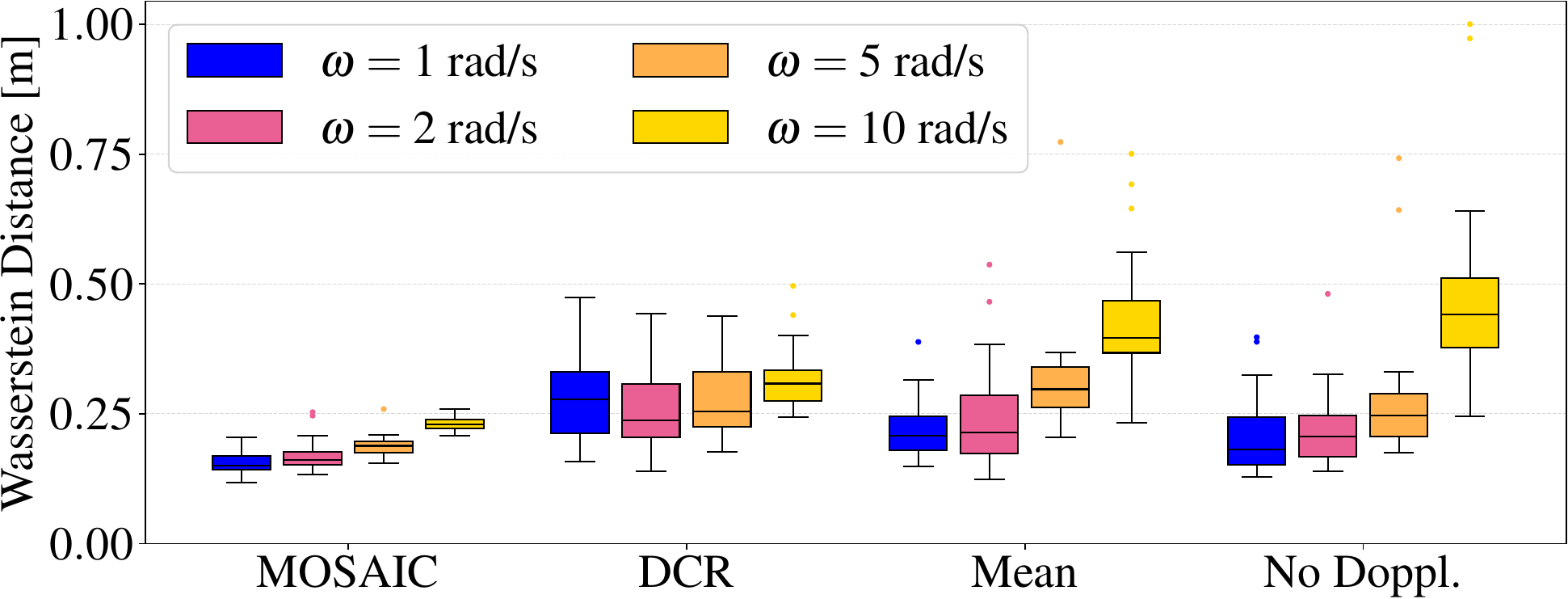}}
		\caption{\rev{Wasserstein distance obtained by \algoname{} and the benchmark algorithms on a target with two moving parts rotating around a common center with angular velocity $\omega \in \{1, 2, 5, 10\}$~rad/s, while jointly translating with velocity vector $\mathbf{v} = [0, -1]^{\mathsf{T}}$. The \ac{ue} density is $1$~\ac{ue}/m$^2$, the \ac{snr} is $20$~dB.}}
		\label{fig:boxplot-rotation}
	\end{center}
 \vspace{-0.3cm}
\end{figure}
}

\subsection{Velocity estimation performance}\label{sec:vel-est}

\begin{figure}[t!]
	\begin{center}   
		\centering
{\includegraphics[width=\columnwidth]{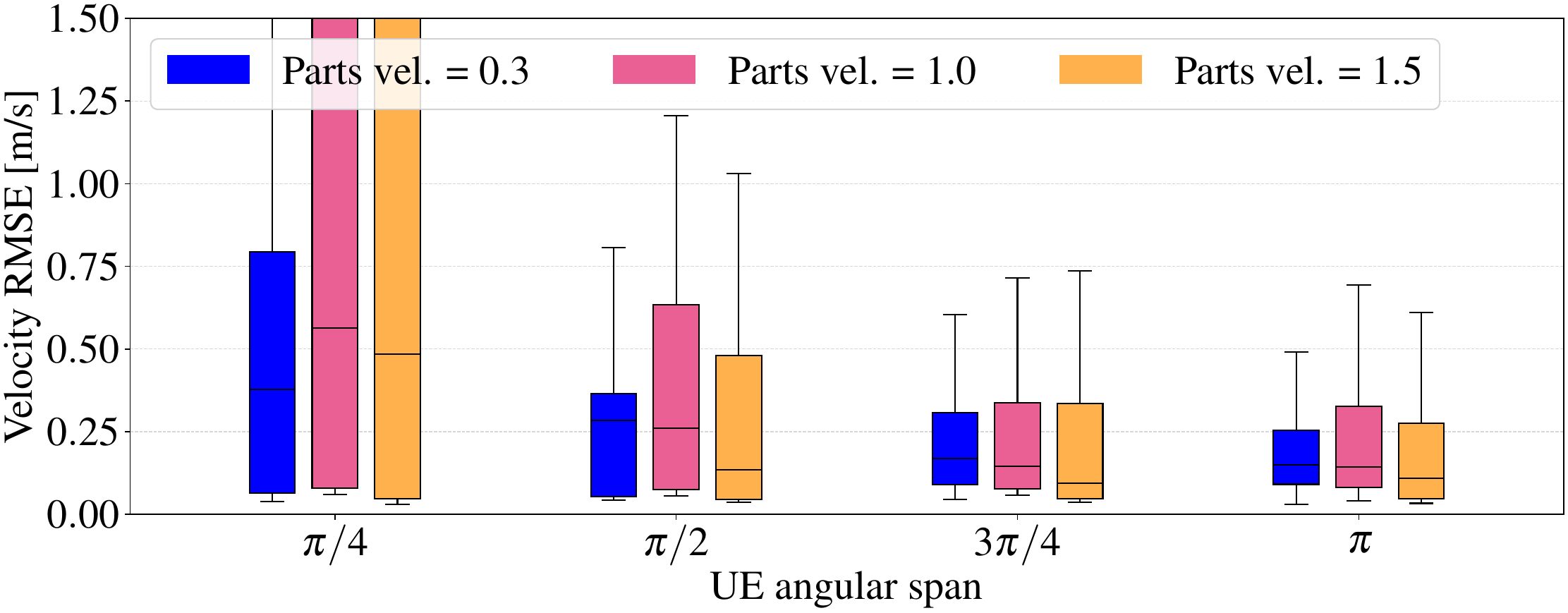}}
		\caption{Velocity \ac{rmse} obtained with \ac{snr}$\, = 20$~dB, $f_0=7$~GHz \ac{ue} span angle ($x$-axis). Different colors indicate the relative velocities of the target parts expressed in m/s.}
		\label{fig:boxplot-vel-rmse}
	\end{center}
 \vspace{-0.3cm}
\end{figure}

\fig{fig:boxplot-vel-rmse} shows the velocity estimation \ac{rmse} obtained by \algoname{} on a target composed of two circular parts, with \ac{snr}$\, =20$~dB, $f_0=7$~GHz.
We vary the relative velocities of the two target parts in $\{\pm 0.3, \pm 1, \pm 1.5\}$ (along the $y$ axis), and the \ac{ue} span angle around the \ac{roi} from $\pi/4$ to $\pi$.

Using a wider \ac{ue} span around the \ac{roi} is beneficial for the velocity estimation, since the velocity vectors of each part are observed (through the Doppler shifts) from multiple points of view.
Moreover, analyzing errors across different relative velocities of the moving parts reveals that higher relative velocities lead to lower median estimation errors but sometimes introduce larger variance.
This shows that \algoname{} estimates target part-specific velocity vectors with dm/s-level accuracy provided that \acp{ue} are sufficiently spread around the \ac{roi}.

\rev{\subsection{Discussion on the limitations}\label{sec:limitations}
In this section, we summarize and discuss the main limitations of \algoname{} that emerge from our numerical results. 

\subsubsection{\ac{ue}-\ac{bs} phase-level synchronization} 
Achieving phase coherence among the \acp{ue} in each cluster is needed for coherent imaging. 
Since such fine-grained synchronization cannot be guaranteed by network protocols, over-the-air methods such as the one presented in \secref{sec:sync} are required.
Moreover, accurate \ac{ue} localization with error lower than $\lambda/4$ is also essential to maintain phase coherence.
Although this is not covered in the present work, methods to mitigate localization errors at the \ac{bs} by registering the images obtained by each \ac{ue}-\ac{bs} pair are available \cite{tagliaferri2024cooperative}.

\subsubsection{Estimating the target curvature radius}
 \algoname{} requires a prior estimate of the maximum curvature radius of the target, $R_{\max}$, to cluster \acp{ue}.
 The value of $R_{\max}$ affects the clustering results and, consequently, the final image quality, as discussed in~\secref{sec:anis-clust}. 
 In particular, our results show that underestimating the curvature radius implies assuming that targets are less anisotropic than in reality, compromising the performance. 
 Conversely, \algoname{} is more robust to overestimating $R_{\max}$ since this leads to a more conservative clustering of the \acp{ue}.

\subsubsection{Velocity estimation accuracy}

Although \algoname{} provides accurate velocity estimation, as shown in \secref{sec:vel-est}, its performance is limited by the angular span of the \acp{ue}.
More than a limitation of the algorithm, this is intrinsic to the geometry of the scenario, since lower \ac{ue} span directly translates into more similar, and hence less useful, observations of the Doppler shift.
This shows that the \ac{ue} distribution plays a key role in the \ac{isac} performance, not only in terms of \ac{ue} density, but also in terms of diversity of the points of view.

}

\section{Concluding remarks}\label{sec:conclusion}

This work introduced \algoname{}, a hierarchical imaging algorithm for composite moving targets in distributed asynchronous \ac{isac} networks combining coherent and non-coherent processing.
We addressed two fundamental limitations of existing methods: \textit{(i)}~the lack of a realistic model of anisotropic target scattering and \textit{(ii)}~the inability to produce images of composite targets with independently moving parts.
A key insight is that fully coherent processing across widely distributed \acp{ue} is not advisable for anisotropic targets outside of a certain coherence interval.
Hence, \algoname{} limits coherent fusion to clusters of closely located \acp{ue}, combining images from different clusters non-coherently, achieving a trade-off between coherent resolution gain and image quality.
Moreover, \algoname{} recasts Doppler shift as an additional degree of freedom for resolving individual target parts, producing focused images for each moving component alongside per-part velocity estimates.
Numerical simulations confirm that \algoname{} outperforms existing methods by 35\% to 45\% in image quality and achieves dm/s-level velocity estimation errors.
\rev{Future work will refine anisotropy models and develop network-level strategies to optimally cluster \acp{ue}.
In addition, the development of a distributed \ac{isac} experimental platform and the assessment of the proposed method under realistic hardware and propagation conditions constitute an important further step for validating the proposed approach.}

\appendices

\section{}\label{sec:app-a}

The Rx signal is obtained by integrating the scattering returns from the whole contour of the target as
\begin{equation}\label{eq:rx-signal-anisotropy}
    y = \int_0^{2 \pi} \zeta(\theta)\varrho(\theta) e^{- j 2\pi f_0D(\theta)/c} R d \theta.
\end{equation}
$\theta_1^*$ is a \textit{stationary point} of the path length $D(\theta)$.
Call its first derivative 
   $D'(\theta) = -R  (\mathbf{u}_1(\theta) + \mathbf{u}_{\rm bs}(\theta))^{\mathsf{T}}\mathbf{t}(\theta)$,
where $\mathbf{t}(\theta)$ is the tangent vector to the target's surface in~$\mathbf{x}(\theta)$ (orthogonal to $\mathbf{n}(\theta)$).
$D'(\theta_1^*) = 0$, since $\mathbf{u}_1(\theta_1^*) + \mathbf{u}_{\rm bs}(\theta_1^*)$ has the same direction as $\mathbf{n}(\theta_1^*)$.
Hence, we approximate the integral in \eq{eq:rx-signal-anisotropy} using the stationary phase principle~\cite{bleistein1975asymptotic} around $\theta_1^*$:
\begin{equation}\label{eq:stat-phase-y}
    y \approx \underbrace{\zeta(\theta_1^*) \sqrt{\frac{c R^2}{f_0 |D''(\theta_1^*)|}}e^{-j  \frac{\pi}{4}\mathrm{sign}(D''(\theta_1^*))} }_{C(\theta_1^*)}\varrho(\theta_1^*)e^{-j 2\pi f_0D(\theta_1^*)/c},
\end{equation}
where $D''(\theta_1^*)$ is the second derivative of the path length computed in $\theta_1^*$, which for $R$ much smaller than the \ac{roi} distance from the \ac{ue} or \ac{bs} equals $D''(\theta_1^*) \approx  2R\cos(\beta^*_1 /2)$.
By compensating for the carrier phase with a term $e^{j2\pi f_0D(\theta)/c}$, similarly to what is done in \eq{eq:BP}, the \ac{bs} obtains the following bistatic image, evaluated at point $\mathbf{x}(\theta)$, 
\begin{align}
    I_1(\theta) =  C(\theta_1^*) \varrho(\theta_1^*) e^{j 2\pi f_0 \left[D(\theta) - D(\theta_1^*)\right]/c}.\label{eq:stat-phase-image}
\end{align}
Finally, \eq{eq:phase_difference_extended_target} is obtained by approximating \eq{eq:stat-phase-image} using the second-order Taylor expansion of $D(\theta)$ around $\theta_1^*$.

\IEEEpeerreviewmaketitle

\bibliography{references}
\bibliographystyle{IEEEtran}
\end{document}

%% file: acronyms.tex
\acrodef{prop}[\textit{MIMORPH}]{MIMO Radio Platform for Heterogeneous wireless systems}

\acrodef{abft}[A-BFT]{Association Beamforming Training}
\acrodef{ack}[ACK]{Acknowledge}
\acrodef{adc}[ADC]{Analog-to-Digital Converter}
\acrodef{aoa}[AoA]{Angle of Arrival}
\acrodef{aod}[AoD]{Angle of Departure}
\acrodef{ap}[AP]{Access Point}
\acrodef{amc}[AMC]{Advanced Mezzanine Card}
\acrodef{awv}[AWV]{Antenna Wave Vector}
\acrodef{axi}[AXI]{Advanced eXtensible Interface}
\acrodef{bs}[BS]{Base Station}
\acrodef{ber}[BER]{Bit Error Rate}
\acrodef{bft}[BFT]{Beamforming Training}
\acrodef{bp}[BP]{Backprojection}
\acrodef{brp}[BRP]{Beam Refinement Phase}
\acrodef{cs}[CS]{Compressed Sensing}
\acrodef{cdf}[CDF]{Cumulative Distribution Function}
\acrodef{cef}[CEF]{Channel Estimation Field}
\acrodef{cfo}[CFO]{Carrier Frequency Offset}
\acrodef{cir}[CIR]{Channel Impulse Response}
\acrodef{cfr}[CFR]{Channel Frequency Response}
\acrodef{csi}[CSI]{Channel State Information}
\acrodef{csirs}[CSI-RS]{CSI-Reference Signal}
\acrodef{cs}[CS]{Compressed Sensing}
\acrodef{cv}[CV]{Constant Velocity}
\acrodef{cnn}[CNN]{Convolutional Neural Network}
\acrodef{cots}[COTS]{Commercial-Off-The-Shelf}
\acrodef{cfar}[CFAR]{Constant False Alarm Rate}
\acrodef{dft}[DFT]{Discrete Fourier Transform}
\acrodef{disac}[DISAC]{Distributed ISAC}
\acrodef{dl}[DL]{Deep Learning}
\acrodef{dma}[DMA]{Direct Memory Access}
\acrodef{dmg}[DMG]{Directional Multi Gigabit}
\acrodef{dti}[DTI]{Data Transfer Interval}
\acrodef{edmg}[EDMG]{Enhanced Directional Multi Gigabit}
\acrodef{ekf}[EKF]{Extended Kalman Filter}
\acrodef{elu}[ELU]{Exponential-Linear Unit}
\acrodef{fmcw}[FMCW]{Frequency-Modulated Continuous-Wave}
\acrodef{fov}[FOV]{Field-of-View}
\acrodef{ft}[FT]{Fourier Transform}
\acrodef{fr2}[FR2]{Frequency Range 2}
\acrodef{fr3}[FR3]{Frequency Range 3}
\acrodef{gpio}[GPIO]{General Purpose Input/Output}
\acrodef{gsps}[GSPS]{Giga-Samples per Second}
\acrodef{har}[HAR]{Human Activity Recognition}
\acrodef{ht}[HT]{High Throughput}
\acrodef{idft}[IDFT]{Inverse Discrete Fourier Transform}
\acrodef{if}[IF]{Intermediate Frequency}
\acrodef{ifs}[IFS]{Inter-Frame Spacing}
\acrodef{iht}[IHT]{Iterative Hard Thresholding}
\acrodef{ista}[ISTA]{Iterative Shrinkage-Thresholding Algorithm}
\acrodef{isac}[ISAC]{Integrated Sensing and Communication}
\acrodef{isafs}[ISAFS]{Iterative Spatial Ambiguity Function Subtraction}
\acrodef{jcs}[JCS]{Joint Communication and Sensing}
\acrodef{jpdaf}[JPDAF]{Joint Probabilistic Data Association Filter}
\acrodef{lo}[LO]{Local Oscillator}
\acrodef{los}[LoS]{Line-of-Sight}
\acrodef{lbm}[LBM]{Loop-Back Memory}
\acrodef{mae}[MAE]{Mean Absolute Error}
\acrodef{mcs}[MCS]{Modulation and Coding Scheme}
\acrodef{md}[$\mu$D]{micro-Doppler}
\acrodef{mimo}[MIMO]{Multiple Input Multiple Output}
\acrodef{mosaic}[MOSAIC]{Moving Object Sensing under Anisotropy via Imaging and Coherent-noncoherent fusion}
\acrodef{mmwave}[mmWave]{Millimeter-Wave}
\acrodef{msps}[MSPS]{Mega-Samples per Second}
\acrodef{mu}[MU]{Multiple User}
\acrodef{MUSIC}[MUSIC]{MUlti SIgnal Classification}
\acrodef{nac}[NAC]{Normalized Auto Correlation}
\acrodef{nco}[NCO]{Numerical Controlled Oscillator}
\acrodef{nlos}[NLoS]{Non-LoS}
\acrodef{nn}[NN]{Neural Network}
\acrodef{nls}[NLS]{Nonlinear Least-Squares}
\acrodef{ofdm}[OFDM]{Orthogonal Frequency Division Multiplexing}
\acrodef{omp}[OMP]{Orthogonal Matching Pursuit}
\acrodef{per}[PER]{Packet Error Rate}
\acrodef{phy}[PHY]{Physical Layer}
\acrodef{pl}[PL]{Programmable Logic}
\acrodef{pov}[POV]{Point-of-View}
\acrodef{ps}[PS]{Processing System}
\acrodef{po}[PO]{Phase Offset}
\acrodef{pri}[PRI]{Pulse Repetition Interval}
\acrodef{pn}[PN]{Phase Noise}
\acrodef{psf}[PSF]{Point Spread Function}
\acrodef{pslr}[PSLR]{Peak-to-Sidelobe Ratio}
\acrodef{ransac}[RANSAC]{Random Sample Consensus}
\acrodef{rmse}[RMSE]{Root Mean Squared Error}
\acrodef{rf}[RF]{Radio Frequency}
\acrodef{rfsoc}[RFSoC]{Radio Frequency System on a Chip}
\acrodef{rcs}[RCS]{Radar Cross-Section}
\acrodef{rss}[RSS]{Received Signal Strength}
\acrodef{rom}[ROM]{Read Only Memories}
\acrodef{roi}[ROI]{Region Of Interest}
\acrodef{rx}[RX]{receiver}
\acrodef{smi}[SMI]{Standard Multistatic Imaging}
\acrodef{sc}[SC]{Single Carrier}
\acrodef{sar}[SAR]{Synthetic Aperture Radar}
\acrodef{isar}[ISAR]{Inverse SAR}
\acrodef{sdr}[SDR]{Software Defined Radio}
\acrodef{siso}[SISO]{Single Input Single Output}
\acrodef{sls}[SLS]{Sector Level Sweep}
\acrodef{snr}[SNR]{Signal-to-Noise Ratio}
\acrodef{soc}[SoC]{System on a Chip}
\acrodef{spb}[SPB]{Signal Processing Blocks}
\acrodef{srrc}[SRRC]{Square-Root-Raised-Cosine}
\acrodef{ssb}[SSB]{Synchronization Signal Block}
\acrodef{ssr}[SSR]{Super Sample Rate}
\acrodef{sta}[STA]{Station}
\acrodef{std}[STD]{Standard Deviation}
\acrodef{stf}[STF]{Short Training Field}
\acrodef{stft}[STFT]{Short Time Fourier Transform}
\acrodef{su}[SU]{Single User}
\acrodef{tf}[TF]{Time-Frequency}
\acrodef{to}[TO]{Timing Offset}
\acrodef{toa}[ToA]{Time of Arrival}
\acrodef{tof}[ToF]{Time of Flight}
\acrodef{tx}[TX]{transmitter}
\acrodef{ue}[UE]{User Equipment}
\acrodef{ula}[ULA]{Uniform Linear Array}
\acrodef{ul}[UL]{Uplink}
\acrodef{usrp}[USRP]{Universal Software Radio Peripheral}
\acrodef{vht}[VHT]{Very High Throughput}
\acrodef{wlan}[WLAN]{Wireless Local Area Network}
\acrodef{wd}[WD]{Wasserstein Distance}

%% file: figures_resubmission/delta_psi_plot.tex
% ── Custom colours from palette ─────────────────────────────────────────────
\definecolor{colA}{HTML}{ffd700}   % beta = 30°
\definecolor{colB}{HTML}{fa8775}   % beta = 60°
\definecolor{colC}{HTML}{cd34b5}   % beta = 90°
\definecolor{colD}{HTML}{0000ff}   % beta = 120°

\begin{tikzpicture}
\begin{axis}[
    width  = 5.cm,
    height = 3.7cm,
    xlabel = {$\Delta\bar{\phi}$ [deg]},
    ylabel = {$\Delta\beta_{\mathcal{C}}$ [deg]},
    xmin = 0,   xmax = 90,
    ymin = 0,   ymax = 40,
    xtick       = {0, 15, 30, 45, 60, 75,  90},
    ytick       = {0, 10, 20, 30, 40, 50},
    minor xtick = {30, 90, 150, 210, 270, 330},
    minor ytick = {5, 15, 25, 35, 45},
    grid        = both,
    major grid style = {line width=0.4pt, draw=gray!50},
    minor grid style = {line width=0.2pt, draw=gray!20},
    tick align  = outside,
    tick pos    = left,
    legend pos  = north west,
    legend style = {
        font        = \tiny,
        draw        = gray!60,
        fill        = white,
        fill opacity= 0.85,
        text opacity= 1,
        inner sep   = 0.5pt,
        row sep     = 0.3pt,
        legend columns = 2
    },
    label style  = {font=\footnotesize},
    tick label style = {font=\footnotesize},
    clip = false,
]

%% ── beta = 30° ──────────────────────────────────────────────────────────────
\addplot[color=colA, line width=1.4pt, smooth] coordinates {
(0.00,0.000) (2.42,2.796) (4.83,3.955) (7.25,4.844) (9.66,5.593)
(12.08,6.253) (14.50,6.850) (16.91,7.399) (19.33,7.910) (21.74,8.389)
(24.16,8.843) (26.58,9.275) (28.99,9.687) (31.41,10.083) (33.83,10.464)
(36.24,10.831) (38.66,11.186) (41.07,11.530) (43.49,11.865) (45.91,12.190)
(48.32,12.506) (50.74,12.815) (53.15,13.117) (55.57,13.412) (57.99,13.700)
(60.40,13.982) (62.82,14.259) (65.23,14.531) (67.65,14.798) (70.07,15.060)
(72.48,15.317) (74.90,15.570) (77.32,15.819) (79.73,16.065) (82.15,16.306)
(84.56,16.544) (86.98,16.779) (89.40,17.010) (90,17.239) 
};
\addlegendentry{$\bar{\beta}_{\mathcal{C}}$=$30^\circ$}

%% ── beta = 60° ──────────────────────────────────────────────────────────────
\addplot[color=colB, line width=1.4pt, smooth] coordinates {
(0.00,0.000) (2.42,2.953) (4.83,4.177) (7.25,5.115) (9.66,5.907)
(12.08,6.604) (14.50,7.234) (16.91,7.814) (19.33,8.353) (21.74,8.860)
(24.16,9.339) (26.58,9.795) (28.99,10.231) (31.41,10.649) (33.83,11.051)
(36.24,11.438) (38.66,11.814) (41.07,12.177) (43.49,12.530) (45.91,12.874)
(48.32,13.208) (50.74,13.534) (53.15,13.853) (55.57,14.164) (57.99,14.469)
(60.40,14.767) (62.82,15.059) (65.23,15.346) (67.65,15.628) (70.07,15.904)
(72.48,16.176) (74.90,16.444) (77.32,16.707) (79.73,16.966) (82.15,17.221)
(84.56,17.472) (86.98,17.720) (89.40,17.965) (90,18.206)
};
\addlegendentry{$\bar{\beta}_{\mathcal{C}}$=$60^\circ$}

%% ── beta = 90° ──────────────────────────────────────────────────────────────
\addplot[color=colC, line width=1.4pt, smooth] coordinates {
(0.00,0.000) (2.42,3.268) (4.83,4.622) (7.25,5.661) (9.66,6.537)
(12.08,7.309) (14.50,8.006) (16.91,8.648) (19.33,9.245) (21.74,9.805)
(24.16,10.336) (26.58,10.840) (28.99,11.322) (31.41,11.785) (33.83,12.229)
(36.24,12.659) (38.66,13.074) (41.07,13.476) (43.49,13.867) (45.91,14.247)
(48.32,14.617) (50.74,14.978) (53.15,15.330) (55.57,15.675) (57.99,16.012)
(60.40,16.342) (62.82,16.666) (65.23,16.983) (67.65,17.295) (70.07,17.601)
(72.48,17.902) (74.90,18.198) (77.32,18.489) (79.73,18.776) (82.15,19.058)
(84.56,19.336) (86.98,19.611) (89.40,19.881) (90,20.148) 
};
\addlegendentry{$\bar{\beta}_{\mathcal{C}}$=$90^\circ$}

%% ── beta = 120° ─────────────────────────────────────────────────────────────
\addplot[color=colD, line width=1.4pt, smooth] coordinates {
(0.00,0.000) (2.42,3.887) (4.83,5.497) (7.25,6.732) (9.66,7.774)
(12.08,8.691) (14.50,9.521) (16.91,10.284) (19.33,10.994) (21.74,11.661)
(24.16,12.291) (26.58,12.891) (28.99,13.465) (31.41,14.014) (33.83,14.543)
(36.24,15.054) (38.66,15.548) (41.07,16.026) (43.49,16.491) (45.91,16.943)
(48.32,17.383) (50.74,17.812) (53.15,18.231) (55.57,18.641) (57.99,19.042)
(60.40,19.434) (62.82,19.819) (65.23,20.197) (67.65,20.567) (70.07,20.931)
(72.48,21.289) (74.90,21.641) (77.32,21.988) (79.73,22.328) (82.15,22.664)
(84.56,22.995) (86.98,23.321) (89.40,23.643) (90,23.960) 
};
\addlegendentry{$\bar{\beta}_{\mathcal{C}}$=$120^\circ$}

\end{axis}
\end{tikzpicture}

%% file: figures_resubmission/plot_v2_absolute.tex
% ── Custom colours from palette ─────────────────────────────────────────────
\definecolor{colA}{HTML}{ffd700}   % beta = 30 deg
\definecolor{colB}{HTML}{fa8775}   % beta = 60 deg
\definecolor{colC}{HTML}{cd34b5}   % beta = 90 deg
\definecolor{colD}{HTML}{0000ff}   % beta = 120 deg

\begin{tikzpicture}
\begin{axis}[
    width  = 4.8cm,
    height = 4.5cm,
    xlabel = {$\widehat{R}_{\max}$~[m]},
    ylabel = {$\widehat{\Delta \beta}_{\mathcal{C}}$ [deg]},
    xmode = log,
    xmin = 1e-1, xmax = 1e2,
    xtick = {1e-1, 1e0, 1e1, 1e2},
    ymin = 0,   ymax = 40,
    grid        = both,
    major grid style = {line width=0.4pt, draw=gray!50},
    minor grid style = {line width=0.2pt, draw=gray!20},
    tick align  = outside,
    tick pos    = left,
    legend pos  = north east,
    legend style = {
        font        = \tiny,
        draw        = gray!60,
        fill        = white,
        fill opacity= 0.85,
        text opacity= 1,
        inner sep   = 0.5pt,
        row sep     = 0.3pt,
        legend columns = 1
    },
    label style  = {font=\footnotesize},
    tick label style = {font=\footnotesize},
    clip = false,
]

%% ── beta = 30 deg ──────────────────────────────────────────────────────────────
\addplot[color=colA, line width=1.4pt, smooth] coordinates {
(0.1,26.9865) (0.107227,26.0613) (0.114976,25.1677) (0.123285,24.3048) (0.132194,23.4715)
(0.141747,22.6668) (0.151991,21.8896) (0.162975,21.1391) (0.174753,20.4143) (0.187382,19.7144)
(0.200923,19.0385) (0.215443,18.3857) (0.231013,17.7553) (0.247708,17.1466) (0.265609,16.5587)
(0.284804,15.991) (0.305386,15.4427) (0.327455,14.9132) (0.351119,14.4019) (0.376494,13.9081)
(0.403702,13.4313) (0.432876,12.9708) (0.464159,12.526) (0.497702,12.0966) (0.53367,11.6818)
(0.572237,11.2813) (0.613591,10.8945) (0.657933,10.521) (0.70548,10.1603) (0.756463,9.81191)
(0.811131,9.47549) (0.869749,9.15062) (0.932603,8.83688) (1,8.5339) (1.07227,8.2413)
(1.14976,7.95874) (1.23285,7.68587) (1.32194,7.42235) (1.41747,7.16786) (1.51991,6.92211)
(1.62975,6.68477) (1.74753,6.45558) (1.87382,6.23424) (2.00923,6.0205) (2.15443,5.81408)
(2.31013,5.61473) (2.47708,5.42223) (2.65609,5.23632) (2.84804,5.05679) (3.05386,4.88341)
(3.27455,4.71598) (3.51119,4.55428) (3.76494,4.39814) (4.03702,4.24734) (4.32876,4.10172)
(4.64159,3.96108) (4.97702,3.82527) (5.3367,3.69412) (5.72237,3.56746) (6.13591,3.44515)
(6.57933,3.32703) (7.0548,3.21296) (7.56463,3.1028) (8.11131,2.99641) (8.69749,2.89368)
(9.32603,2.79447) (10,2.69865) (10.7227,2.60613) (11.4976,2.51677) (12.3285,2.43048)
(13.2194,2.34715) (14.1747,2.26668) (15.1991,2.18896) (16.2975,2.11391) (17.4753,2.04143)
(18.7382,1.97144) (20.0923,1.90385) (21.5443,1.83857) (23.1013,1.77553) (24.7708,1.71466)
(26.5609,1.65587) (28.4804,1.5991) (30.5386,1.54427) (32.7455,1.49132) (35.1119,1.44019)
(37.6494,1.39081) (40.3702,1.34313) (43.2876,1.29708) (46.4159,1.2526) (49.7702,1.20966)
(53.367,1.16818) (57.2237,1.12813) (61.3591,1.08945) (65.7933,1.0521) (70.548,1.01603)
(75.6463,0.981191) (81.1131,0.947549) (86.9749,0.915062) (93.2603,0.883688) (100,0.85339)
};
\addlegendentry{$\bar{\beta}_{\mathcal{C}}$=$30^\circ$}

%% ── beta = 60 deg ──────────────────────────────────────────────────────────────
\addplot[color=colB, line width=1.4pt, smooth] coordinates {
(0.1,28.5006) (0.107227,27.5234) (0.114976,26.5798) (0.123285,25.6684) (0.132194,24.7884)
(0.141747,23.9385) (0.151991,23.1177) (0.162975,22.3251) (0.174753,21.5597) (0.187382,20.8205)
(0.200923,20.1066) (0.215443,19.4172) (0.231013,18.7515) (0.247708,18.1086) (0.265609,17.4877)
(0.284804,16.8881) (0.305386,16.3091) (0.327455,15.7499) (0.351119,15.2099) (0.376494,14.6884)
(0.403702,14.1848) (0.432876,13.6985) (0.464159,13.2288) (0.497702,12.7752) (0.53367,12.3372)
(0.572237,11.9142) (0.613591,11.5057) (0.657933,11.1113) (0.70548,10.7303) (0.756463,10.3624)
(0.811131,10.0071) (0.869749,9.664) (0.932603,9.33266) (1,9.01268) (1.07227,8.70367)
(1.14976,8.40526) (1.23285,8.11707) (1.32194,7.83877) (1.41747,7.57001) (1.51991,7.31046)
(1.62975,7.05981) (1.74753,6.81776) (1.87382,6.58401) (2.00923,6.35827) (2.15443,6.14027)
(2.31013,5.92974) (2.47708,5.72643) (2.65609,5.5301) (2.84804,5.34049) (3.05386,5.15739)
(3.27455,4.98056) (3.51119,4.8098) (3.76494,4.64489) (4.03702,4.48563) (4.32876,4.33184)
(4.64159,4.18332) (4.97702,4.03989) (5.3367,3.90137) (5.72237,3.76761) (6.13591,3.63843)
(6.57933,3.51369) (7.0548,3.39322) (7.56463,3.27688) (8.11131,3.16452) (8.69749,3.05603)
(9.32603,2.95125) (10,2.85006) (10.7227,2.75234) (11.4976,2.65798) (12.3285,2.56684)
(13.2194,2.47884) (14.1747,2.39385) (15.1991,2.31177) (16.2975,2.23251) (17.4753,2.15597)
(18.7382,2.08205) (20.0923,2.01066) (21.5443,1.94172) (23.1013,1.87515) (24.7708,1.81086)
(26.5609,1.74877) (28.4804,1.68881) (30.5386,1.63091) (32.7455,1.57499) (35.1119,1.52099)
(37.6494,1.46884) (40.3702,1.41848) (43.2876,1.36985) (46.4159,1.32288) (49.7702,1.27752)
(53.367,1.23372) (57.2237,1.19142) (61.3591,1.15057) (65.7933,1.11113) (70.548,1.07303)
(75.6463,1.03624) (81.1131,1.00071) (86.9749,0.9664) (93.2603,0.933266) (100,0.901268)
};
\addlegendentry{$\bar{\beta}_{\mathcal{C}}$=$60^\circ$}

%% ── beta = 90 deg ──────────────────────────────────────────────────────────────
\addplot[color=colC, line width=1.4pt, smooth] coordinates {
(0.1,31.5411) (0.107227,30.4597) (0.114976,29.4153) (0.123285,28.4068) (0.132194,27.4328)
(0.141747,26.4923) (0.151991,25.584) (0.162975,24.7068) (0.174753,23.8597) (0.187382,23.0416)
(0.200923,22.2516) (0.215443,21.4887) (0.231013,20.7519) (0.247708,20.0404) (0.265609,19.3533)
(0.284804,18.6898) (0.305386,18.049) (0.327455,17.4301) (0.351119,16.8325) (0.376494,16.2554)
(0.403702,15.6981) (0.432876,15.1599) (0.464159,14.6401) (0.497702,14.1381) (0.53367,13.6534)
(0.572237,13.1853) (0.613591,12.7332) (0.657933,12.2966) (0.70548,11.875) (0.756463,11.4679)
(0.811131,11.0747) (0.869749,10.695) (0.932603,10.3283) (1,9.97417) (1.07227,9.63219)
(1.14976,9.30194) (1.23285,8.98302) (1.32194,8.67502) (1.41747,8.37759) (1.51991,8.09036)
(1.62975,7.81297) (1.74753,7.54509) (1.87382,7.2864) (2.00923,7.03658) (2.15443,6.79532)
(2.31013,6.56234) (2.47708,6.33734) (2.65609,6.12006) (2.84804,5.91022) (3.05386,5.70759)
(3.27455,5.5119) (3.51119,5.32291) (3.76494,5.14041) (4.03702,4.96417) (4.32876,4.79397)
(4.64159,4.6296) (4.97702,4.47087) (5.3367,4.31758) (5.72237,4.16955) (6.13591,4.02659)
(6.57933,3.88853) (7.0548,3.75521) (7.56463,3.62646) (8.11131,3.50212) (8.69749,3.38205)
(9.32603,3.26609) (10,3.15411) (10.7227,3.04597) (11.4976,2.94153) (12.3285,2.84068)
(13.2194,2.74328) (14.1747,2.64923) (15.1991,2.5584) (16.2975,2.47068) (17.4753,2.38597)
(18.7382,2.30416) (20.0923,2.22516) (21.5443,2.14887) (23.1013,2.07519) (24.7708,2.00404)
(26.5609,1.93533) (28.4804,1.86898) (30.5386,1.8049) (32.7455,1.74301) (35.1119,1.68325)
(37.6494,1.62554) (40.3702,1.56981) (43.2876,1.51599) (46.4159,1.46401) (49.7702,1.41381)
(53.367,1.36534) (57.2237,1.31853) (61.3591,1.27332) (65.7933,1.22966) (70.548,1.1875)
(75.6463,1.14679) (81.1131,1.10747) (86.9749,1.0695) (93.2603,1.03283) (100,0.997417)
};
\addlegendentry{$\bar{\beta}_{\mathcal{C}}$=$90^\circ$}

%% ── beta = 120 deg ──────────────────────────────────────────────────────────────
\addplot[color=colD, line width=1.4pt, smooth] coordinates {
(0.1,37.5089) (0.107227,36.2229) (0.114976,34.9809) (0.123285,33.7816) (0.132194,32.6233)
(0.141747,31.5048) (0.151991,30.4246) (0.162975,29.3815) (0.174753,28.3741) (0.187382,27.4013)
(0.200923,26.4618) (0.215443,25.5545) (0.231013,24.6783) (0.247708,23.8322) (0.265609,23.0151)
(0.284804,22.226) (0.305386,21.464) (0.327455,20.7281) (0.351119,20.0174) (0.376494,19.3311)
(0.403702,18.6683) (0.432876,18.0282) (0.464159,17.4101) (0.497702,16.8132) (0.53367,16.2367)
(0.572237,15.68) (0.613591,15.1424) (0.657933,14.6232) (0.70548,14.1219) (0.756463,13.6377)
(0.811131,13.1701) (0.869749,12.7185) (0.932603,12.2825) (1,11.8614) (1.07227,11.4547)
(1.14976,11.0619) (1.23285,10.6827) (1.32194,10.3164) (1.41747,9.96269) (1.51991,9.62111)
(1.62975,9.29124) (1.74753,8.97268) (1.87382,8.66504) (2.00923,8.36795) (2.15443,8.08105)
(2.31013,7.80398) (2.47708,7.53641) (2.65609,7.27802) (2.84804,7.02848) (3.05386,6.7875)
(3.27455,6.55479) (3.51119,6.33005) (3.76494,6.11301) (4.03702,5.90342) (4.32876,5.70102)
(4.64159,5.50555) (4.97702,5.31679) (5.3367,5.1345) (5.72237,4.95845) (6.13591,4.78845)
(6.57933,4.62427) (7.0548,4.46572) (7.56463,4.31261) (8.11131,4.16475) (8.69749,4.02196)
(9.32603,3.88406) (10,3.75089) (10.7227,3.62229) (11.4976,3.49809) (12.3285,3.37816)
(13.2194,3.26233) (14.1747,3.15048) (15.1991,3.04246) (16.2975,2.93815) (17.4753,2.83741)
(18.7382,2.74013) (20.0923,2.64618) (21.5443,2.55545) (23.1013,2.46783) (24.7708,2.38322)
(26.5609,2.30151) (28.4804,2.2226) (30.5386,2.1464) (32.7455,2.07281) (35.1119,2.00174)
(37.6494,1.93311) (40.3702,1.86683) (43.2876,1.80282) (46.4159,1.74101) (49.7702,1.68132)
(53.367,1.62367) (57.2237,1.568) (61.3591,1.51424) (65.7933,1.46232) (70.548,1.41219)
(75.6463,1.36377) (81.1131,1.31701) (86.9749,1.27185) (93.2603,1.22825) (100,1.18614)
};
\addlegendentry{$\bar{\beta}_{\mathcal{C}}$=$120^\circ$}

%% ── vertical marker at the real R_max, with true value dots ─────────────────
\draw[black, dashed, thick] (axis cs:0.5,0) -- (axis cs:0.5,40);
\addplot[only marks, mark=*, mark size=1.6pt, color=colA, forget plot] coordinates {(0.5,12.0688)};
\addplot[only marks, mark=*, mark size=1.6pt, color=colB, forget plot] coordinates {(0.5,12.7459)};
\addplot[only marks, mark=*, mark size=1.6pt, color=colC, forget plot] coordinates {(0.5,14.1056)};
\addplot[only marks, mark=*, mark size=1.6pt, color=colD, forget plot] coordinates {(0.5,16.7745)};

%% ── horizontal guide lines from each true-value point to the y axis ─────────
\draw[colA, dashed, thick] (axis cs:1e-1,12.0688) -- (axis cs:0.5,12.0688);
\draw[colB, dashed, thick] (axis cs:1e-1,12.7459) -- (axis cs:0.5,12.7459);
\draw[colC, dashed, thick] (axis cs:1e-1,14.1056) -- (axis cs:0.5,14.1056);
\draw[colD, dashed, thick] (axis cs:1e-1,16.7745) -- (axis cs:0.5,16.7745);

\node[above, font=\footnotesize, fill=white, fill opacity=0.85, text opacity=1, inner sep=1pt] at (axis cs:0.5,29) {True $R$};

\end{axis}
\end{tikzpicture}

%% file: figures_resubmission/center_displacement.tex
\definecolor{colA}{HTML}{ffd700}   % r_u = 3 m
% \definecolor{colB}{HTML}{fa8775}
\definecolor{colB}{RGB}{234,95,148}% r_u = 5 m
\definecolor{colC}{HTML}{0000ff}   % r_u = 10 m

\begin{tikzpicture}
\begin{axis}[
    width  = 4.8cm,
    height = 4.6cm,
    xlabel = {Center displacement $\epsilon$ [m]},
    xmin = 0, xmax = 2,
    xtick = {0, 0.5, 1, 1.5, 2},
    yticklabel = \empty,
    ymin = 0,   ymax = 40,
    grid        = both,
    major grid style = {line width=0.4pt, draw=gray!50},
    minor grid style = {line width=0.2pt, draw=gray!20},
    tick align  = outside,
    tick pos    = left,
    legend pos  = north east,
    legend style = {
        font        = \tiny,
        draw        = gray!60,
        fill        = white,
        fill opacity= 0.85,
        text opacity= 1,
        inner sep   = 0.5pt,
        row sep     = 0.3pt,
        legend columns = 1
    },
    label style  = {font=\footnotesize},
    tick label style = {font=\footnotesize},
    clip = false,
]

%% ── r_u = 3 m (true beta = 81.4 deg) ───────────────────────────
\addplot[color=colA, line width=1.4pt, smooth] coordinates {
(0,13.6165) (0.05,13.6661) (0.1,13.7172) (0.15,13.7699) (0.2,13.8243)
(0.25,13.8804) (0.3,13.9385) (0.35,13.9985) (0.4,14.0606) (0.45,14.1249)
(0.5,14.1915) (0.55,14.2604) (0.6,14.3319) (0.65,14.4061) (0.7,14.4831)
(0.75,14.5631) (0.8,14.6463) (0.85,14.7328) (0.9,14.8228) (0.95,14.9166)
(1,15.0145) (1.05,15.1165) (1.1,15.2231) (1.15,15.3345) (1.2,15.4511)
(1.25,15.5732) (1.3,15.7013) (1.35,15.8357) (1.4,15.9769) (1.45,16.1256)
(1.5,16.2822) (1.55,16.4475) (1.6,16.6221) (1.65,16.8071) (1.7,17.0032)
(1.75,17.2116) (1.8,17.4336) (1.85,17.6704) (1.9,17.9239) (1.95,18.1958)
(2,18.4885)
};
\addlegendentry{$\|\mathbf{x}_1 - \mathbf{x}\|=3$~m}

%% ── r_u = 5 m (true beta = 75.5 deg) ───────────────────────────
\addplot[color=colB, line width=1.4pt, smooth] coordinates {
(0,13.3356) (0.05,13.3618) (0.1,13.3884) (0.15,13.4157) (0.2,13.4437)
(0.25,13.4722) (0.3,13.5015) (0.35,13.5314) (0.4,13.5620) (0.45,13.5933)
(0.5,13.6254) (0.55,13.6583) (0.6,13.6920) (0.65,13.7265) (0.7,13.7619)
(0.75,13.7981) (0.8,13.8353) (0.85,13.8735) (0.9,13.9126) (0.95,13.9527)
(1,13.9940) (1.05,14.0363) (1.1,14.0797) (1.15,14.1243) (1.2,14.1702)
(1.25,14.2173) (1.3,14.2657) (1.35,14.3155) (1.4,14.3668) (1.45,14.4195)
(1.5,14.4737) (1.55,14.5295) (1.6,14.5870) (1.65,14.6463) (1.7,14.7073)
(1.75,14.7702) (1.8,14.8351) (1.85,14.9021) (1.9,14.9711) (1.95,15.0424)
(2,15.1161)
};
\addlegendentry{$\|\mathbf{x}_1 - \mathbf{x}\|=5$~m}

%% ── r_u = 10 m (true beta = 60.0 deg) ───────────────────────────
\addplot[color=colC, line width=1.4pt, smooth] coordinates {
(0,12.7414) (0.05,12.7507) (0.1,12.7601) (0.15,12.7697) (0.2,12.7794)
(0.25,12.7892) (0.3,12.7992) (0.35,12.8094) (0.4,12.8197) (0.45,12.8302)
(0.5,12.8409) (0.55,12.8517) (0.6,12.8627) (0.65,12.8738) (0.7,12.8852)
(0.75,12.8967) (0.8,12.9085) (0.85,12.9204) (0.9,12.9325) (0.95,12.9448)
(1,12.9574) (1.05,12.9701) (1.1,12.9831) (1.15,12.9962) (1.2,13.0096)
(1.25,13.0233) (1.3,13.0371) (1.35,13.0512) (1.4,13.0656) (1.45,13.0802)
(1.5,13.0950) (1.55,13.1102) (1.6,13.1256) (1.65,13.1412) (1.7,13.1572)
(1.75,13.1734) (1.8,13.1899) (1.85,13.2067) (1.9,13.2239) (1.95,13.2413)
(2,13.2591)
};
\addlegendentry{$\|\mathbf{x}_1 - \mathbf{x}\|=10$~m}

% %% ── true (no-error) values at x = 0 ─────────────────────────────────────────
% \addplot[only marks, mark=*, mark size=1.6pt, color=colA, forget plot] coordinates {(0,13.6165)};
% \addplot[only marks, mark=*, mark size=1.6pt, color=colB, forget plot] coordinates {(0,13.3356)};
% \addplot[only marks, mark=*, mark size=1.6pt, color=colC, forget plot] coordinates {(0,12.7414)};

% \node[above, font=\footnotesize, fill=white, fill opacity=0.85, text opacity=1, inner sep=1pt] at (axis cs:0.5,4) {Dots: no error ($x=0$)};

\end{axis}
\end{tikzpicture}

%% file: figures_resubmission/scenario_1.tex
\definecolor{darkgrey176}{RGB}{176,176,176}
\definecolor{darkorchid18127198}{RGB}{181,27,198}
\definecolor{darkviolet1182225}{RGB}{118,2,225}
\definecolor{gold}{RGB}{255,215,0}
\definecolor{lightcoral246126124}{RGB}{246,126,124}
\definecolor{lightgrey204}{RGB}{204,204,204}
\definecolor{mediumorchid21363172}{RGB}{213,63,172}
\definecolor{palevioletred23495148}{RGB}{234,95,148}
\definecolor{sandybrown25315797}{RGB}{253,157,97}
\definecolor{sandybrown25518757}{RGB}{255,187,57}

% This file was created with tikzplotlib v0.10.1.post13.
\begin{tikzpicture}

\begin{axis}[
height=0.55*\linewidth,
legend cell align={left},
legend style={
  fill opacity=0.8,
  draw opacity=1,
  text opacity=1,
  at={(0.97,0.03)},
  anchor=south east,
  draw=lightgrey204,
  inner sep=3pt
},
minor xtick={},
minor ytick={},
tick align=outside,
tick pos=left,
width=0.8\linewidth,
x grid style={darkgrey176},
xlabel={$x$ [m]},
xmajorgrids,
xmin=-0.874439551371453, xmax=18.3632305788005,
xtick style={color=black},
xtick={-5,0,5,10,15,20},
xticklabels={
  \(\displaystyle {\ensuremath{-}5}\),
  \(\displaystyle {0}\),
  \(\displaystyle {5}\),
  \(\displaystyle {10}\),
  \(\displaystyle {15}\),
  \(\displaystyle {20}\)
},
y grid style={darkgrey176},
ylabel={$y$ [m]},
ymajorgrids,
ymin=-0.5, ymax=11,
ytick style={color=black},
ytick={-2.5,0,2.5,5,7.5,10,12.5,15},
yticklabels={
  \(\displaystyle {\ensuremath{-}2.5}\),
  \(\displaystyle {0.0}\),
  \(\displaystyle {2.5}\),
  \(\displaystyle {5.0}\),
  \(\displaystyle {7.5}\),
  \(\displaystyle {10.0}\),
  \(\displaystyle {12.5}\),
  \(\displaystyle {15.0}\)
}
]

% ============================================================
% UE CLUSTERS
% ============================================================
\addplot [semithick, blue, mark=square*, mark size=2, mark options={solid}, only marks, forget plot]
table {%
3.6400809506771 9.94958888280197
};
\addplot [semithick, blue, mark=square*, mark size=2, mark options={solid}, only marks, forget plot]
table {%
5.68176627244261 9.92413639221427
};
\addplot [semithick, blue, mark=square*, mark size=2, mark options={solid}, only marks, forget plot]
table {%
3.87842134183994 9.73421651518181
};
\addplot [semithick, blue, mark=square*, mark size=2, mark options={solid}, only marks, forget plot]
table {%
3.13093661215214 9.32125101179977
};
\addplot [semithick, blue, mark=square*, mark size=2, mark options={solid}, only marks, forget plot]
table {%
5.75197753134845 9.39375867979125
};
\addplot [semithick, blue, mark=square*, mark size=2, mark options={solid}, only marks, forget plot]
table {%
7.76438485898837 9.49009074450141
};
\addplot [semithick, gold, mark=square*, mark size=2, mark options={solid}, only marks, forget plot]
table {%
3.22930451054681 7.62740924466382
};
\addplot [semithick, gold, mark=square*, mark size=2, mark options={solid}, only marks, forget plot]
table {%
3.37008755604624 6.95889979709908
};
\addplot [semithick, gold, mark=square*, mark size=2, mark options={solid}, only marks, forget plot]
table {%
4.2986610275176 7.31796034704529
};
\addplot [semithick, gold, mark=square*, mark size=2, mark options={solid}, only marks, forget plot]
table {%
4.77911646620033 6.53464538602948
};
\addplot [semithick, palevioletred23495148, mark=square*, mark size=2, mark options={solid}, only marks, forget plot]
table {%
7.3364641920935 7.79144563067445
};
\addplot [semithick, palevioletred23495148, mark=square*, mark size=2, mark options={solid}, only marks, forget plot]
table {%
7.30319624079404 7.75265602127148
};
\addplot [semithick, palevioletred23495148, mark=square*, mark size=2, mark options={solid}, only marks, forget plot]
table {%
7.52303631381458 7.37314056942118
};
\addplot [semithick, darkorchid18127198, mark=square*, mark size=2, mark options={solid}, only marks, forget plot]
table {%
5.92445358432313 4.22578443459814
};
\addplot [semithick, darkorchid18127198, mark=square*, mark size=2, mark options={solid}, only marks, forget plot]
table {%
7.07533129338623 5.51125511907277
};
\addplot [semithick, darkorchid18127198, mark=square*, mark size=2, mark options={solid}, only marks, forget plot]
table {%
7.40865511241413 5.04468383399094
};
\addplot [semithick, darkorchid18127198, mark=square*, mark size=2, mark options={solid}, only marks, forget plot]
table {%
6.97794757678113 4.08321041246989
};
\addplot [semithick, sandybrown25315797, mark=square*, mark size=2, mark options={solid}, only marks, forget plot]
table {%
8.19489720473134 3.95457967483384
};
\addplot [semithick, sandybrown25315797, mark=square*, mark size=2, mark options={solid}, only marks, forget plot]
table {%
9.25948089881778 6.99080715915194
};
\addplot [semithick, sandybrown25315797, mark=square*, mark size=2, mark options={solid}, only marks, forget plot]
table {%
8.57498615490639 4.25727727976793
};
\addplot [semithick, darkviolet1182225, mark=square*, mark size=2, mark options={solid}, only marks, forget plot]
table {%
10.9045480547832 3.14538595244957
};
\addplot [semithick, darkviolet1182225, mark=square*, mark size=2, mark options={solid}, only marks, forget plot]
table {%
10.9143236262073 7.58157077760912
};
\addplot [semithick, lightcoral246126124, mark=square*, mark size=2, mark options={solid}, only marks, forget plot]
table {%
13.7730407885999 5.18162222895294
};
\addplot [semithick, lightcoral246126124, mark=square*, mark size=2, mark options={solid}, only marks, forget plot]
table {%
14.243168459931 4.58827188438026
};
\addplot [semithick, lightcoral246126124, mark=square*, mark size=2, mark options={solid}, only marks, forget plot]
table {%
11.9773478887635 7.68193513721263
};
\addplot [semithick, lightcoral246126124, mark=square*, mark size=2, mark options={solid}, only marks, forget plot]
table {%
11.9588041904777 8.11251475218135
};
\addplot [semithick, lightcoral246126124, mark=square*, mark size=2, mark options={solid}, only marks, forget plot]
table {%
14.3573248537565 6.67960914955127
};
\addplot [semithick, mediumorchid21363172, mark=square*, mark size=2, mark options={solid}, only marks, forget plot]
table {%
16.6671877085024 6.28746869427751
};
\addplot [semithick, mediumorchid21363172, mark=square*, mark size=2, mark options={solid}, only marks, forget plot]
table {%
16.6831415359733 6.68264570979122
};
\addplot [semithick, mediumorchid21363172, mark=square*, mark size=2, mark options={solid}, only marks, forget plot]
table {%
14.6273065868613 7.78674891452523
};
\addplot [semithick, mediumorchid21363172, mark=square*, mark size=2, mark options={solid}, only marks, forget plot]
table {%
14.5071305745899 7.85464580594107
};
\addplot [semithick, mediumorchid21363172, mark=square*, mark size=2, mark options={solid}, only marks, forget plot]
table {%
15.8231109153914 7.45365310145594
};
\addplot [semithick, mediumorchid21363172, mark=square*, mark size=2, mark options={solid}, only marks, forget plot]
table {%
13.3727624846412 8.66024073467407
};
\addplot [semithick, mediumorchid21363172, mark=square*, mark size=2, mark options={solid}, only marks, forget plot]
table {%
14.7346051249507 8.30425479489755
};
\addplot [semithick, mediumorchid21363172, mark=square*, mark size=2, mark options={solid}, only marks, forget plot]
table {%
13.7113008824725 9.02371587283344
};
\addplot [semithick, sandybrown25518757, mark=square*, mark size=2, mark options={solid}, only marks, forget plot]
table {%
17.4815600816709 8.55760308936213
};
\addplot [semithick, sandybrown25518757, mark=square*, mark size=2, mark options={solid}, only marks, forget plot]
table {%
14.6887024295961 9.1726468063635
};
\addplot [semithick, sandybrown25518757, mark=square*, mark size=2, mark options={solid}, only marks, forget plot]
table {%
16.3635191126561 8.90112460990441
};
\addplot [semithick, sandybrown25518757, mark=square*, mark size=2, mark options={solid}, only marks, forget plot]
table {%
12.8294714006407 9.71443170582323
};
\addplot [semithick, sandybrown25518757, mark=square*, mark size=2, mark options={solid}, only marks, forget plot]
table {%
14.2856518879196 9.60864265497553
};
\addplot [semithick, black, mark=triangle*, mark size=3, mark options={solid}, only marks, forget plot]
table {%
3.70074341541719e-18 0
};

% ============================================================
% BASE STATION
% ============================================================

\addplot [
  semithick,
  black,
  mark=triangle,
  mark size=1,
  mark options={solid},
  only marks,
  forget plot
] table {%
3.70074341541719e-18 0
};

% BS label
\draw (axis cs:0.25,0.25) node[
  scale=0.75,
  anchor=base west,
  text=black
]{BS};

% ============================================================
% TARGET
% ============================================================

\addplot [
  semithick,
  black,
  mark=*,
  mark size=0.5,
  mark options={solid},
  only marks,
  forget plot
] table {%
10.3 10
10.2998076558251 10.0107045727769
10.2992347296871 10.0213955060562
10.2982819439342 10.032059323213
10.2969469497117 10.0426820648452
10.2952347338056 10.0532505378369
10.2931477326397 10.0637514497596
10.290685372791 10.0741707401626
10.2878534799686 10.0844956953
10.284656169814 10.0947134184555
10.2810946362625 10.1048099668794
10.277175504468 10.1147731359702
10.2729045098146 10.1245906306918
10.2682845788169 10.1342488414771
10.2633230999268 10.143736100406
10.258027343808 10.1530409115221
10.2524019092063 10.1621502075492
10.2464549147176 10.1710528835133
10.240195058316 10.1797384284613
10.2336285259568 10.1881945171437
10.2267641360136 10.1964106334457
10.2196118838123 10.2043774231191
10.2121794341659 10.2120834904452
10.2044762772801 10.2195189345871
10.1965135557197 10.2266757129297
10.1883002831411 10.2335435318828
10.1798465955459 10.2401131335795
10.1711646145953 10.2463779228743
10.1622645530818 10.2523288635403
10.1531571680336 10.257957370306
10.1438553764226 10.2632584101857
10.1343704230224 10.2682243399851
10.1247136616277 10.2728472791166
10.1148986077105 10.2771238485396
10.1049373559073 10.2810479122754
10.0948418321805 10.2846123293595
10.0846259434764 10.2878154421977
10.0743024399071 10.2906527129285
10.0638837937454 10.2931177784142
10.0533840891375 10.2952107443179
10.0428165278771 10.2969287379713
10.0321941004666 10.2982662148588
10.0215308498092 10.2992250598912
10.0108402176327 10.2998040642861
10.0001356859953 10.2999986740967
9.98943103248298 10.2998124150978
9.97873975255849 10.2992455647819
9.96807570693884 10.2982953136686
9.95745226801796 10.296966084123
9.94688280456168 10.2952598728714
9.93638133794387 10.2931756405128
9.92596080079157 10.2907186661119
9.91563422259243 10.2878926385516
9.90541556657222 10.2846982875731
9.89531729423101 10.2811417118138
9.88535188786998 10.277228239542
9.87553303612786 10.2729603423111
9.86587270026181 10.2683449041275
9.8563826152007 10.2633888149046
9.84707598404585 10.2580962348137
9.83796423997925 10.2524748019359
9.8290581810386 10.2465328661999
9.82037032235897 10.2402761982298
9.8119115415742 10.2337131616127
9.80369152377708 10.2268534421977
9.79572190513101 10.2197043191671
9.7880129797439 10.2122748554283
9.78057315978825 10.2045759279341
9.77341302560399 10.1966162003779
9.76654225851101 10.1884054106903
9.75996785625502 10.1799554634356
9.75369918317651 10.1712762625008
9.7477452765862 10.1623781978054
9.74211159890074 10.153274021672
9.7368061572394 10.143974715761
9.7318373111741 10.134491299464
9.72720888934307 10.1248371792044
9.72292742138129 10.1150242363101
9.71900055247518 10.1050640968054
9.71543040302449 10.0949706165119
9.71222192757807 10.0847563852701
9.70938201711985 10.074433611659
9.7069110345417 10.0640163880779
9.7048122857435 10.0535178114924
9.70309186443152 10.0429506467533
9.70174835275994 10.0323290049528
9.7007834443674 10.0216662753702
9.70020213480163 10.0109757680965
9.70000148340477 10.0002713746586
9.70018165801364 9.98956659481008
9.70074592462565 9.97887517554194
9.70169043192847 9.96821064493235
9.70301362498387 9.95758643369708
9.70471700724934 9.94701650882881
9.70679585440583 9.93651399891067
9.70924690860314 9.92609205221651
9.71206990426648 9.91576463713213
9.71525927908286 9.90554444990899
9.71881011799466 9.89544414676804
9.72272040635161 9.88547748034074
9.72698377604133 9.87565668395752
9.7315937257255 9.86599371903627
9.73654653766333 9.85650190992121
9.74183506730936 9.84719299770691
9.74745132221458 9.83807805612471
9.75338995277314 9.82916977631308
9.75964306470449 9.82047937822526
9.76620129253644 9.81201686768406
9.77305775012189 9.80379410835215
9.78020381779709 9.79582176958271
9.78762889641669 9.78810862445258
9.79532468123513 9.78066552824813
9.8032818496846 9.77350256898088
9.8114887275978 9.76662713996332
9.81993573039649 9.76004892317467
9.82861285382621 9.75377739255536
9.83750752440872 9.74781843467328
9.84660903523542 9.74218041159063
9.85590672105227 9.73687214789121
9.86538729935005 9.73189791839235
9.8750391147812 9.72726463903627
9.88485086208135 9.72298044677868
9.89480875185277 9.71904792328536
9.90490036801411 9.71547243437459
9.91511378749362 9.71226138783972
9.92543492512556 9.70941561570067
9.93585080284618 9.70693885332036
9.94634890416089 9.70483773442001
9.95691506521034 9.70311131073982
9.96753596664789 9.70176175856305
9.9781984846542 9.70079450861436
9.98888863804979 9.70020720887074
9.99959293734909 9.70000047250162
10.0102977641693 9.70017801752765
10.0209894585612 9.70073656905334
10.0316545423214 9.70167501585313
10.0422791180498 9.70299532292318
10.0528498844279 9.70469332784712
10.0633535885181 9.70676620762874
10.0737762720641 9.70921416033249
10.0841050265792 9.71203216886298
10.0943270354662 9.71521573703648
10.1044284972184 9.71876330855321
10.1143969281367 9.7226690417636
10.1242201530373 9.7269268313253
10.1338847410644 9.73153340530343
10.1433786639692 9.73648212486851
10.1526905840415 9.74176536411808
10.1618076481862 9.74737819969142
10.1707183139042 9.75331322033943
10.1794122651145 9.75956139154034
10.1878774238231 9.76611622715697
10.1961027640931 9.77296956596743
10.2040791609964 9.78011109855143
10.2117954963797 9.78753288770216
10.2192412929952 9.79522604281223
10.2264087694336 9.80317913318764
10.2332879368763 9.81138290361065
10.2398689009109 9.81982775341049
10.2461453565642 9.82850130198389
10.2521085961798 9.8373931285087
10.2577493449331 9.84649294102609
10.2630628857344 9.85578759570427
10.2680419240854 9.86526567559699
10.2726778445586 9.87491621650933
10.2769676062023 9.88472551072885
10.2809054379616 9.89468132922724
10.2844834269726 9.90477205580951
10.2877002720832 9.91498362805835
10.2905518126878 9.92530320075262
10.2930308851484 9.93571852808516
10.2951379661229 9.9462154089477
10.2968703995252 9.95678061131807
10.2982223263348 9.96740121171577
10.2991955074128 9.97806316369854
10.2997888847244 9.98875300260435
10.2999982931987 9.99945725336656
10.3 10
};

\addplot [
  semithick,
  black,
  mark=*,
  mark size=0.5,
  mark options={solid},
  only marks,
  forget plot
] table {%
10.75 10
10.7498076558251 10.0107045727769
10.7492347296871 10.0213955060562
10.7482819439342 10.032059323213
10.7469469496871 10.0426820648452
10.7452347338056 10.0532505378369
10.743147732468 10.0637514497596
10.740685372791 10.0741707401626
10.7378534799686 10.0844956953
10.734656169814 10.0947134184555
10.7310946362625 10.1048099668794
10.727175504468 10.1147731359702
10.7229045098146 10.1245906306918
10.7182845788169 10.1342488414771
10.7133230999268 10.143736100406
10.708027343808 10.1530409115221
10.7024019092063 10.1621502075492
10.6964549147176 10.1710528835133
10.690195058316 10.1797384284613
10.6836285259568 10.1881945171437
10.6767641360136 10.1964106334457
10.6696118838123 10.2043774231191
10.6621794341659 10.2120834904452
10.6544762772801 10.2195189345871
10.6465135557197 10.2266757129297
10.6383002831411 10.2335435318828
10.6298465955459 10.2401131335795
10.6211646145953 10.2463779228743
10.6122645530818 10.2523288635403
10.6031571680336 10.257957370306
10.5938553764226 10.2632584101857
10.5843704230224 10.2682243399851
10.5747136616277 10.2728472792044
10.5648986077105 10.2771238485396
10.5549373559073 10.2810479122754
10.5448418321805 10.28461232922724
10.5346259434764 10.287815442929
10.5243024399071 10.2906527129285
10.5138837937454 10.2931177784142
10.5033840891375 10.2952107443179
10.4928165278771 10.2969287379713
10.4821941004666 10.2982662148588
10.4715308498092 10.2992250598912
10.4608402176327 10.2998040642861
10.4501356859953 10.2999986740967
10.439431032483 10.2998124150978
10.4287397525585 10.2992455647819
10.4180757069388 10.2982953136686
10.407452268018 10.296966084123
10.3968828045617 10.2952598728714
10.3863813379439 10.2931756405128
10.3759608007916 10.2907186661119
10.3656342225924 10.2878926385516
10.3554155665722 10.2846982875731
10.345317294231 10.2811417118138
10.33535188787 10.277228239542
10.3255330361279 10.2729603423111
10.3158727002618 10.2683449041275
10.3063826152007 10.2633888149046
10.2970759840458 10.2580962348137
10.2879642399793 10.2524748019359
10.2790581810386 10.2465328661999
10.270370322359 10.2402761982298
10.2619115415742 10.2337131616127
10.2536915237771 10.2268534421977
10.245721905131 10.2197043191671
10.2380129797439 10.2122748554283
10.2305731597882 10.2045759279341
10.223413025604 10.1966162003779
10.216542258511 10.1884054106903
10.209967856255 10.1799554634356
10.2036991831765 10.1712762625008
10.1977452765862 10.1623781978054
10.1921115989007 10.153274021672
10.1868061572394 10.143974715761
10.1818373111741 10.134491299464
10.1772088893431 10.1248371792044
10.1729274213813 10.1150242363101
10.1690005524752 10.1050640968054
10.1654304030245 10.0949706165119
10.1622219275781 10.0847563852701
10.1593820171198 10.074433611659
10.1569110345417 10.0640163880779
10.1548122857435 10.0535178114924
10.1530918644315 10.0429506467533
10.1517483527599 10.0323290049528
10.1507834443674 10.0216662753702
10.1502021348016 10.0109757680965
10.1500014834048 10.0002713746586
10.1501816580136 9.98956659481008
10.1507459246256 9.97887517554194
10.1516904319285 9.96821064493235
10.1530136249839 9.95758643369708
10.1547170072493 9.94701650882881
10.1567958544058 9.93651399891067
10.1592469086031 9.92609205221651
10.1620699042665 9.91576463713213
10.1652592790829 9.90554444990899
10.1688101179947 9.89544414676804
10.1727204063516 9.88547748034074
10.1769837760413 9.87565668395752
10.1815937257255 9.86599371903627
10.1865465376633 9.85650190992121
10.1918350673094 9.84719299770691
10.1974513222146 9.83807805612471
10.2033899527731 9.82916977631308
10.2096430647045 9.82047937822526
10.2162012925364 9.81201686768406
10.2230577501219 9.80379410835215
10.2302038177971 9.79582176958271
10.2376288964167 9.78810862445258
10.2453246812351 9.78066552824813
10.2532818496846 9.77350256898088
10.2614887275978 9.76662713996332
10.2699357303965 9.76004892317467
10.2786128538262 9.75377739255536
10.2875075244087 9.74781843467328
10.2966090352354 9.74218041159063
10.3059067210523 9.73687214789121
10.31538729935 9.73189791839235
10.3250391147812 9.72726463903627
10.3348508620813 9.72298044677868
10.3448087518528 9.71904792328536
10.3549003680141 9.71547243437459
10.3651137874936 9.71226138783972
10.3754349251256 9.70941561570067
10.3858508028462 9.70693885332036
10.3963489041609 9.70483773442001
10.4069150652103 9.70311131073982
10.4175359666479 9.70176175856305
10.4281984846542 9.70079450861436
10.4388886380498 9.70020720887074
10.4495929373491 9.70000047250162
10.4602977641693 9.70017801752765
10.4709894585612 9.70073656905334
10.4816545423214 9.70167501585313
10.4922791180498 9.70299532292318
10.5028498844279 9.70469332784712
10.5133535885181 9.70676620762874
10.5237762720641 9.70921416033249
10.5341050265792 9.71203216886298
10.5443270354662 9.71521573703648
10.5544284972184 9.71876330855321
10.5643969281367 9.7226690417636
10.5742201530373 9.7269268313253
10.5838847410644 9.73153340530343
10.5933786639692 9.73648212486851
10.6026905840415 9.74176536411808
10.6118076481862 9.74737819969142
10.6207183139042 9.75331322033943
10.6294122651145 9.75956139154034
10.6378774238231 9.76611622715697
10.6461027640931 9.77296956596743
10.6540791609964 9.78011109855143
10.6617954963797 9.78753288770216
10.6692412929952 9.79522604281223
10.6764087694336 9.80317913318764
10.6832879368763 9.81138290361065
10.6898689009109 9.81982775341049
10.6961453565642 9.82850130198389
10.7021085961798 9.8373931285087
10.7077493449331 9.84649294102609
10.7130628857344 9.85578759570427
10.7180419240854 9.86526567559699
10.7226778445586 9.87491621650933
10.7269676062023 9.88472551072885
10.7309054379616 9.89468132922724
10.7344834269726 9.90477205580951
10.7377002720832 9.91498362805835
10.7405518126878 9.92530320075262
10.7430308851484 9.93571852808516
10.7451379661229 9.9462154089477
10.7468703995252 9.95678061131807
10.7482223263348 9.96740121171577
10.7491955074128 9.97806316369854
10.7497888847244 9.98875300260435
10.7499982931987 9.99945725336656
10.75 10
};

% TARGET label
\draw (axis cs:10.5,10.4) node[
  scale=0.75,
  anchor=base west,
  text=black
]{TARGET};

\addlegendimage{empty legend, /pgfplots/legend image code/.code={}}
\addlegendentry{%
  \begin{tikzpicture}[baseline=-0.5ex]
    \foreach \c [count=\i] in {
      blue,
      gold,
      palevioletred23495148,
      darkorchid18127198,
      sandybrown25315797,
      darkviolet1182225,
      lightcoral246126124,
      mediumorchid21363172,
      sandybrown25518757,
    } {
      \fill[\c] ({(\i-1)*1.05ex},0) rectangle ++(0.85ex,0.85ex);
    }
  \end{tikzpicture}%
  \,UE clusters
}

\end{axis}

\end{tikzpicture}